\documentclass{jfm}

\usepackage{amsmath}
\usepackage{amssymb}
\usepackage{graphicx}
\usepackage{psfrag}
\usepackage{epsfig}
\usepackage{subfig}
\usepackage{multirow}
\usepackage[title]{appendix}
\usepackage[normalem]{ulem}

\begin{document}

\shorttitle{Dynamic slip wall model for LES}
\shortauthor{H. J. Bae, A. Lozano-Dur\'an, S. T. Bose and P. Moin}

\title{Dynamic slip wall model for large-eddy simulation}
\author{Hyunji Jane Bae\aff{1,2} \corresp{\email{hjbae@stanford.edu}},
Adri\'an Lozano-Dur\'an\aff{1}, Sanjeeb T. Bose\aff{2,3} \and Parviz
Moin\aff{1}}

\affiliation{\aff{1}Center for Turbulence Research, Stanford University
\aff{2}Institute for Computational and Mathematical Engineering,
Stanford University
\aff{3}Cascade Technologies Inc.}

\date{\today}
\maketitle

\begin{abstract} 

Wall modelling in large-eddy simulation (LES) is necessary to overcome
the prohibitive near-wall resolution requirements in
high-Reynolds-number turbulent flows. Most existing wall models rely
on assumptions about the state of the boundary layer and require
\emph{a priori} prescription of tunable coefficients. They also impose
the predicted wall stress by replacing the no-slip boundary condition
at the wall with a Neumann boundary condition in the wall-parallel
directions while maintaining the no-transpiration condition in the
wall-normal direction. In the present study, we first
motivate and analyse the Robin (slip) boundary condition with
transpiration (nonzero wall-normal velocity) in the context of
wall-modelled LES.  The effect of the slip boundary
condition on the one-point statistics of the flow is investigated in
LES of turbulent channel flow and flat-plate turbulent boundary layer.
It is shown that the slip condition provides a framework to compensate
for the deficit or excess of mean momentum at the wall. Moreover, the
resulting nonzero stress at the wall alleviates the well-known problem
of the wall-stress under-estimation by current subgrid-scale (SGS)
models \citep{Jimenez2000}.  Secondly, we discuss the
requirements for the slip condition to be used in conjunction with
wall models and derive the equation that connects the slip boundary
condition with the stress at the wall. Finally, a
dynamic procedure for the slip coefficients is formulated, providing a
dynamic slip wall model free of \emph{a priori} specified
coefficients. The performance of the proposed dynamic wall model is
tested in a series of LES of turbulent channel flow at varying
Reynolds numbers, non-equilibrium three-dimensional transient
channel flow, and zero-pressure-gradient flat-plate turbulent
boundary layer. The results show that the dynamic wall model is able
to accurately predict one-point turbulence statistics for various flow
configurations, Reynolds numbers, and grid resolutions.

\end{abstract} 

\maketitle

%=====================================================================
\section{Introduction}\label{sec:intro}
%=====================================================================

% EXPAND INTRO. TO ADD:
%  - more discussion about what are the bc? Ghosal1995
%  - ...

% Introduce LES and the problem near the wall

% motivation
The near-wall resolution requirement to accurately resolve the
boundary layer in wall-bounded flows remains a pacing item in
large-eddy simulation (LES) for high-Reynolds-number engineering
applications. \citet{Choi2012} estimated that the number of grid
points necessary for a wall-resolved LES scales as $\Rey^{13/7}$,
where $\Rey$ is the characteristic Reynolds number of the problem.
The computational cost is still excessive for many practical problems,
especially for external aerodynamics, despite the favourable
comparison to the $\Rey^{37/14}$ scaling required for direct numerical
simulation (DNS), where all the relevant scales of motion are resolved.

% necessity wall models
By modelling the near-wall flow such that only the large-scale motions
in the outer region of the boundary layer are resolved, the grid-point
requirement for wall-modelled LES scales at most linearly with
increasing Reynolds number. Therefore, wall-modelling stands as the
most feasible approach compared to wall-resolved LES or DNS. Several
strategies for modelling the near-wall region have been explored in
the past, and most of them are effectively applied by replacing the
no-slip boundary condition in the wall-parallel directions by a
Neumann condition. This fact is motivated by the observation that,
with the no-slip condition, most subgrid scale models do not provide
the correct stress at the wall when the near-wall layer is not
resolved by the grid \citep{Jimenez2000}.

% existing wall models
Examples of the most popular and well-known wall models are the
traditional wall-stress models (or approximate boundary conditions),
and detached eddy simulation (DES) and its variants. Approximate
boundary condition models compute the wall stress using either the law
of the wall \citep{Deardorff1970,Schumann1975,Piomelli1989,Yang2015}
or the Reynolds-averaged Navier-Stokes (RANS) equations
\citep{Balaras1996,Wang2002,Chung2009,Kawai2013,Park2014}.  DES
\citep{Spalart1997} combines RANS equations close to the wall and LES
in the outer layer, with the interface between RANS and LES domains
enforced implicitly through the change in the turbulence model. The
reader is referred to \citet{Piomelli2002}, \citet{Cabot2000},
\citet{Larsson2015}, and \citet{Bose2017} for a more comprehensive
review of wall-stress models and to \citet{Spalart2009} for a review
of DES.

% Bose
One of the most important limitations of the models above is that they
depend on pre-computed parameters and/or assume explicitly or
implicitly a particular law for the mean velocity profile close to the
wall. Recently this has been challenged by \citet{Bose2014} with a
dynamic slip wall model that is free of any \emph{a priori} specified
coefficients. In addition, the no-transpiration condition used in most
wall models was replaced by a Robin boundary condition in the
wall-normal direction. The present study extends the
work by \citet{Bose2014} and is divided in two parts. In the first
part, we investigate the use of the slip boundary condition at the
wall for the three velocity components in the context of wall-modelled
LES. The motivation for the use of this boundary condition is
corroborated both theoretically and through detailed \emph{a priori}
tests of filtered velocity fields. We then assess whether this
condition is physically advantageous compared to other boundary
conditions when the LES grid resolution is insufficient to accurately
resolve the near-wall region.  Additionally, sensitivities of the slip
boundary condition with respect to Reynolds number, grid resolution,
and SGS model in actual LES are explored. In the second part of the
paper, we discuss the requirements for constructing wall models based
on the slip condition and propose a dynamic procedure independent of
any \emph{a priori} tunable parameters, consistent with the slip
boundary condition, and based on the invariance of wall stress under
test filtering.

% sec:slipbc
%    sec:theory 
%    sec:apriori
%    sec:slip_constraint
%
% sec:first_order_stat
%    sec:numerics
%    sec:control
%    sec:pred_log
%    sec:UV_RMS
%    sec:SGS
%    sec:transpiration
%
% sec:dynamicWM
%    sec:previous
%    sec:new_model
%    sec:test_cases_wm
%    
% sec:results_wm
%    sec:2d_channel_results
%    sec:3d_channel_results
%
% sec:conclusions
%

% organization
The paper is organised as follows. In section \ref{sec:slipbc}, we
present and motivate the suitability of the slip boundary condition
for LES by considering the behaviour of the filtered velocities at the
wall. \emph{A priori} testing is performed on filtered DNS data to
test the validity of the analysis. In section
\ref{sec:first_order_stat}, we perform a set of turbulent channel LES
with the slip boundary condition and study the effect of the slip
parameters, choice of SGS model, grid resolution, and Reynolds number
on the one-point statistics such as the mean and root-mean-squared
(rms) velocity profiles.  A dynamic procedure is presented in section
\ref{sec:dynamicWM}, and its performance is evaluated in section
\ref{sec:results_wm} for LES of two-dimensional and non-equilibrium
three-dimensional transient turbulent channel flows, and
zero-pressure-gradient flat-plate turbulent boundary layer. Finally,
conclusions are offered in section \ref{sec:conclusions}.

%====================================================================%
\section{Slip boundary condition with transpiration}
\label{sec:slipbc}
%====================================================================%

\begin{figure} 
\centerline{\includegraphics[width=0.8\textwidth]{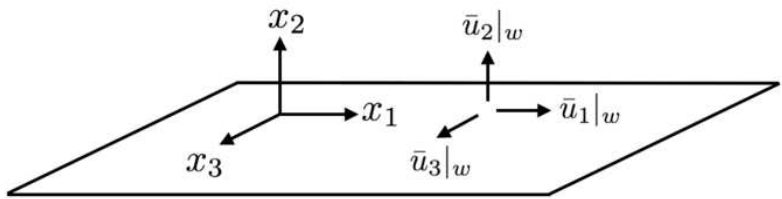}} 
\caption{\label{fig:slip_bc} Sketch of the slip boundary condition
with transpiration ($\bar{u}_2|_w \neq 0$) for a flat wall.}
\end{figure}

%% Slip boundary condition
We define the slip boundary condition with transpiration as 
\begin{equation}
\bar{u}_i|_w = l_i \left.\frac{\partial \bar{u}_i}{\partial
n}\right|_w + v_i,\quad i=1,2,3, \label{eq:slip_bc}
\end{equation}
where repeated indices do not imply summation. The indices $i = 1,2,3$
denote the streamwise, wall-normal, and spanwise
spatial directions represented by $x_1,x_2$ and $x_3$, respectively,
$u_i$ are the flow velocities, $\bar{(\cdot)}$ is the resolved LES
field (or filter operation), $n$ is the wall-normal direction, and
$(\cdot)|_w$ indicates quantities evaluated at the wall.
The grid or filter size will be denoted as $\Delta_i$ for
the respective directions. We define $l_i$ to be the slip lengths and
$v_i$ the slip velocities. In general, both the slip lengths and
velocities are functions of space and time. A sketch of the slip
boundary condition for a flat wall is given in figure
\ref{fig:slip_bc}.
%
%-----------------------------------------------------------%
%-----------------------------------------------------------%
%
% summary of section
In this section, we provide theoretical motivation for the slip
boundary condition in the context of filtered velocity components and
inspect the validity of (\ref{eq:slip_bc}) using \emph{a priori}
testing of filtered DNS data. The physical implications of the slip
lengths are also investigated in terms of the mean velocity profile
and rms velocity fluctuations. Some preliminary results of this
section can be found in \citet{Bae2016}.

%====================================================================%
\subsection{Theoretical motivation}\label{sec:theory}
%====================================================================%

\begin{figure} 
\centerline{
\includegraphics[width=0.5\textwidth]{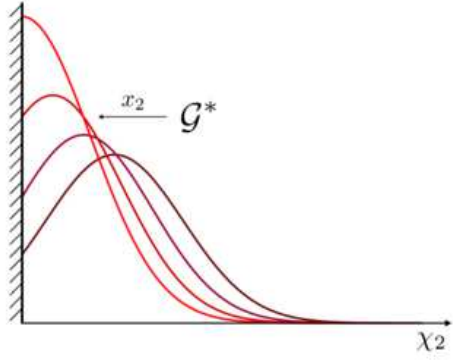} }
\caption{\label{fig:kernels}Sketch of the change in effective
kernel approaching the wall.}
\end{figure}

Let us consider a wall-bounded flow. For the Navier-Stokes equations,
the velocity at the wall is given by the no-slip boundary condition
\citep{Stokes1901}
\begin{equation}
\left.u_i\right|_w = 0,\quad i = 1,2,3.
\end{equation}
If we interpret LES as the solution of the filtered Navier-Stokes
equations \citep{Leonard1975}, the filtering operation in the
wall-normal direction will result, in general, in non-zero velocities
at the wall. For wall-resolved LES, where the effective filter size
near the wall is small, $\left.\bar{u}_i\right|_w$ can still be
approximated by the no-slip boundary condition \citep{Ghosal1995}.
However, when the filter size is large or the near wall resolution is
coarse, such as in wall-modelled LES, a modified wall boundary
condition different from the usual no-slip is required for
the three velocity components. Consider a one-dimensional
symmetric filter kernel $\mathcal{G}(\chi)$ with nonzero filter size
$\Delta_{\mathcal{G}}$ defined by its second moment 
\begin{equation}
\Delta_{\mathcal{G}}^2 =
\int_{-\infty}^{\infty}\mathcal{G}({\chi'}){\chi'^2}\,\mathrm{d}{\chi'}.
\end{equation}
Then, far from the wall ($x_2\gg \Delta_\mathcal{G}$), where $x_2=0$
is the location of the wall, the filtered velocity field can be
computed as
\begin{equation}
\bar{u}_i(\boldsymbol{x}) =
\int_{-\infty}^{\infty}\int_{-\infty}^{\infty}
\int_{-\infty}^{\infty}\mathcal{G}(-{x_1}+{\chi_1'})\mathcal{G}(-{x_2}+{\chi_2'})\mathcal{G}(-{x_3}+{\chi_3'})u_i(\boldsymbol{\chi'})\,\mathrm{d}\boldsymbol{\chi'},
\end{equation}
where $\boldsymbol{x}=(x_1,x_2,x_3)$ and $\boldsymbol{\chi'}=
(\chi_1',\chi_2',\chi_3')$. However, in the near-wall region, the
filter kernel in the wall-normal direction has a functional dependence
on the wall-normal distance, which becomes prevalent for
$x_2\rightarrow 0$, as depicted in figure \ref{fig:kernels}.  The new
filter operator restricted by the wall then becomes
\begin{equation}
\bar{u}_i(\boldsymbol{x}) =
\int_{-\infty}^{\infty}\int_{0}^{\infty}
\int_{-\infty}^{\infty}\mathcal{G}(-{x_1}+{\chi_1'})\mathcal{G}^*(-{x_2}+{\chi_2'};x_2)\mathcal{G}(-{x_3}+{\chi_3'})u_i(\boldsymbol{\chi'})\,\mathrm{d}\boldsymbol{\chi'}
\end{equation}
%\begin{equation}
%\bar{u}_i(x_2) =
%\int_{0}^{\infty}\mathcal{G}^*(x_2,-x_2+\chi')u_i(\chi')\,\mathrm{d}\chi',
%\end{equation} 
%
where
\begin{equation}
\mathcal{G}^*(\chi;x_2) =
\frac{\mathcal{G}(\chi)}{\int_{-x_2}^{\infty}\mathcal{G}(\varphi)\,\mathrm{d}\varphi}
\end{equation}
is the effective (rescaled) kernel at a wall-normal distance $x_2$.
%
%-----------------------------------------------------------%
%-----------------------------------------------------------%
%

Assuming a no-slip boundary condition for the unfiltered velocity and
approximating the velocity profile near the wall as a Taylor expansion
$u_i(\boldsymbol{x}) = \sum_{m=1}^{p} a_m x_2^m$, the filtered
velocity can be expressed as
\begin{equation}
\bar{u}_i|_w = \sum_{m=1}^{p} a_m M_{\mathcal{G}^*_0}^{(m)},
\label{eq:vel_moment}
\end{equation}
where $M_{\mathcal{G}^*_0}^{(m)}$ is the $m$-th moment of
$\mathcal{G}^*(\chi;0)$, the effective filter at the wall. From
(\ref{eq:vel_moment}), it can be shown that  
\begin{equation}
\left.\frac{\partial\bar{u}_i}{\partial x_2}\right|_w = \sum_{m=1}^{p}
-2a_m\mathcal{G}(0) M_{\mathcal{G}_0^*}^{(m)} +
ma_mM_{\mathcal{G}_0^*}^{(m-1)}.
\label{eq:grad_moment}
\end{equation}
Since $M_{\mathcal{G}_0^*}^{(m)}\sim \Delta_\mathcal{G}^m $ and
$\mathcal{G}(0)\sim\Delta_\mathcal{G}^{-1}$, we can use 
(\ref{eq:vel_moment}) and (\ref{eq:grad_moment}) to obtain a
second-order approximation of the boundary condition of the form
\begin{equation}
\bar{u}_i|_w = l \left.\frac{\partial\bar{u}_i}{\partial
x_2}\right|_w, 
\label{eq:eq_l}
\end{equation}
where $l = M_{\mathcal{G}_0^*}^{(1)}/\left(1-2\mathcal{G}(0)
M_{\mathcal{G}_0^*}^{(1)}\right)$.

This boundary condition is exact for a linear velocity
profile but is expected to deteriorate as the linear approximation is
no longer valid for $x_2 < \Delta_\mathcal{G}$. In this case, the
second- and higher-order terms excluded in (\ref{eq:eq_l}) may result in
different slip lengths and extra terms for each velocity component as
in (\ref{eq:slip_bc}) in order to achieve an accurate representation
of the flow at the wall. The particular expressions for $l_i$ and
$v_i$ depend formally on the filter
shape, size, and instantaneous configuration of the filtered velocity
vector at the wall. 

Equation (\ref{eq:eq_l}) motivates the use of the slip boundary
condition for wall-modelled LES. Nevertheless, it is important to
highlight a few remarks regarding the derivation and the consistency
of the slip condition. The first observation is that in the case of
explicitly-filtered LES \citep{Lund1995,Lund2003,Bose2012,Bae2017} with a
well-defined filter operator, the filter size is a given function of
the wall-normal distance, and the slip lengths and velocities can be
computed explicitly.  However, in the present study, we focus on traditional
implicitly-filtered LES, where the filter operator is not distinctly
defined and, consequently, neither is the filter size, typically
assumed to be proportional to the grid size. This supposed relation
between the filter and grid sizes is not always valid
\citep{Lund2003,Silvis2016} and worsens close to the wall.
Therefore, in the near wall region, it is reasonable to assume that
the effective filter size is an unknown function of the wall-normal
distance, and the slip lengths and velocities must be modelled as they
cannot be computed explicitly. As a final remark, note that
commutation of the filter and derivative operators is necessary to
formally derive the LES equations which, in turn, entails a
constant-in-space filter size or a filter operator that is constructed
to be commutative \citep{Marsden2002}. This condition is not met by
(\ref{eq:slip_bc}), but given that the filter size for
implicitly-filtered LES is an unknown function of space, we also
neglect terms arising from commutation errors.

%====================================================================%
\subsection{A priori evaluation} \label{sec:apriori}
%====================================================================%

% data
\emph{A priori} testing of the slip boundary condition is conducted to
assess the validity of (\ref{eq:slip_bc}) in
filtered DNS data of turbulent channel flow from
\citet{DelAlamo2004}, \citet{Hoyas2006}, and \citet{Lozano-Duran2014} at
friction Reynolds numbers of $\Rey_\tau \approx 950,\, 2000$ and
$4200$, respectively.

% definitions
In the following, $u_\tau$ is the friction velocity, $\nu$ is the
kinematic viscosity, and the channel half-height is denoted by
$\delta$. Wall units are defined in terms of $\nu$ and $u_\tau$ and
denoted by the superscript $+$ and outer units in terms
of $\delta$ and $u_\tau$. In each case, DNS velocity vector
is filtered in the three spatial directions with a box-filter with filter size equal to
$\Delta_1 \times \Delta_2 \times \Delta_3$. The resulting filtered
data contain $\left.\bar{u}_i\right|_w$ and $\left.\partial
\bar{u}_i/\partial x_2 \right|_w$, which can be used to test the
accuracy of (\ref{eq:slip_bc}) by computing their joint probability
density function (PDF).
% slip lengths for differential filter
%
%-----------------------------------------------------------%
\begin{figure} 
\centerline{ 
\psfrag{X}[cc]{$\bar{u}_1^+|_w$}
\psfrag{Y}[bc]{$\partial \bar{u}_1^+/\partial (x_2^+)|_w$}
\subfloat[]{{\includegraphics[width=0.47\textwidth]{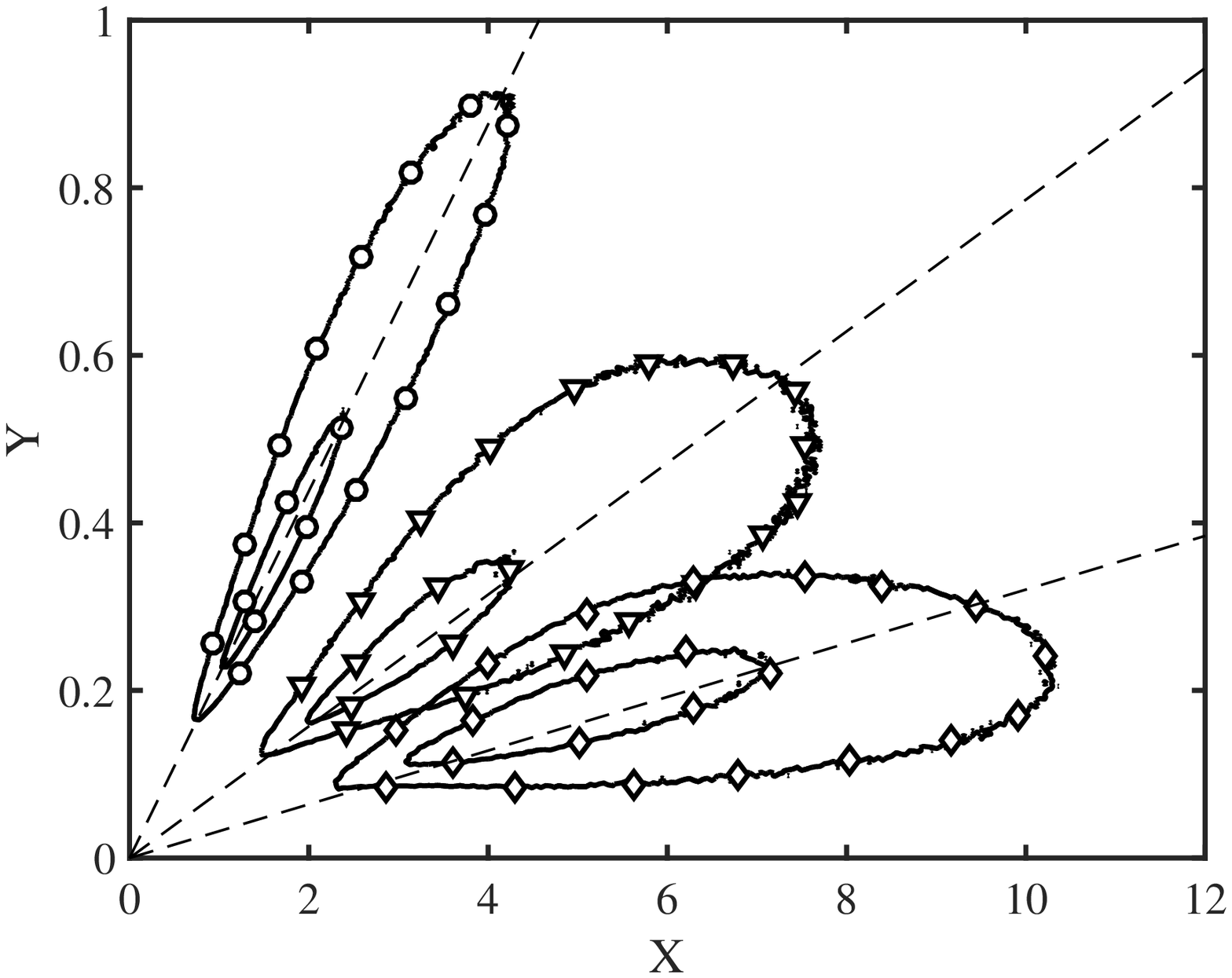}}} 
\hspace{0.2cm} 
\psfrag{X}[cc]{$\bar{u}_2^+|_w$}
\psfrag{Y}[bc]{$\partial \bar{u}_2^+/\partial (x_2^+)|_w$}
\subfloat[]{{\includegraphics[width=0.47\textwidth]{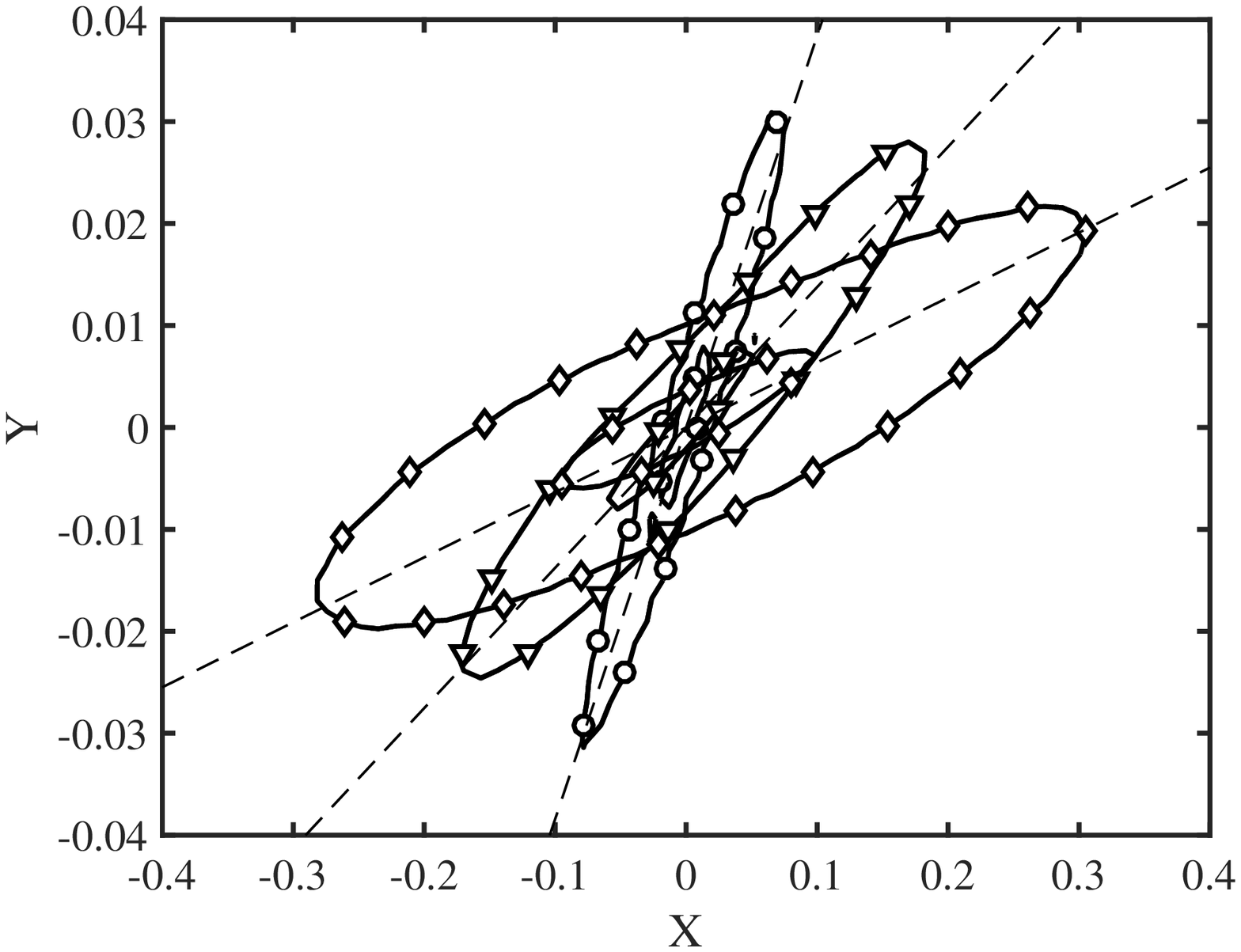}}}} 
\centerline{ 
\psfrag{X}[cc]{$\bar{u}_3^+|_w$}
\psfrag{Y}[bc]{$\partial \bar{u}_3^+/\partial (x_2^+)|_w$}
\subfloat[]{{\includegraphics[width=0.47\textwidth]{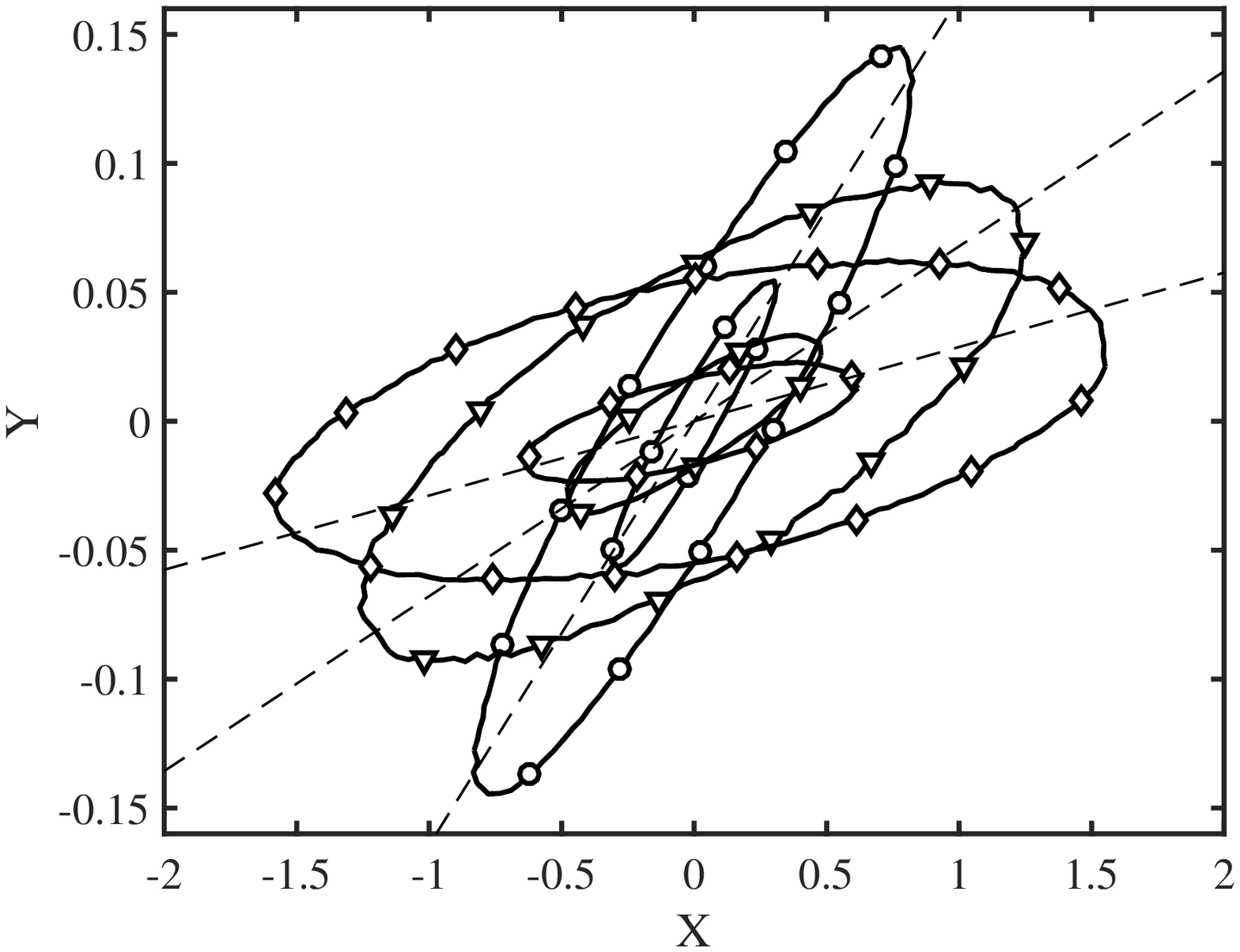}}} 
\hspace{0.2cm} 
\psfrag{X}[cc]{$\Rey_\tau$}
\psfrag{Y}[bc]{$l_1/\delta$}
\subfloat[]{{\includegraphics[width=0.47\textwidth]{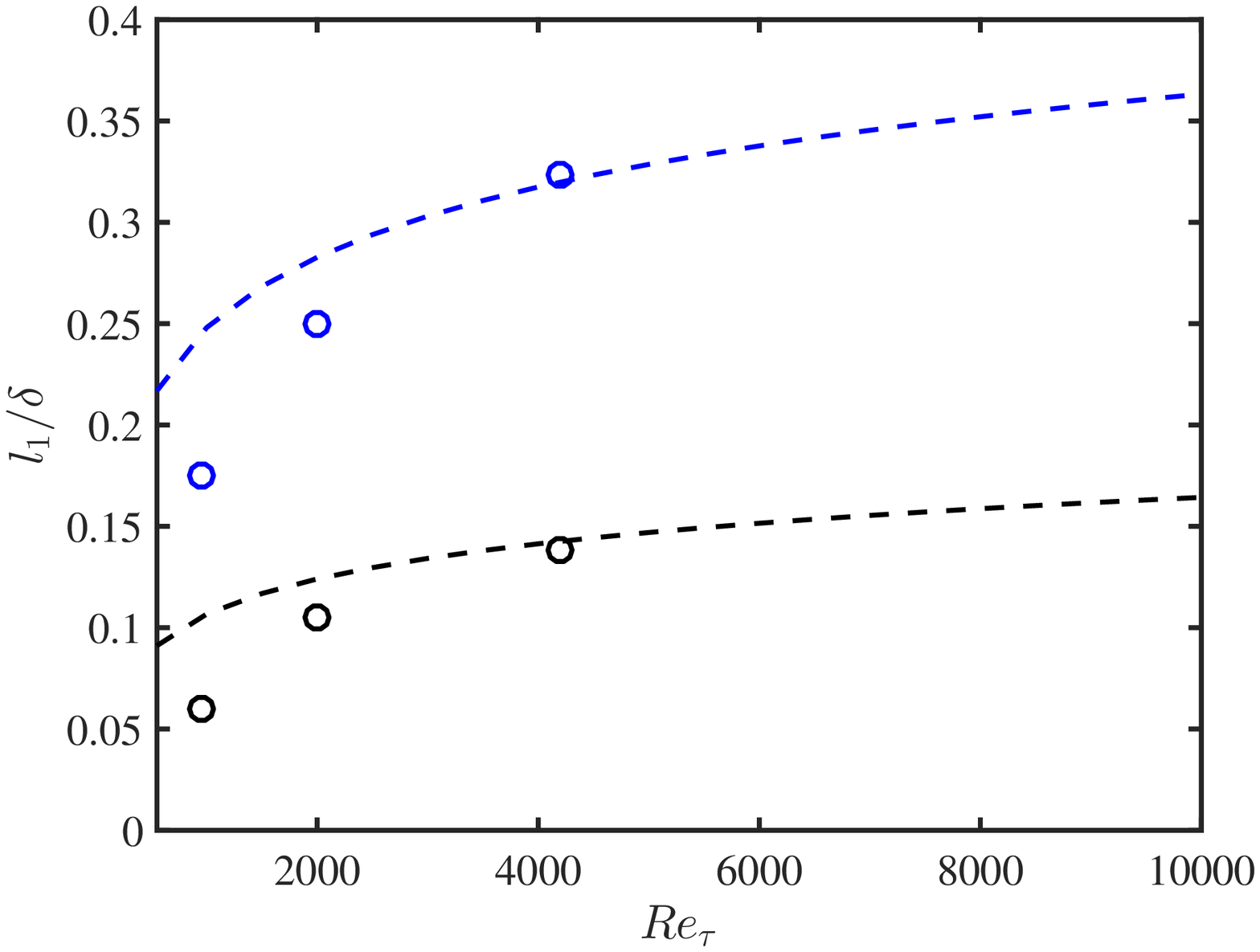}}} }
\caption{Joint PDF of (a)
$\left.\bar{u}_1\right|_w$ and $\left.\partial \bar{u}_1/\partial
x_2\right|_w$, (b) $\left.\bar{u}_2\right|_w$ and $\left.\partial
\bar{u}_2/\partial x_2\right|_w$, and (c) $\left.\bar{u}_3\right|_w$
and $\left.\partial \bar{u}_3/\partial x_2\right|_w$ for box-filtered
DNS with $\Delta_i = 0.01\delta$ ($\circ$), $0.02\delta$
($\triangledown$), and $0.03\delta$ ($\lozenge$) at $\Rey_\tau \approx
950$. For each probability distribution the contours are 50\% and
95\%. The straight-dashed lines are obtained by the least squares
fitting to the joint PDF. (d) $l_1$ dependence on
$Re_\tau$ with $\Delta_2 = 0.05\delta$ (black) and $\Delta_2 =
0.1\delta$ (blue) calculated from box-filtered DNS channel flow data
($\circ$), and estimation from box-filtered logarithmic layer
approximation $l_1/\delta =
\Delta_2/(2\delta)\left[\log(Re_\tau\Delta_2/(2\delta))-1\right]+\kappa
B\Delta_2/(2\delta)$ with  $\kappa = 0.41$ and $B = 5.3$, (dashed line).
%$l_1$ dependence on $Re_\tau$ for$\Delta = 0.1\delta$. Plots (a--c)
%are in wall units.  
\label{fig:pdf}} 
\end{figure}
%-----------------------------------------------------------%

% slip length for box filter
The joint PDF for
$\Rey_\tau\approx 950$ with $\Delta_1=\Delta_2=\Delta_3$ are plotted
in figure \ref{fig:pdf}(a--c). The results show, on average, a linear
correlation between $\left.\bar{u}_i\right|_w$ and $\left.\partial
\bar{u}_i / \partial x_2 \right|_w$, which supports the suitability of
the slip boundary condition given in (\ref{eq:slip_bc}) with $v_i=0$.
However, the spread of the joint PDFs increases with increasing filter
size. A trend similar to that of figure \ref{fig:pdf}(a--c) appears
when increasing the Reynolds number for a constant filter size
$\Delta_i/\delta$ (not shown), which implies that the
second-order approximation deteriorates as the filter size increases
in wall units. However, despite this scaling, a linear relationship
between $\bar{u}_i|_w$ and $\partial \bar{u}_i/\partial n|_w$ is still
satisfied on average, and it will be shown in section
\ref{sec:first_order_stat} that this is enough to obtain accurate
predictions of the mean velocity profile in actual LES.

% slip length for Re
Finally, figure \ref{fig:pdf}(d) shows the streamwise slip length
$l_1$ as a function of Reynolds number for a constant
$\Delta_i/\delta$. The slip length was computed as the average ratio
of $\bar{u}_1|_w$ and $\left.\partial\bar{u}_1/\partial x_2\right|_w$.
The plot shows that the dependence of the streamwise slip length on
Reynolds number is stronger for smaller $\Rey_\tau$.
This behaviour can be explained by the relative thickness of the
filter size and the buffer layer.  When the ratio is of $O(1)$, the
filtered velocity at the wall takes into account contributions from
the buffer layer, and $l_1$ is expected to be sensitive to changes in
Reynolds number.  However, when the buffer layer is a small fraction
of the filter size, most of the contribution to $l_1$ comes from the
logarithmic layer, which has a universal behaviour with $\Rey_\tau$.
Neglecting the effect of the buffer layer, the approximate functional
dependence of the slip length on Reynolds number can be estimated from
the logarithmic velocity profile, 
\begin{equation} 
\frac{\langle u_1 \rangle}{u_\tau} = \frac{1}{\kappa}
\log\left( x_2^+ \right) + B, \label{eq:log_law} 
\end{equation}
where $\langle\cdot\rangle$ denotes ensemble average in the
homogeneous directions and time, $\kappa$ is the von K\'arm\'an
constant, and $B$ is the intercept constant. In the limit of high
Reynolds numbers, the box-filtered streamwise velocity and its
wall-normal derivative can be estimated by assuming a logarithmic law
in the entire near-wall region and integrating (\ref{eq:log_law}).
This gives an approximation for the average streamwise slip length
\begin{equation} 
\frac{l_1}{\delta} \sim \frac{\langle \bar{u}_1/u_\tau
\rangle}{\partial\langle \bar{u}_1/u_\tau \rangle /
\partial(x_2/\delta)} \sim \frac{\Delta_2}{2\delta}
\left[\log\left(\Rey_\tau\frac{\Delta_2}{2\delta}\right)-1\right] +
\kappa B \frac{\Delta_2}{2\delta} 
\end{equation}
(dashed lines in figure \ref{fig:pdf}d), which predicts a
weak $\Rey_\tau$ dependence for large Reynolds numbers. It is
important to remark that $l_1$ from the figure is only an estimation
from \emph{a priori} testing and the particular values are not
expected to work in an actual LES, although we expect the trends to be
relevant. 

%====================================================================%
\subsection{Consistency constraints on the slip parameters}
\label{sec:slip_constraint}
%====================================================================%

In an actual LES implementation, the choice of $l_i$ and $v_i$ must
comply with the symmetries of the flow. Moreover, it is also necessary
to satisfy on average the impermeability constraint of the wall to
preserve the physics of the flow (more details are offered in section
\ref{sec:transpiration}). Therefore, the slip boundary condition for a
plane channel flow should fulfil
\begin{equation} 
\langle \bar{u}_i|_w\rangle = \left\langle\left.
l_i\frac{\partial\bar{u}_i}{\partial n}\right|_w\right\rangle +
\langle v_i\rangle = 0,\quad i=2,3.  
\label{eq:avg_bc} 
\end{equation}
Equation (\ref{eq:avg_bc}) can be further simplified by assuming $l_i$
and $v_i$ to be constant in the homogeneous directions. Since $\langle
\bar{u}_i|_w \rangle = 0$ and
$\left.\left\langle{\partial\bar{u}_i}/{\partial x_2}\right|_w
\right\rangle = 0$ for $i=2,3$, $\langle v_2\rangle$ and $\langle
v_3\rangle$ must be set to zero. We can also set $\langle v_1\rangle
=0$ without loss of generality, since its average effect can be
absorbed by moving the frame of reference at constant uniform
velocity. Then, the slip boundary condition consistent with the
symmetries of the channel is of the form
\begin{equation} 
\left.\bar{u}_i\right|_w = l_i \left.\frac{\partial
\bar{u}_i}{\partial n}\right|_w.  
\label{eq:channel} 
\end{equation}

When the flow is no longer homogeneous in $x_1$, as in a spatially
developing flat-plate boundary layer, the above arguments based on the
symmetry of the channel do not hold. Then, the slip velocity $v_i$ can
be used to impose zero mean mass flow through the walls and ensure
that the boundary behaves, on average, as a non-permeable wall. Since
the  mass flow through a flat wall is only affected by $v_2$, we can
still set $v_1$ and $v_3$ to zero for simplicity. 

%====================================================================%
\section{Effect of the slip boundary condition on one-point
statistics} \label{sec:first_order_stat}
%====================================================================%

It was argued in section \ref{sec:slipbc} that the most general form
of the Robin boundary condition, given by (\ref{eq:slip_bc}), should
replace the no-slip condition in wall-modelled LES. In this section,
we investigate the effects of $l_i$ and $v_i$ on the one-point
statistics of LES of plane turbulent channel flow and flat-plate boundary
layer. Our conclusions will be numerically corroborated by considering
$l_i$ and $v_i$ as free parameters in an LES with slip boundary
condition at the wall. A dynamic procedure to compute these
parameters will be given in section \ref{sec:dynamicWM}.

%====================================================================%
\subsection{Numerical experiments}\label{sec:numerics}
%====================================================================%

\begin{table} 
\begin{center} 
\setlength{\tabcolsep}{12pt}
\begin{tabular}{l c c c c c c} 
Case        & SGS model & $\Rey_\tau$    & $\Delta_i^+$ & 
$\Delta_i/\delta$       & $l_1/\delta$   & $l_2/\delta$ \\ 
\hline \hline 
DSM-2000    & DSM       & 2003           & 100          & 
0.050                   & 0.008          & 0.008        \\ 
DSM-2000-s1 & DSM       & 2003           & 100          &
0.050                   & 0.008          & 0.004        \\ 
DSM-2000-s2 & DSM       & 2003           & 100          & 
0.050                   & 0.004          & 0.008        \\ 
DSM-2000-s3 & DSM       & 2003           & 100          & 
0.050                   & 0.097          & 0.045        \\ 
\hline 
DSM-2000-c1 & DSM       & 2003           & 154          & 
0.077                   & 0.008          & 0.008        \\ 
DSM-2000-c2 & DSM       & 2003           & 200          & 
0.100                   & 0.008          & 0.008        \\ 
\hline 
DSM-950     & DSM       & 934            & 46           & 
0.050                   & 0.008          & 0.008        \\ 
DSM-4200    & DSM       & 4179           & 210          & 
0.050                   & 0.008          & 0.008        \\ 
\hline 
AMD-2000    & AMD       & 2003           & 100          & 
0.050                   & 0.008          & 0.008        \\ 
SM-2000     & SM        & 2003           & 100          & 
0.050                   & 0.008          & 0.008        \\ 
NM-2000     & NM        & 2003           & 100          & 
0.050                   & 0.008          & 0.008        \\ 
\hline 
\end{tabular} 
\end{center}
\caption{Tabulated list of cases. The numerical experiments are
labelled following the convention [SGS model]-[$\Rey_\tau$](-[other
cases]). SGS models used are the dynamic Smagorinsky model (DSM),
constant coefficient Smagorinsky model (SM), anisotropic
minimum-dissipation model (AMD), and no model (NM). Grid resolutions
different from the baseline case are noted by c1 and c2. Three
additional cases with different slip length than the baseline case are
labelled s1, s2, and s3. See text for details.} 
\label{tab:cases}
\end{table}

% numerics
We perform a set of plane turbulent channel LES listed
in table \ref{tab:cases}. The simulations are computed with a
staggered second-order finite difference \citep{Orlandi2000} and a
fractional-step method \citep{Kim1985} with a third-order Runge-Kutta
time-advancing scheme \citep{Wray1990}.  The code has been validated
in previous studies in turbulent channel flows
\citep{Lozano2016,Bae2018} and flat-plate boundary layers
\citep{Lozano-Duran2018a}.  The size of the channel is $2\pi \delta\times
2\delta\times \pi \delta$ in the streamwise, wall-normal, and spanwise
directions, respectively.  It has been shown that this domain size is
sufficient to accurately predict one-point statistics for $\Rey_\tau$
up to $4200$ \citep{Lozano-Duran2014}.  Periodic boundary conditions
are imposed in the streamwise and spanwise directions.
The eddy viscosity $\nu_t$ is computed at the cell
centres and the values at the wall are obtained by assuming Neumann
boundary conditions for the discretised $\nu_t$, which is motivated by
the fact that for coarse grid resolutions the SGS contribution at the
wall must be non-zero. The
channel flow is driven by imposing a constant mean pressure gradient,
and all simulations were run for at least 100 eddy turnover times,
defined as $\delta/u_\tau$, after transients.

% slip bc
The slip boundary condition from (\ref{eq:channel}) is used on the top
and bottom walls. We have tested the variability of $l_i$ in time by
oscillating $l_i(t)$ with different amplitudes and frequencies around
a given mean. The frequency of the oscillation considered were 0.5, 1,
and 2 times the natural frequency given by the size of the grid and
$u_\tau$, and the amplitudes imposed were up to 0.5 times the value of
the mean. The different cases resulted in almost identical one-point
statistics as those obtained with a constant $l_i$ of the same mean
with the relative difference in the resulting wall stress
below 0.5\% for all cases. Thus, $l_i$ will be fixed to a constant
value in both homogeneous directions and time for the remainder of
the section.

% base case
We take as a baseline case the friction Reynolds number
$\Rey_\tau\approx 2000$ with a uniform grid resolution of
$128\times40\times64$ in the $x_1$, $x_2$, and $x_3$ directions,
respectively. The grid size in outer units is $0.050\delta$ in the
three directions, and follows the recommendations by 
\citet{Chapman1979} for resolving the large eddies in the outer
portion of the boundary-layer. The baseline SGS model used is the
dynamic Smagorinsky model \citep[DSM,][]{Germano1991,Lilly1992}. The
baseline slip lengths are $l_i=0.008\delta$, $i= 1,2,3$ for reasons
given in section \ref{sec:control}.  Three additional cases with
different slip lengths, which are given in the first
group of table \ref{tab:cases}, are used to study the effect of $l_i$
on the one-point statistics.

% grid resolution sensitivity
To study the effects of slip boundary condition on grid
resolution, we define two meshes with $82\times26\times42$
and $64\times20\times32$ grid points distributed
uniformly in each direction, which correspond to a uniform grid size
of $0.077\delta$ and $0.100\delta$,
respectively as listed in the second group of table
\ref{tab:cases}. The resolutions were chosen such that the first
interior point lies in the logarithmic region and is far from the
inner-wall peak of the streamwise rms velocity. The intention is to
avoid capturing (even partially) the dynamic cycle in the buffer
layer, since the wall-normal lengths of the near-wall vortices and
streaks scale in viscous units, and that scaling is incompatible with
the computational efficiency pursued in wall-modelled LES. The range
of grid resolutions is limited due to the fact that the outer layer
still needs to be resolved by the LES.  However, it will be shown in
section \ref{sec:SGS} that the selected range of grid resolutions is
sufficient to show the sensitivity of the slip lengths to the grid
resolution. 

% Re and SGS model sensitivity
To investigate the effect of the Reynolds number we will consider
three cases DSM-950, DSM-2000, and DSM-4200, which
constitute the third group of table \ref{tab:cases}. The sensitivity
to the SGS model will be assessed by comparing results from DSM,
constant-coefficient Smagorinsky model \citep[SM,][]{Smagorinsky1963}
without a damping function at the wall, the anisotropic
minimum-dissipation model \citep[AMD,][]{Rozema2015}, and cases without
an SGS model (NM), given in the fourth group of table
\ref{tab:cases}.  Finally, LES results will be compared with DNS data
from \citet{DelAlamo2004,Hoyas2006} and \citet{Lozano-Duran2014}. 

%====================================================================%
\subsection{Control of the wall stress and optimal slip lengths}
\label{sec:control}
%====================================================================%

% optimal li and connection with wall stress
%We define the optimal slip lengths as those providing the correct mean
%wall stress. 

\begin{figure} 
\vspace{0.2cm} 
\centerline{ 
\subfloat[]{
\psfrag{X}[cc]{$x_2/\delta$} 
\psfrag{Y}[bc]{$\langle u_1^+\rangle$}
\includegraphics[width=0.48\textwidth]{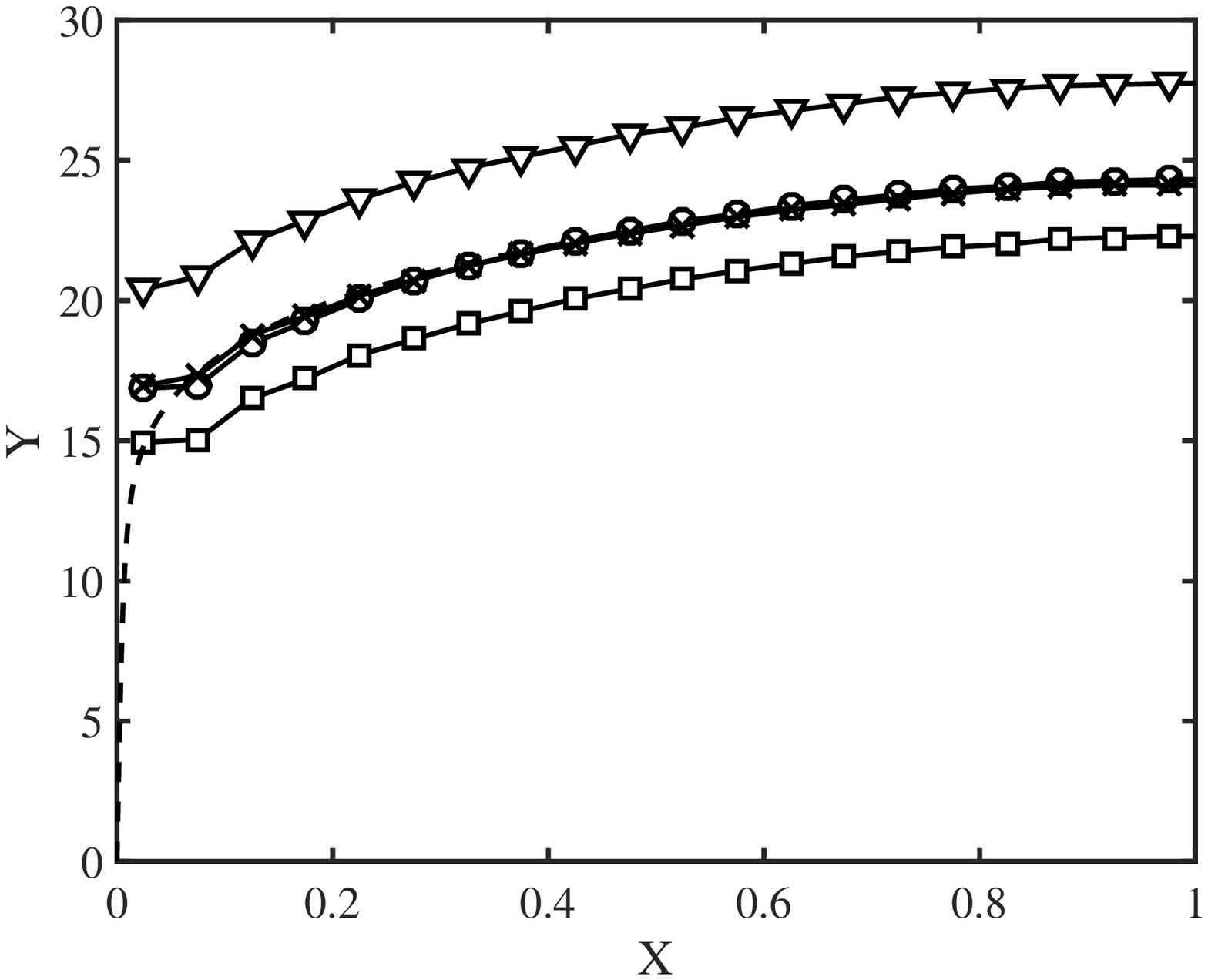} }
\hspace{0.2cm} 
\subfloat[]{ 
\psfrag{X}[cc]{$x_2^+$}
\psfrag{Y}[bc]{$\langle u_1^+\rangle$}
\includegraphics[width=0.48\textwidth]{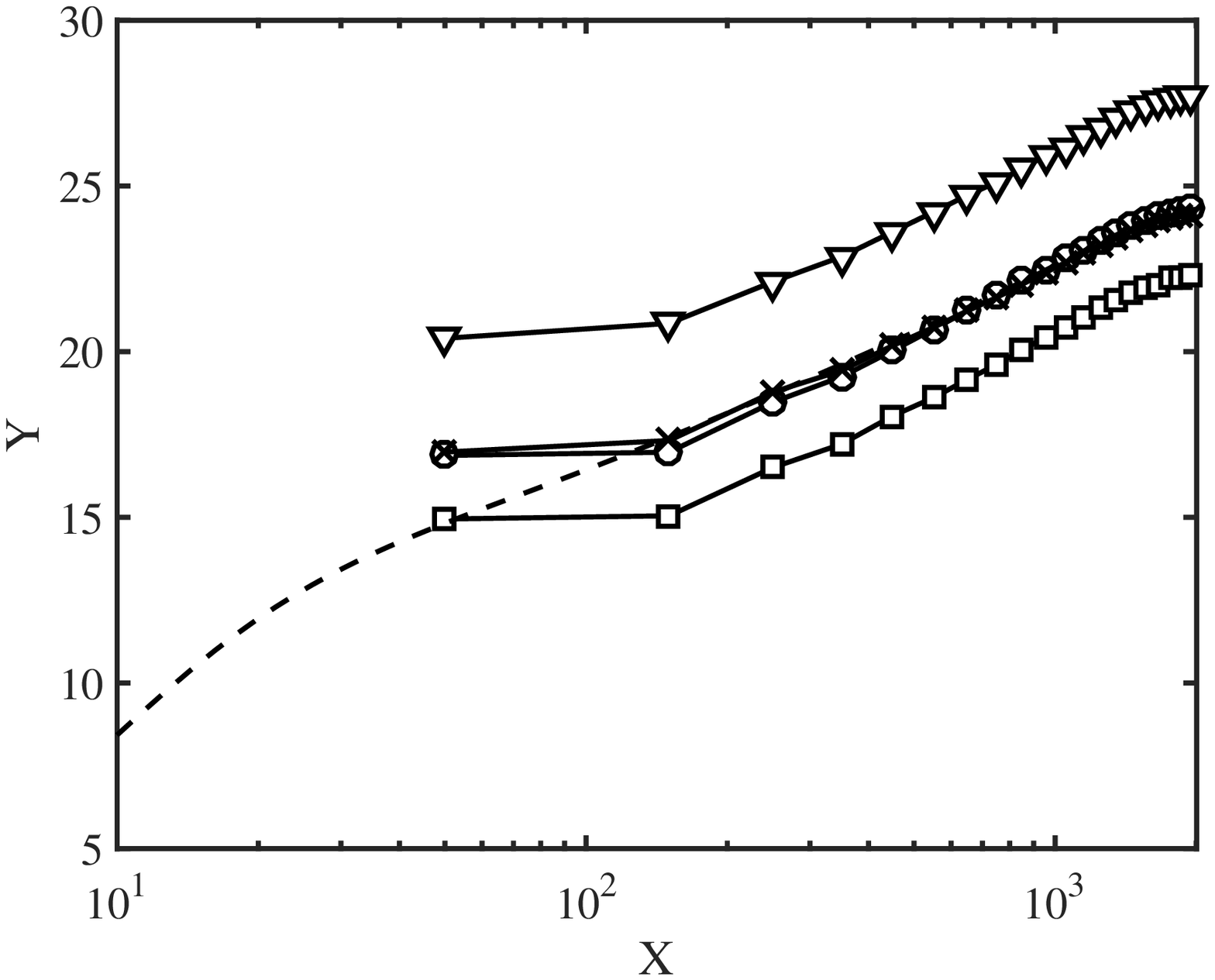} } }
\caption{ Mean streamwise velocity profile as a function of (a) outer
units and (b) wall units for DSM-2000
$(l_1,l_2)=(0.008\delta,0.008\delta)$, $\circ$; DSM-2000-s1
$(l_1,l_2)=(0.008\delta,0.004\delta)$, $\triangledown$; DSM-2000-s2
$(l_1,l_2)=(0.004\delta,0.008\delta)$, $\square$; DSM-2000-s3
$(l_1,l_2)=(0.097\delta,0.045\delta)$, $\times$;
and DNS (dashed line) \label{fig:mean_u_slip}} 
\end{figure}

In an LES of channel flow with the slip
boundary condition (\ref{eq:channel}), the wall stress is
given by
\begin{equation} 
\langle\tau_w\rangle = \nu\left.\left\langle\frac{\partial
\bar{u}_1}{\partial{x_2}}\right|_w\right\rangle - \langle
\bar{u}_1\bar{u}_2|_w\rangle -
\left.\left\langle\tau^\text{SGS}_{12}\right|_w\right\rangle,
\label{eq:wall_stress} 
\end{equation}
where $\tau_w$ is the stress at the wall, $\tau_{12}^\text{SGS}$ is
the tangential SGS stress tensor, and $\langle \bar{u}_1\bar{u}_2|_w\rangle$ is
the result of the non-zero velocity provided by the slip condition.
The slip lengths can be explicitly introduced by substituting
$\bar{u}_1\bar{u}_2$ from
(\ref{eq:channel}) such that
\begin{equation} 
\langle \tau_w \rangle = \nu\left.\left \langle \frac{\partial
\bar{u}_1}{\partial{x_2}}\right|_w\right\rangle -\left.\left\langle
l_1l_2\frac{\partial\bar{u}_1}{\partial
x_2}\frac{\partial\bar{u}_2}{\partial x_2}\right|_w\right\rangle
-\left.\left\langle\tau^\text{SGS}_{12}\right|_w\right\rangle,
\label{eq:l_2} 
\end{equation}
where $\tau^\text{SGS}_{12}$ may also depend on the slip lengths.
%The optimal slip lengths are effectively computed by
%running an LES with slip boundary condition using Eq. (\ref{eq:l_2})
%to obtain the slip lengths providing the correct stress at each time
%step, then averaging the values in time. Note that we have the freedom
%to impose the ratio of $l_1$ and $l_2$, and the choice of optimal slip
%lengths are not unique.
Therefore, the wall stress (and hence the
mass flow) can be controlled by the proper choice of slip lengths.
This is an important property of the slip boundary condition, and it
is illustrated in figure \ref{fig:mean_u_slip}. For coarse LES with
no-slip boundary conditions, the near-wall region cannot be accurately
computed due to the inadequacy of the current SGS models
and large numerical errors in the near-wall region, even if the
resolution is sufficient to resolve the outer layer eddies. This
mainly results in under- or over-predictions of the wall stress, among
other effects, and the shift of mean velocity with respect to DNS.
%
%-----------------------------------------------------------%
%-----------------------------------------------------------%
%
Figure \ref{fig:mean_u_slip} shows the mean streamwise velocity
profile for cases DSM-2000 and DSM-2000-s[1--3].  The results
reveal that increasing $l_1$ (at constant $l_2$) moves up the mean
velocity profile by 8\%, while increasing $l_2$ (at constant $l_1$)
have the opposite effect and decreases the mean by 15\%.
%
% l3 not very important
Although not shown, it was tested that varying $l_3$ has a
second-order effect on the mean velocity profile when compared to
changes of the same order in $l_1$ and $l_2$. For
instance, the change in mean velocity profile is 1.2\% when $l_3$ is
changed from $0.008\delta$ to $0.004\delta$ for case DSM-2000. The
result is not totally unexpected since $\bar{u}_1$ and $\bar{u}_2$ are
active components of the mean streamwise momentum balance in a channel
flow (see equation \ref{eq:l_2}), while $\bar{u}_3$ enters only
indirectly.  All calculations in the present study have been performed
with $l_3$ equal to $l_1$.

%It is important to remark that this control of the mean velocity
%profile cannot always be achieved with the traditional Neumann
%boundary condition. For example, when the contribution of the SGS
%model to the wall stress is larger than the correct wall stress, the
%only possible solution is to impose an unphysical gradient at the
%wall, $\left.\partial\overline{u}_1/\partial x_2\right|_w < 0$.

% physical explanation
The observations in figure \ref{fig:mean_u_slip} may be explained in
terms of the mean streamwise momentum balance at the wall and non-zero
streamwise slip. With respect to the former, increasing $l_2$ enhances
the $\langle\bar{u}_1\bar{u}_2\rangle$ contribution at the wall, which
is translated into a lower mean velocity profile due to the higher
momentum drain at the boundaries. The same argument applies when
increasing $l_1$.  However, higher $l_1$ also implies larger 
slip in $x_1$, which overcomes the previous momentum drain,
and the
resulting net effect is an increase of the mean mass flow. For the
laminar Poiseuille flow with the slip boundary condition, the shift in
the mean velocity profile can be computed analytically and is shown to
be  proportional to $l_1$.
%
%\begin{figure} \centerline{
%\subfloat[]{{\includegraphics[width=0.5\textwidth]{./figs/V_slip_sketch}
%}}
%\subfloat[]{{\includegraphics[width=0.5\textwidth]{./figs/U_slip_sketch}
%}} }
%
%\caption{\label{fig:sketch}Sketch of the effects of (a) non-zero
%streamwise and wall-normal slip lengths on the mean Reynolds shear
%stress and (b) non-zero streamwise slip length on the mean velocity
%profile.} \end{figure}

% existance of non-unique solution
The duality between the streamwise and wall-normal slip lengths makes
possible to always achieve the correct wall stress by an appropriate
selection of $(l_1,l_2)$, which we define as optimal slip
lengths. The optimal slip lengths are effectively computed by running
an LES with slip boundary condition using (\ref{eq:l_2}) with $\tau_w
= \tau_w^\text{DNS}$ as a constraint, and then averaging in time the values
obtained for $l_1$ and $l_2$. Note that we have the freedom to impose
the ratio $l_1/l_2$, and the optimal slip lengths are not
unique. Two examples are shown in figure \ref{fig:mean_u_slip}. It is
also important to remark that the control of the mean velocity profile
is not possible in general without wall-normal transpiration.
In particular, if an LES with $l_1/\delta=l_2/\delta=0$
(no-slip) already over-predicts the mean velocity profile with respect
to DNS, the only possible outcome of increasing $l_1/\delta$ while
maintaining $l_2/\delta=0$ is a positive shift of the mean velocity
profile. Since our experience shows that negative values of $l_1$ will
result in an unstable solution, the conclusion is that the correct
mean velocity profile cannot be achieved in this case unless $l_2 \neq
0$, making the wall-normal slip length indispensable.

%====================================================================%
\subsection{Prediction of the logarithmic layer}\label{sec:pred_log}
%====================================================================%

A second observation from figure \ref{fig:mean_u_slip} is that the
shape of the mean velocity profile remains roughly constant for
different slip lengths, and changes in $l_1$ and $l_2$ are mainly
responsible for a shift along the mean velocity axis. We would like to
connect the previous observation with the classic logarithmic profile
for the mean streamwise velocity given in (\ref{eq:log_law}).
Assuming that the filter operation does not alter the logarithmic
shape of $\langle \bar{u}_1 \rangle$ for the typical filter sizes (or
grid resolutions), (\ref{eq:log_law}) should also hold for LES.
However, it is not clear whether this would be the case for an actual
LES. For example, Millikan's asymptotic matching argument
\citep{Millikan1938} requires a scale separation that tends to
infinity as the Reynolds number increases, which is not the case in
wall-modelled LES as the length scales are fixed in outer units.
Other arguments, such as the Prandtl's mixing length hypothesis
\citep{Prandtl1925} would suggest that the correct wall-normal mixing
of the flow should be obtained in order to recover the logarithmic
profile. Alternatively, from the point of view of Townsend's attached
eddy hypothesis \citep{Townsend1980}, the flow from the LES should be
populated by a self-similar hierarchy of eddies with sizes
proportional to the wall distance and the proper number of eddies per
unit area. In all cases, the SGS model plays a non-negligible role in
fulfilling these conditions, especially at high Reynolds numbers and
coarse grids. As a consequence, not all SGS models are expected to
recover the correct shape of the mean velocity profile, and this is
further discussed in section \ref{sec:SGS}.

The prominent role of the SGS model in the correct representation of
the logarithmic layer can be seen from the integrated mean streamwise
momentum balance for filtered velocities 
\begin{equation} 
\langle\bar{u}^+_1\rangle (x_2^+) =
\underbrace{\langle\bar{u}^+_1|_w\rangle}_{\sim B}+
\underbrace{x_2^+\left(1-\frac{x_2}{2\delta}\right) + \int_0^{x_2^+}
\langle \bar{u}_1^+\bar{u}_2^+ + \tau^{\text{SGS}+}_{12}\rangle
dx_2'^{+}}_{\sim 1/\kappa\log(x_2^+)}. 
\label{eq:shift} 
\end{equation}
By comparing the structure of (\ref{eq:log_law}) and (\ref{eq:shift}),
it is reasonable to hypothesise that the slip boundary condition
mainly influences the intercept $B$, which is independent of $x_2$,
while the SGS model controls the $x_2$-dependent slope $1/\kappa$,
related to the wall-normal mixing of the flow by the attached eddies.
Note that this is not strictly the case, and some coupling is expected
between all terms in (\ref{eq:shift}). For example, the value of the
integrand at the wall will depend on the slip lengths.

%===================================================================%
\subsection{Turbulence intensities and Reynolds stress contribution}
\label{sec:UV_RMS}
%====================================================================%

\begin{figure}
%\centerline{ \psfrag{X}[cc]{$x_2/\delta$} \psfrag{Y}[bc]{$\langle
%u_1^{'2+}\rangle^{1/2}$}
%\subfloat[]{{\includegraphics[width=0.48\textwidth]{./figs/urms_sl}
%}} \hspace{0.1cm} \psfrag{X}[cc]{$x_2/\delta$}
%\psfrag{Y}[bc]{$\langle u_2^{'2+}\rangle^{1/2}$}
%\subfloat[]{{\includegraphics[width=0.48\textwidth]{./figs/vrms_sl}
%}} }
\centerline{ 
\psfrag{X}[cc]{$x_2/\delta$} 
\psfrag{Y}[bc]{$\langle u_i'^{2+}\rangle^{1/2}$}
\subfloat[]{{\includegraphics[width=0.48\textwidth]{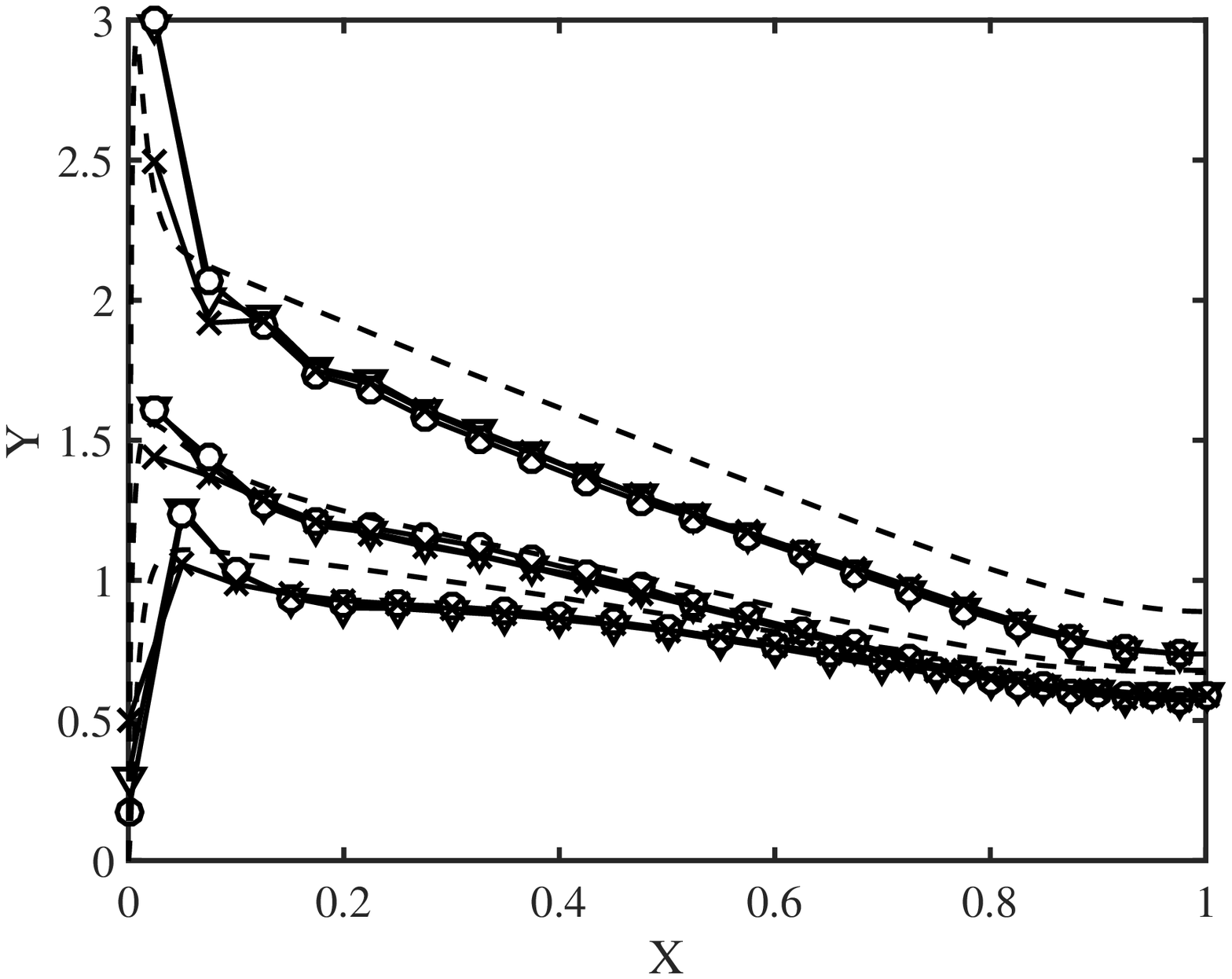}}} 
\hspace{0.1cm} 
\psfrag{X}[cc]{$x_2/\delta$} 
\psfrag{Y}[bc]{$\langle u_1^+u_2^+\rangle$}
\subfloat[]{{\includegraphics[width=0.48\textwidth]{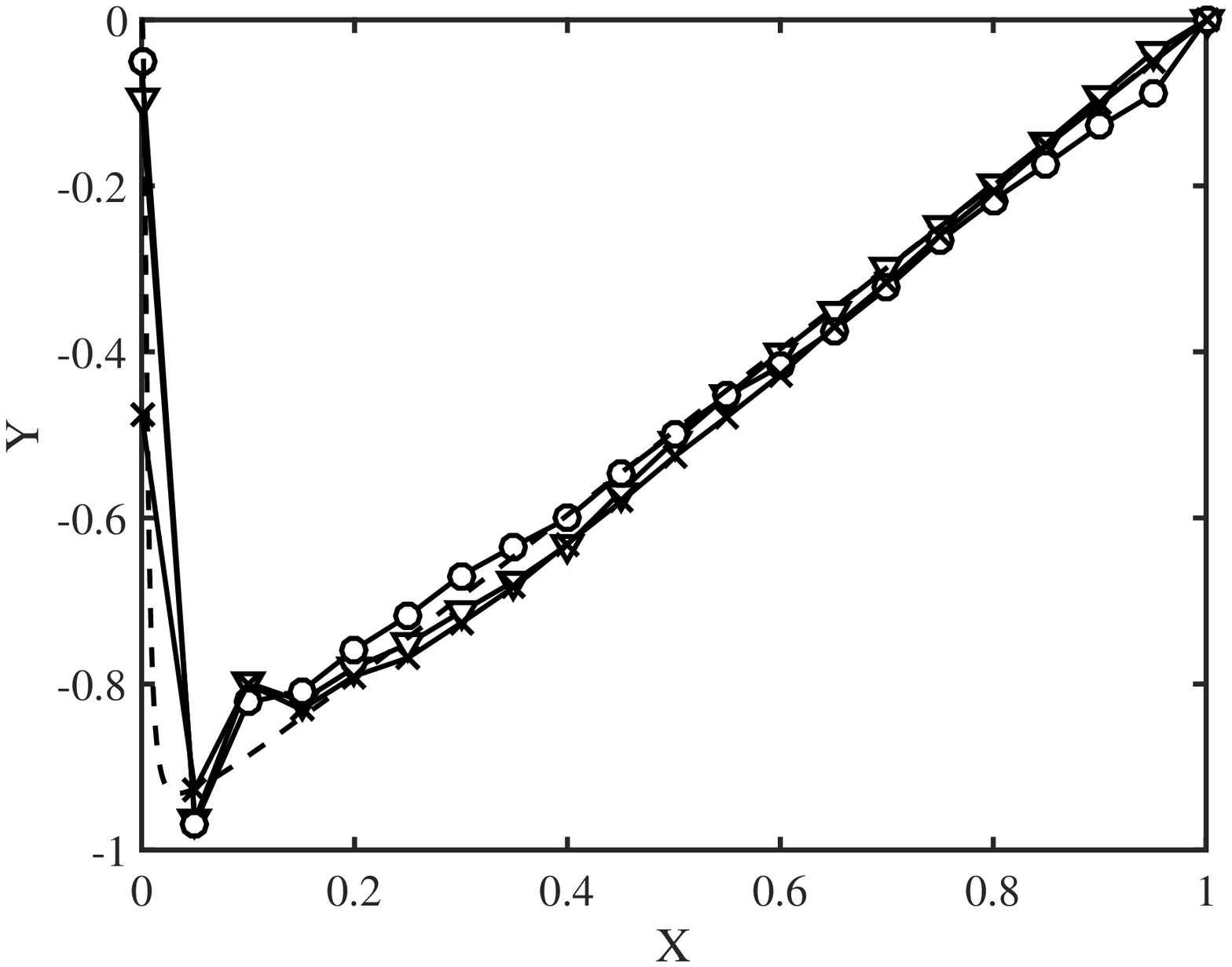} }} }
\caption{(a) Streamwise, spanwise and wall-normal rms velocity
fluctuations (from top to bottom), and (b) mean Reynolds stress
contribution for DSM-2000
$(l_1,l_2)=(0.008\delta,0.008\delta)$, $\circ$; DSM-2000-s1
$(l_1,l_2)=(0.008\delta,0.004\delta)$, $\triangledown$; DSM-2000-s3
$(l_1,l_2)=(0.097\delta,0.045\delta)$, $\times$; and DNS (dashed line) 
\label{fig:RMS_uv}}
\end{figure}

% RMS 
The sensitivity of $\langle u_i'^{2}\rangle^{1/2}$ to the choice
of slip lengths is examined in figure \ref{fig:RMS_uv}(a).  Three
cases DSM-2000, DSM-2000-s1, and DSM-2000-s3 are considered, two of
them supplying the correct mean velocity profile. The rms velocities
are insensitive, especially away from the wall, even when the slip
lengths are such that the mean velocity profile does not match that of
DNS. The most noticeable difference is observed at the wall, where
larger $l_1$ results in smaller rms values. The consequence is that
even if the mean velocity profile matches that of DNS for an optimal 
$(l_1,l_2)$, some pairs are preferred in order to avoid the near-wall
under- and over-shoot in the rms velocity fluctuations. A more
comprehensive study of the near-wall turbulent intensities with the
slip boundary condition can be found in \citet{Bae2018}.
%
%-----------------------------------------------------------%
%-----------------------------------------------------------%

% mean stress
Figure \ref{fig:RMS_uv}(b) stresses another important property of the
slip condition.  As was the case for rms velocities, the changes in
the mean tangential Reynolds stresses are negligible to varying values
of $l_1$ and $l_2$, except at the walls. For the case with larger
$l_1$, the wall-stress contribution from the
$\langle\bar{u}_1\bar{u}_2\rangle$ is roughly 50\%, and the remaining
stress is then carried by the SGS and viscous terms. This can be
considered an advantage compared to the classic no-transpiration
condition since the SGS model, usually known to under-predict the wall
stress \citep{Jimenez2000}, is not constrained to account for the
resolved non-zero $\bar{u}_i \bar{u}_j$ at the wall.

\begin{figure} 
\centerline{ 
\psfrag{X}[cc]{$x_1/\delta$}
\psfrag{Y}[bc]{$x_3/\delta$}
\subfloat[]{{\includegraphics[width=0.5\textwidth]{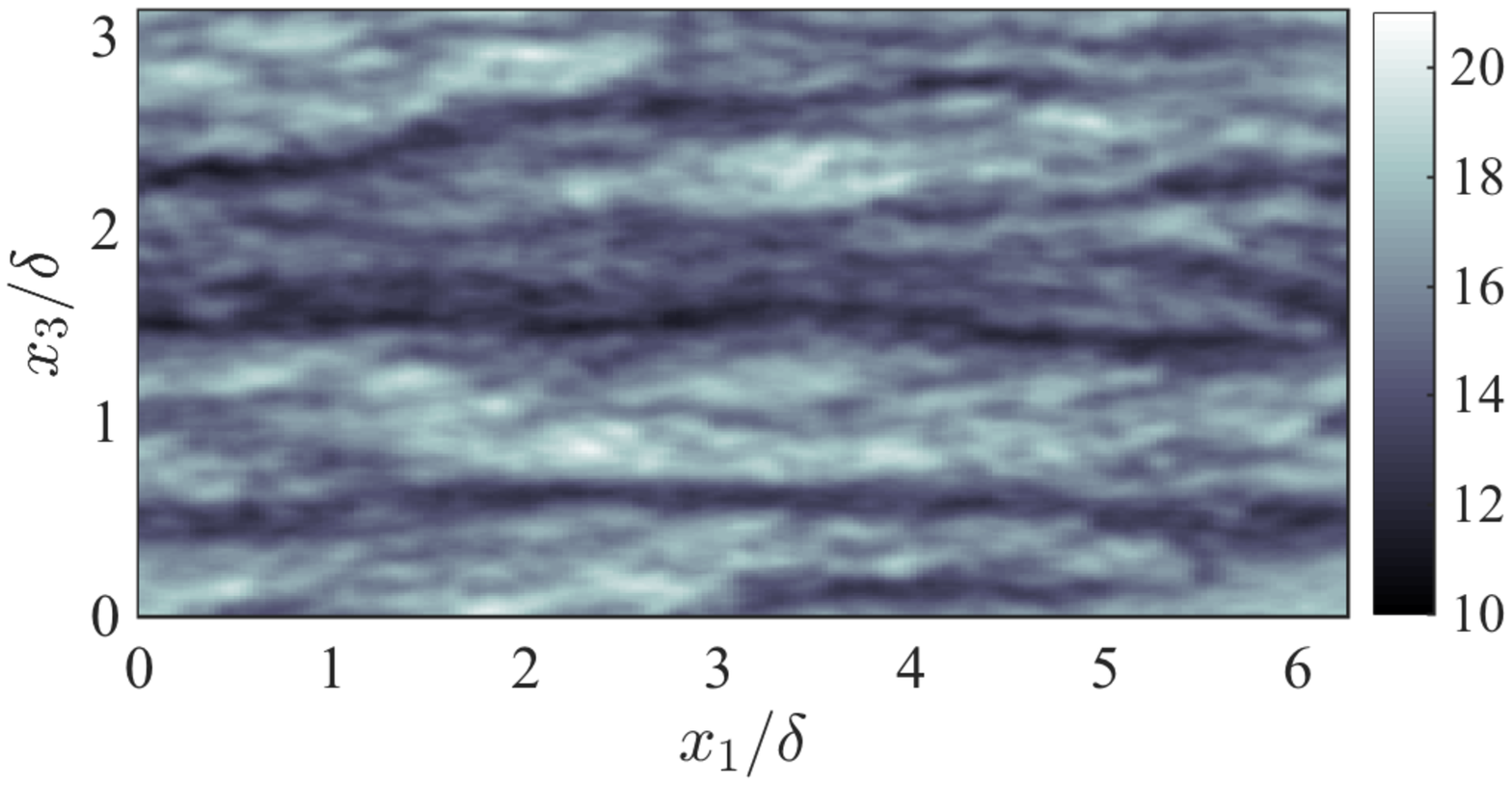}}} 
\psfrag{X}[cc]{$x_1/\delta$} 
\psfrag{Y}[bc]{$x_3/\delta$}
\subfloat[]{{\includegraphics[width=0.5\textwidth]{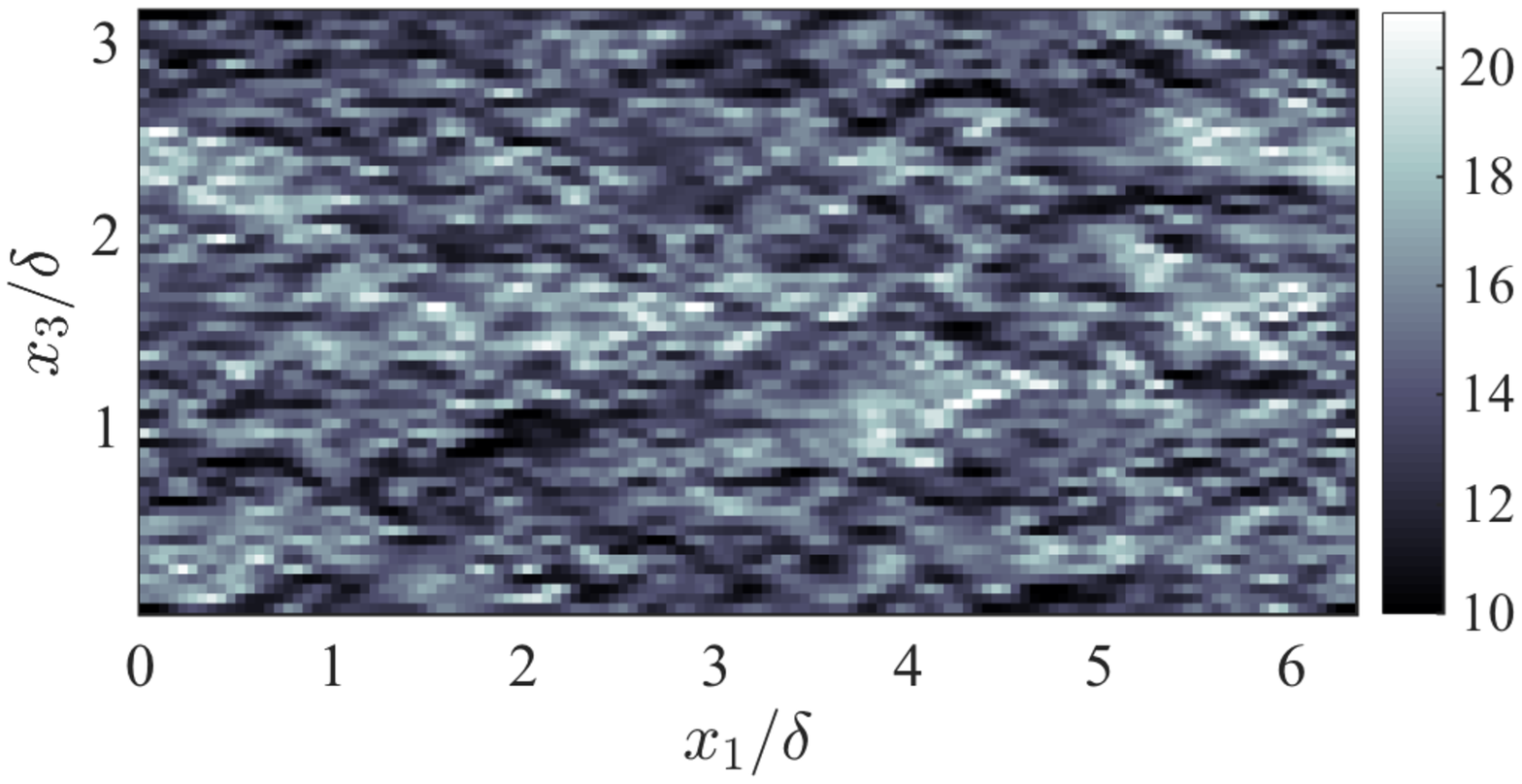}}} }
\caption{Instantaneous snapshot of the streamwise velocity at the wall
for (a) box-filtered DNS ($\Rey_\tau\approx2000$), and (b)
wall-modelled LES (DSM-2000) of channel flow for the $x_1-x_3$ plane.
For both cases, the filter or grid size is $\Delta_i/\delta = 0.050$,
$i=1,2,3$. Colours indicates velocity in wall units.
\label{fig:wall_snapshot}} \end{figure}

Finally, the structure of the streamwise velocity at the wall for
filtered DNS and wall-modelled LES is shown in figure
\ref{fig:wall_snapshot}. The filtered DNS data was obtained by
box-filtering the streamwise velocity with filter size $\Delta_i =
0.050\delta$, $i=1,2,3$, that coincides with the LES grid resolution. Although
our analysis is qualitative, the figures show that despite the
comparable intensities, the filtered DNS is organised into more
elongated streaks. Also note that for a constant $l_1$, the slip
boundary condition forces the velocity and its wall-normal derivative
to have the same structure close to the wall that is inconsistent
with box-filtered DNS. This suggests that an accurate representation
of the flow structure at the wall is neither expected nor necessary in
order to obtain accurate predictions of the low-order flow statistics
far from the wall. This is consistent with previous studies indicating
that the outer layer dynamics are relatively independent of the
near-wall cycle
\citep{DelAlamo2006,Flores2006,Mizuno2013,Jimenez2013,Lozano-Duran2014_2,Dong2017}.
%
%-----------------------------------------------------------%
%-----------------------------------------------------------%

%====================================================================%
\subsection{Sensitivity to SGS model, Reynolds number, and grid
resolution} \label{sec:SGS}
%====================================================================%

In this section, we study the effect of the SGS model, Reynolds
number, and grid resolution on the mean velocity profile for the slip
boundary condition. The discussion is necessary for understanding the
most relevant sensitivities of wall models based on the slip boundary
condition.

\begin{figure} 
\centerline{ 
\psfrag{X}[cc]{$x_2/\delta$}
\psfrag{Y}[bc]{$\langle u_1^+\rangle$}
\subfloat[]{{\includegraphics[width=0.48\textwidth]{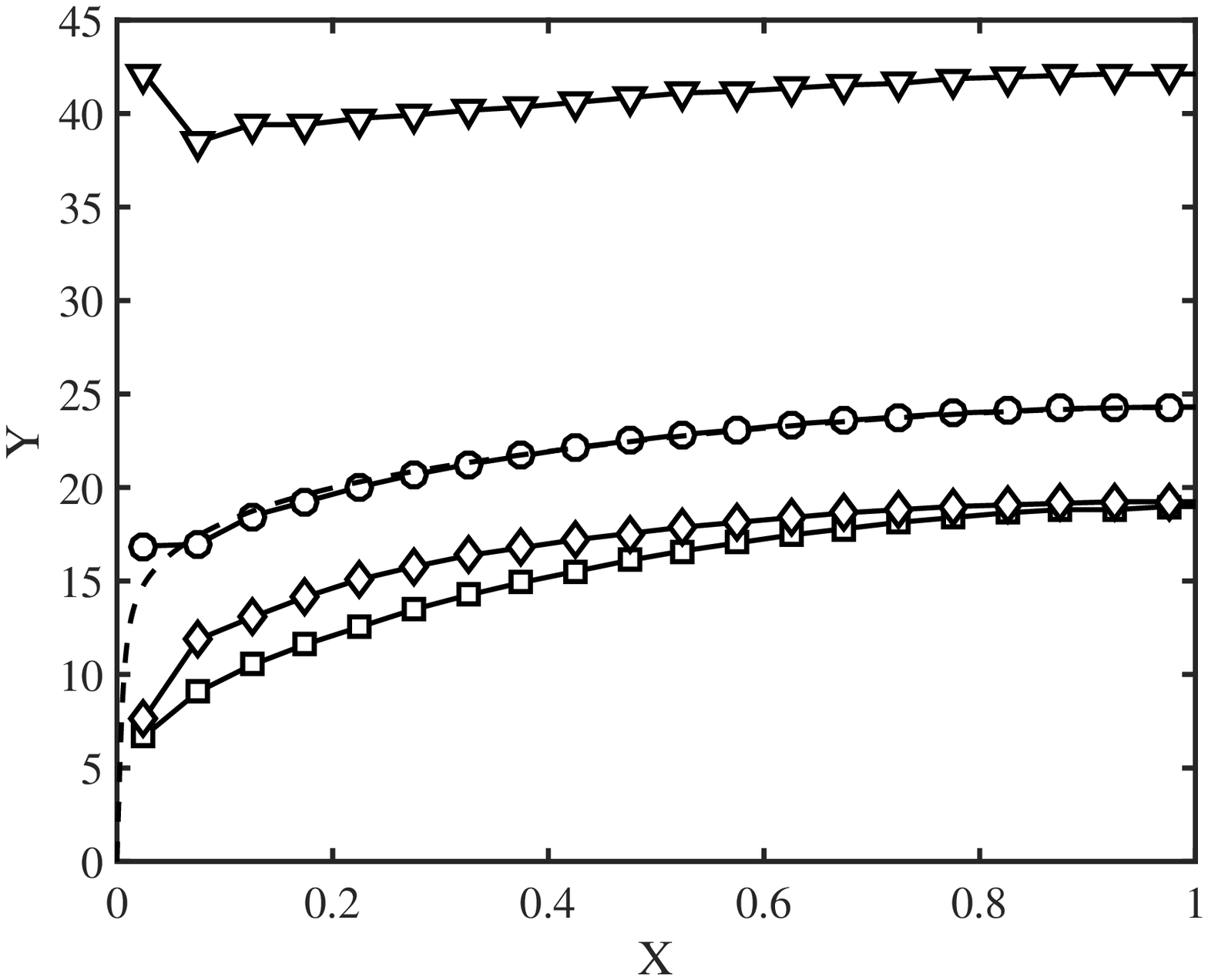}}} 
\hspace{0.3cm} 
\psfrag{X}[cc]{$x_2^+$} 
\psfrag{Y}[bc]{$\langle u_1^+ - \Delta u\rangle$}
\subfloat[]{{\includegraphics[width=0.48\textwidth]{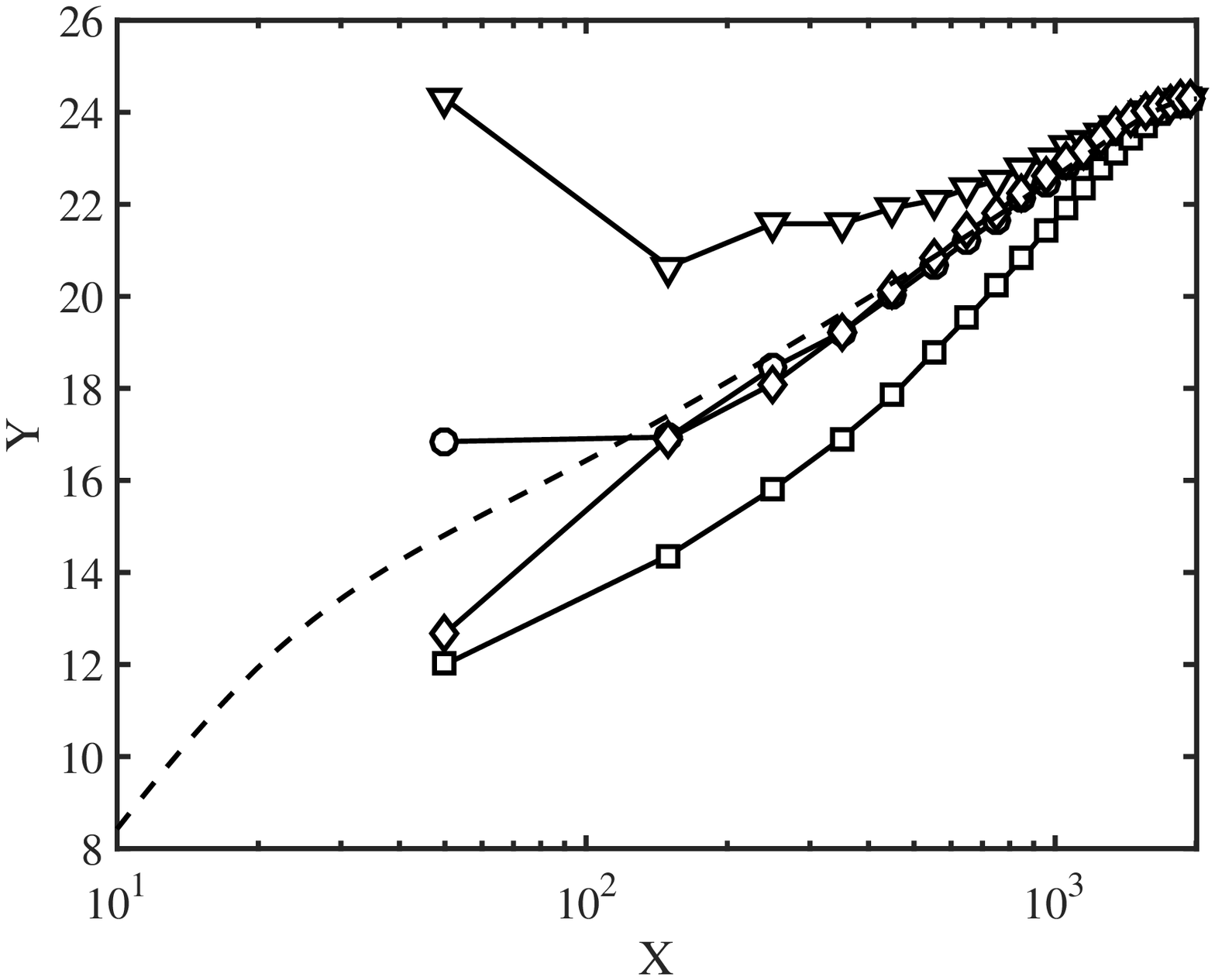}}} }
\caption{(a) Effect of SGS models on the mean velocity profile for
$l_i=0.008\delta$. (b) The mean velocity profiles have been
shifted to compare the shapes of the mean velocity profile, where the
shift is given by $\Delta u = u_1^+(\delta)-u_1^{+\text{DNS}}(\delta)$.
Dynamic Smagorinsky model ($\circ$), constant coefficient Smagorinsky
model ($\square$), anisotropic minimum-dissipation model ($\lozenge$),
and no model ($\triangledown$) are given for the turbulent channel
with $\Rey_\tau\approx2000$.  DNS (dashed line).  \label{fig:SGS}}
\end{figure}

Figure \ref{fig:SGS} shows the sensitivity of the mean velocity to
different SGS models for DSM-2000, SM-2000, AMD-2000, and NM-2000. In
all of the cases, the slip lengths are fixed and equal to
$0.008\delta$ such that the velocity profile at the centre of the
channel for DSM-2000 matches the DNS data.  Note that
this particular choice is arbitrary, and that alternative values of
slip lengths could be selected to find the best match between SM-2000,
AMD-2000, or NM-2000 and DNS. However, the results below are
independent of this choice, since the relative shift between cases is
barely affected. The results in figure \ref{fig:SGS} reveal that not
only the shape but also the mean mass flow, and thus the
optimal slip lengths for each SGS model, are impacted by the
SGS model at grid resolutions typical of wall-modelled LES.
%
%-----------------------------------------------------------%
%-----------------------------------------------------------%
%
Regarding the shape of $\langle\bar{u}_1\rangle$ (figure
\ref{fig:SGS}b), for low-dissipation SGS models (e.g. NM), the flow
becomes more turbulent, causing the mean velocity profile to flatten
due to the enhanced mixing. On the other hand, for highly-dissipative
SGS models (e.g. SM), the shape approaches a parabolic profile, that
is closer to the laminar solution. 

The effect of each SGS model on the mass flow rate in wall units can
be understood by considering the definition of the friction velocity,
\begin{equation} 
u_\tau^2 = -\langle \bar{u}_1\bar{u}_2 |_w\rangle + \left\langle \nu
\frac{\partial \bar{u}_1}{\partial x_2} \biggr|_w\right\rangle +
\left\langle \nu_t \frac{\partial \bar{u}_1}{\partial x_2}
\biggr|_w\right\rangle.
\label{eq:utau} 
\end{equation}
For a channel flow driven by a constant mass flow rate, the last term
in (\ref{eq:utau}) is zero for LES without an SGS model, which results
in lower $u_\tau$ and therefore a positive shift of the
mean velocity profile scaled in wall units.  For non-zero eddy
viscosity, the mean SGS stress at the wall will contribute to increase
$u_\tau$, creating a negative shift in the mean velocity
profile in wall units. The actual impact of the SGS model on $u_\tau$
is more intricate due to the coupling between $\nu_t$ and the flow
velocities.  However, the qualitative behaviour of $u_\tau$ described
above still holds.

Conversely, the effects on the mean velocity profile can also be
explained for a channel flow driven by a constant pressure gradient.
In this case, the left-hand side of (\ref{eq:utau}) is fixed.
Hence, variations in the SGS stress at the wall must be
compensated by variations in the Reynolds and viscous stress terms. We
have observed that these changes are balanced by the viscous stress,
$\nu\left.\partial\bar{u}_1/\partial x_2\right|_w$, rather than by the
Reynolds stress term. The variation in the mass flow can then be
understood through the slip boundary condition, where larger
$\left.\partial\bar{u}_1/\partial x_2\right|_w$ implies a larger slip
at the wall and, hence, higher mass flow.

%-----------------------------------------------------------%
\begin{figure} 
\centerline{ 
\psfrag{X}[cc]{$x_2/\delta$}
\psfrag{Y}[bc]{$\langle \nu_t\rangle/\nu$}
\subfloat[]{{\includegraphics[width=0.48\textwidth]{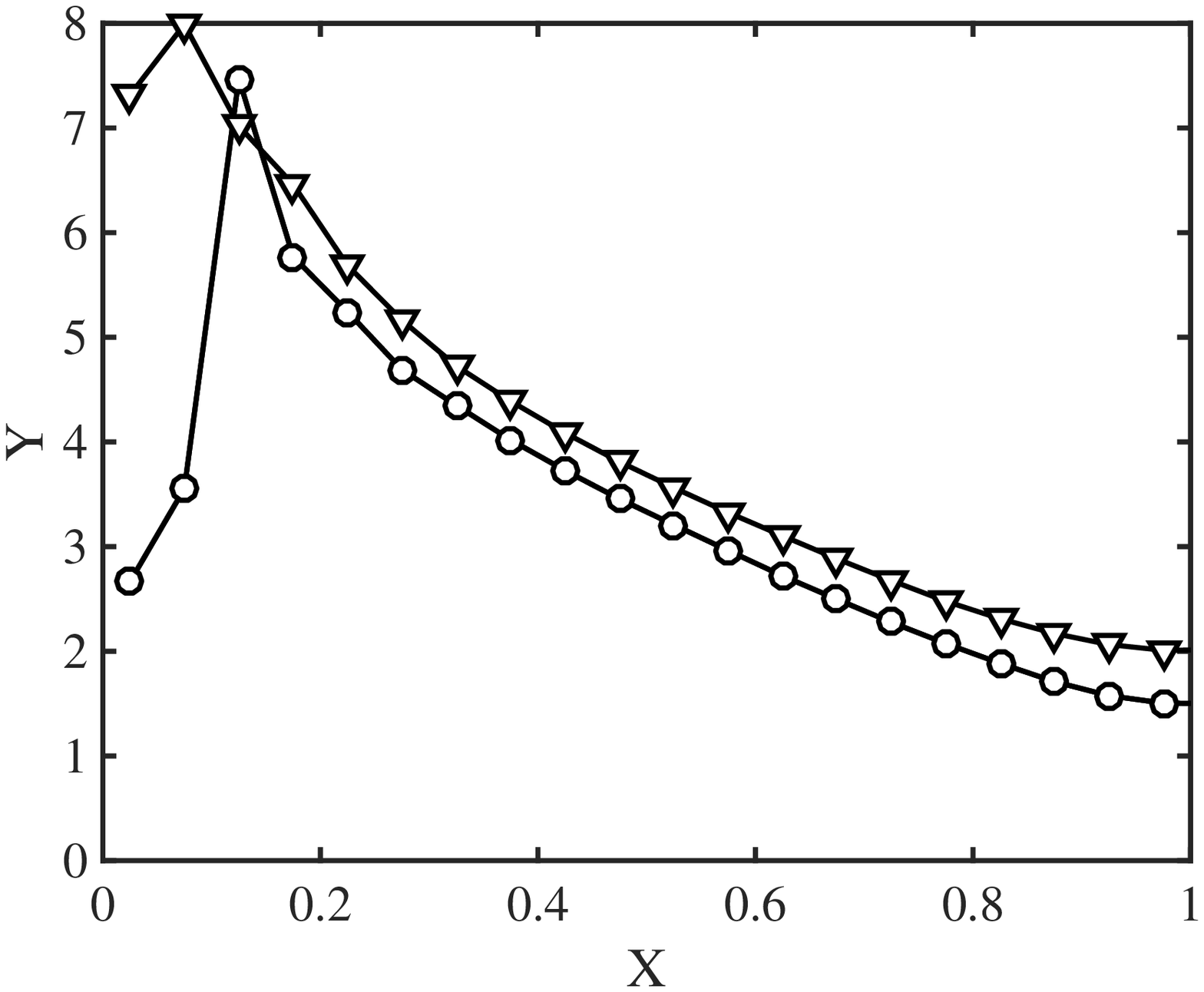}}} 
\hspace{0.3cm} 
\psfrag{X}[cc]{$x_2/\delta$} 
\psfrag{Y}[bc]{$\langle \nu_t\rangle/\nu$}
\subfloat[]{{\includegraphics[width=0.48\textwidth]{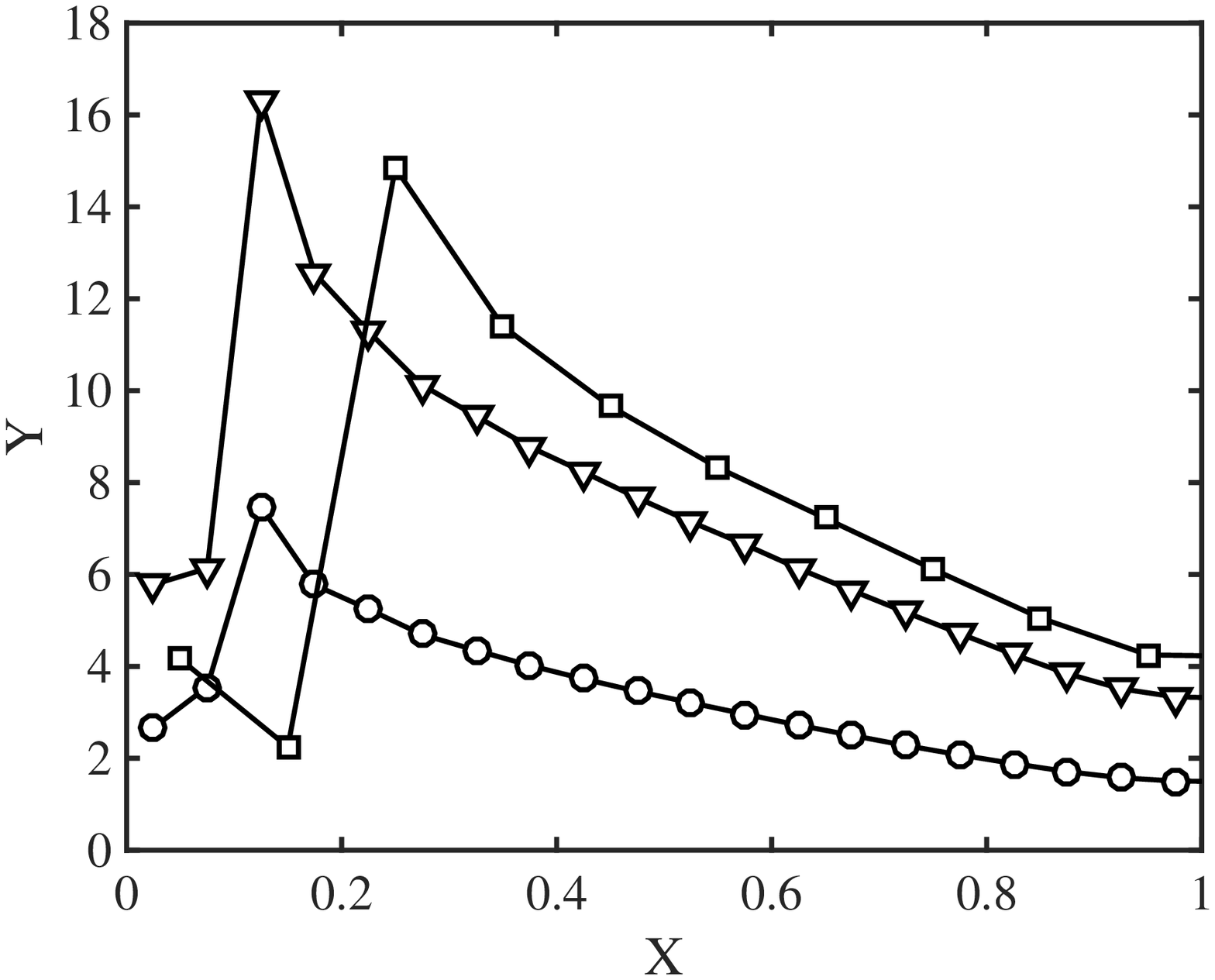}}} 
}
\caption{Eddy viscosity, $\nu_t$, as a function of wall-normal height with $l_i = 0.008\delta$
for (a) DSM-2000 ($\circ$) and AMD-2000 ($\triangledown$), and
(b) DSM-2000 ($\circ$), DSM-4200 ($\triangledown$), and DMS-2000-c2 ($\square$).
\label{fig:nu_t}}
\end{figure}
%-----------------------------------------------------------%
%
Figure \ref{fig:nu_t}(a) shows the eddy viscosity as a function of
wall-normal height for various SGS models. In particular, the eddy
viscosity of DSM and AMD model are comparable far from the wall,
consistent with the shape of the mean velocity profile in the outer
region shown in figure \ref{fig:SGS}(b). However, $\nu_t$ differs
notably for DSM and AMD model in the first two grid points off the
wall, leading to the differences in mean velocity profile observed in
figure \ref{fig:SGS}(a). Hence, the results above are indicative of
the fact that different SGS models demand different optimal slip
lengths.

\begin{figure} 
\centerline{ 
\psfrag{X}[cc]{$x_2/\delta$}
\psfrag{Y}[bc]{$\langle u_1^+\rangle$}
\subfloat[]{{\includegraphics[width=0.48\textwidth]{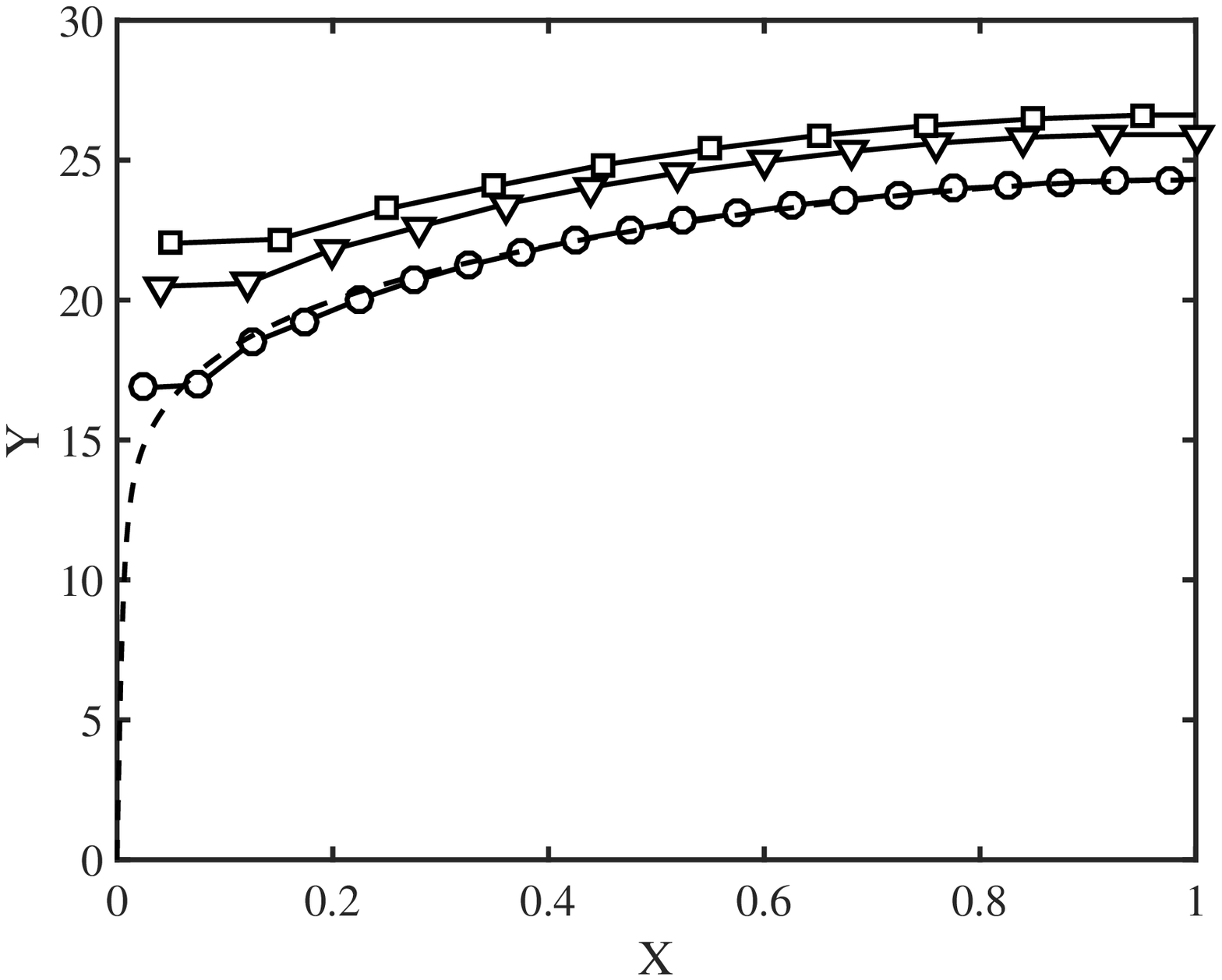}}} 
\hspace{0.3cm} 
\psfrag{X}[cc]{$x_2/\delta$} 
\psfrag{Y}[bc]{$\langle u_1^+\rangle$}
\subfloat[]{{\includegraphics[width=0.48\textwidth]{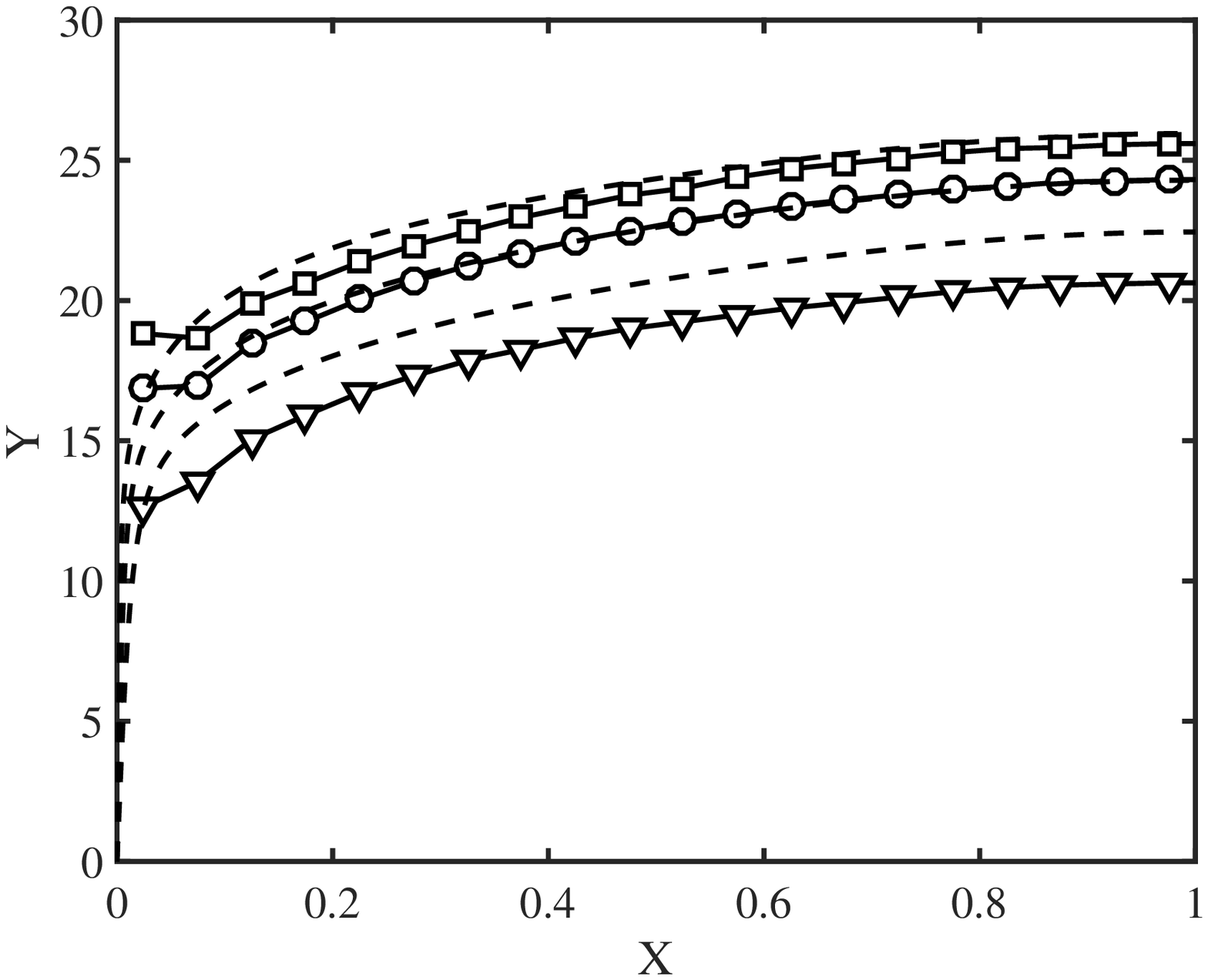}}} }
\caption{Effect of (a) the grid resolution, and (b) Reynolds number on
the mean velocity profile for $l_i = 0.008\delta$. (a)
$\Delta_i/\delta = 0.050$ ($\circ$), $0.077$($\triangledown$), and
$0.100$($\square$). (b) $\Rey_\tau\approx950$ ($\triangledown$),
$\Rey_\tau\approx2000$ ($\circ$), $\Rey_\tau\approx4200$, ($\square$). DNS
(dashed line).  \label{fig:grid_RE}} 
\end{figure}

The grid resolution and Reynolds number sensitivity are studied in
figure \ref{fig:grid_RE}, again maintaining constant slip lengths.
Regarding the resolution, coarsening the grid increases the mass flow.
This phenomenon is also observed in LES with no-slip boundary
condition and, in the present case, is probably related to an
inconsistency between the choice of slip lengths and the wall-normal
momentum flux provided by the SGS model. Figure \ref{fig:grid_RE}(b)
shows a weak dependence of the mean velocity profile on the Reynolds
number. The most notable observation is the under-estimation of the
mass flow for the lowest Reynolds number, but overall the optimal slip
lengths are quite insensitive to $\Rey_\tau$, consistent with our
analysis in section \ref{sec:apriori} regarding the effect of
$Re_\tau$.
%
%-----------------------------------------------------------%
%-----------------------------------------------------------%
%
Finally, increasing the Reynolds number or coarsening
the grid resolution augments  the eddy viscosity for DSM
(figure \ref{fig:nu_t}b), which is consistent with the expected
behaviour from  SGS models.

%====================================================================%
\subsection{The role of slip velocity in imposing zero mean mass flow
through the walls} \label{sec:transpiration}
%====================================================================%

% intro: flat plate and notation
As discussed in section \ref{sec:slip_constraint}, we require the slip
velocities to guarantee no net mass flow through the walls for flows
which are inhomogeneous in the wall-parallel direction. The requirement
is demonstrated here in an LES of zero-pressure-gradient flat-plate turbulent
boundary layer.

% simulation
The numerical method is similar to that of the channel flow presented
in section \ref{sec:numerics} with the
exception of the boundary conditions and the Poisson solver, which was
modified to take into account the non-periodic boundary conditions in
the streamwise direction. The simulation ranges from 
$\Rey_\theta\approx1000$ to 10 000, where $\Rey_\theta$ is the
Reynolds number based on the momentum thickness.  This range is
comparable to the boundary layer simulation by
\citet{Sillero2013} that will be used for comparisons.

% inflow and bc
The slip boundary condition from (\ref{eq:slip_bc}) is used at 
the wall, located at $x_2=0$. In the top plane, we
impose $u_1 = U_\infty$ (free-stream velocity), $u_3=0$, and $u_2$
estimated from the known experimental growth of the displacement
thickness for the corresponding range of Reynolds numbers as in
\citet{Jimenez2010}.  This controls the average streamwise pressure
gradient, whose nominal value is set to zero. The turbulent inflow is
generated by the recycling scheme of \citet{Lund1998}, in which the
velocities from a reference downstream plane, $x_\text{ref}$, are used
to synthesise the incoming turbulence. The reference plane is located
well beyond the end of the inflow region to avoid spurious feedback
\citep{Nikitin2007,Simens2009}. In our case,
$x_\text{ref}/\theta_0 = 890$, where $\theta_0$ is the
momentum thickness at the inlet. 
A convective boundary condition is applied at the
outlet with convective velocity $U_\infty$ \citep{Pauley1990} and
small corrections to enforce global mass conservation
\citep{Simens2009}. The spanwise direction is periodic.
% dimensions, resolution, and time
The length, height and width of the simulated box are
$L_x = 1060\theta_\text{avg}$, $L_y =
18\theta_\text{avg}$ and $L_z = 35\theta_\text{avg}$, where
$\theta_\text{avg} \approx 2.12\theta_0$ denotes the momentum
thickness averaged along the streamwise coordinate. This domain size
is similar to those used in previous studies
\citep{Schlatter2010,Jimenez2010,Sillero2013}. The streamwise and
spanwise resolutions are $\Delta_1/\delta = 0.05$
($\Delta_1^+ = 118$) and $\Delta_3/\delta = 0.04$ ($\Delta_3^+ =
84.3$) at $\Rey_\theta\approx6500$.  The number of wall-normal grid
points per boundary layer thickness is chosen to be $\sim 20$ at
the inlet, which is in line with the channel flow simulations in the
previous sections. The grid is slightly stretched in the wall-normal
direction with minimum $\Delta_2/\delta = 0.01$
($\Delta_2^+ = 20.8$). All computations were run with CFL=0.5 and for
50 washouts after transients. 
%Figure \ref{fig:bl_snapshot} shows instantaneous snapshots of the
%three velocities in the $x_1$-$x_2$ plane.
%
%\begin{figure} 
%\centerline{
%\includegraphics[width=\textwidth]{./figs/BL_snapshot} }
%
%\caption{Instantaneous snapshot of the streamwise (top), wall-normal
%(middle), and spanwise (bottom) velocities for wall-modelled LES of
%flat-plate boundary layer.  \label{fig:bl_snapshot}} 
%\end{figure}

% Cf -- slip lengths, addition of v_i term
The slip lengths are computed to match the empirical friction
coefficient, $C_f$, from \citet{White2006}. The connection between the
slip parameters and the friction coefficient is
\begin{equation} 
\frac{1}{2} U_\infty^2 \langle C_f \rangle =
\nu\left.\left \langle \frac{\partial
\bar{u}_1}{\partial{x_2}}\right|_w\right\rangle -\left.\left\langle
l_1l_2\frac{\partial\bar{u}_1}{\partial
x_2}\frac{\partial\bar{u}_2}{\partial x_2}\right|_w\right\rangle
+\left.\left\langle l_1v_2\frac{\partial \bar{u}_1}{\partial
x_2}\right|_w\right\rangle
-\left.\left\langle\tau^\text{SGS}_{12}\right|_w\right\rangle,
\label{eq:l_2_bl} 
\end{equation}
which is equivalent to (\ref{eq:wall_stress}) with the slip boundary
condition applied to the $u_1u_2$ term. The slip lengths are now a
function of the streamwise coordinate to take into account the
inhomogeneity of the flow in $x_1$. In order to ensure numerical
stability, exponential filtering in time with filter size
$0.2\delta/u_\tau$ was applied to the slip lengths in addition to
averaging in the homogeneous direction.  Equation (\ref{eq:l_2_bl}) is
key to guarantee the correct wall stress, but we have the freedom to
impose two more conditions to fully determine $l_1$, $l_2$ and $l_3$.
For simplicity, we set $l_1 = l_2 = l_3 = l$, and compute
$l$ from (\ref{eq:l_2_bl}) with the correct $C_f$ prescribed. The
value of $v_2$ is computed at
each time step to ensure global zero mean mass flow through the wall such
that
\begin{equation} 
v_2(t + \Delta t) = - \langle u_2(x_1,0,x_3,t) \rangle_{w}, 
\label{eq:v_2_bl} 
\end{equation}
where $\Delta t$ is the time step and $\langle \cdot \rangle_{w}$
denotes average over the entire wall. The
average $v_2$ obtained is of the order of $10^{-3}U_\infty$. The slip
velocities $v_1$ and $v_3$ in (\ref{eq:slip_bc}) are set to zero.

\begin{figure} 
\centerline{
\includegraphics[width=0.8\textwidth]{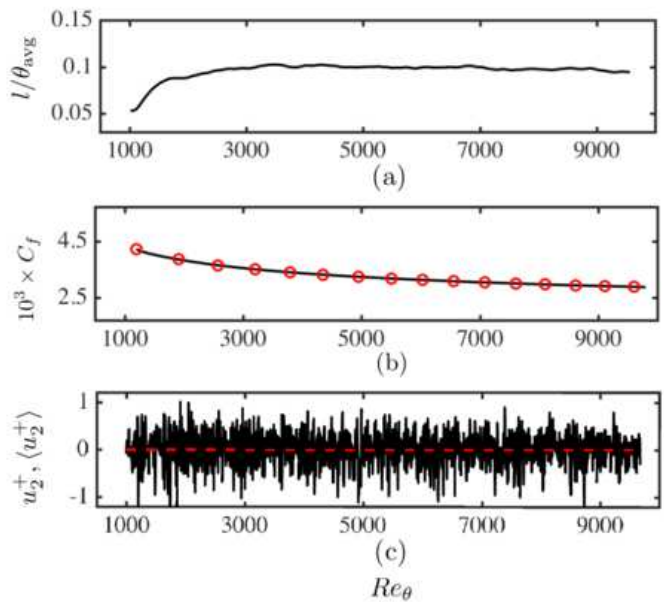}}
\caption{ (a) Mean slip lengths $l$ normalised by
$\theta_{\text{avg}}$, the average momentum thickness, (b) the
friction coefficient from the LES with slip boundary condition (black
---) and the empirical friction coefficient from \citet{White2006}
(red $\circ$), and (c) the instantaneous (black solid line) and
time-averaged (red dashed line) wall-normal velocities
at the wall as a function of $\Rey_\theta$.
\label{fig:bl_Re_theta}} 
\end{figure}

% results
Figure \ref{fig:bl_Re_theta}(a) shows the resulting mean slip lengths
computed to produce the target $C_f$, which is successfully
achieved as shown in figure \ref{fig:bl_Re_theta}(b). It is important
to stress again that one of the main differences of the boundary layer
case with respect to the channel flow is the necessity of a nonzero
$v_2$ term from (\ref{eq:slip_bc}) in order to guarantee that the wall
behaves as a no-transpiration boundary on average. We have implemented
a global condition (constant-in-space $v_2$, equation \ref{eq:v_2_bl})
that does not prevent instantaneous local mass flow at a particular
streamwise location as seen in figure \ref{fig:bl_Re_theta}(c).
However, the mean mass flow remains locally close to zero for all
streamwise locations.
%
%-----------------------------------------------------------%
%-----------------------------------------------------------%

\begin{figure} 
\centerline{ 
\psfrag{X}[cc]{$x_2^+$}
\psfrag{Y}[bc]{$\langle u_1^+\rangle$}
\subfloat[]{\includegraphics[width=0.48\textwidth]{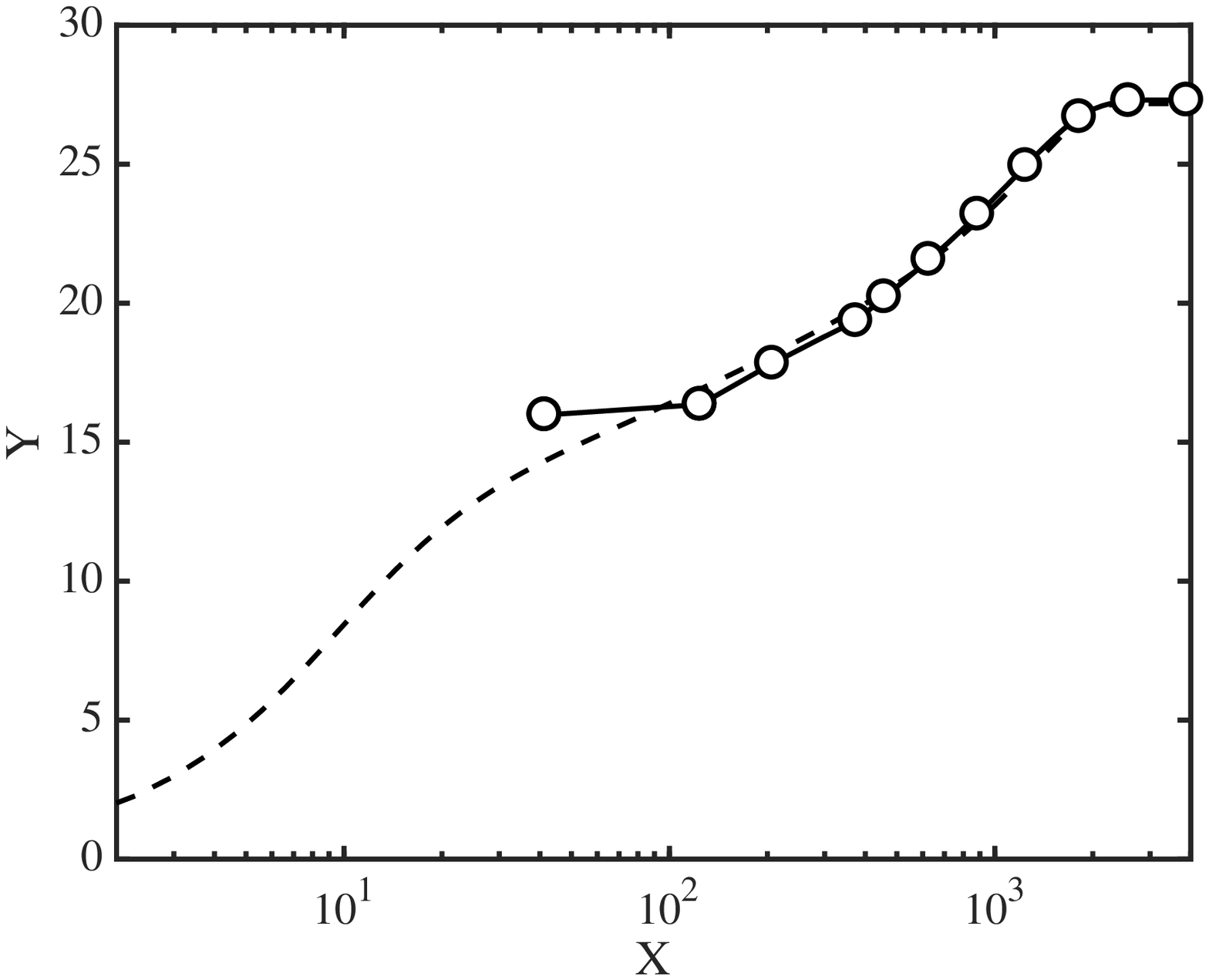}}
\hspace{0.2cm} 
\psfrag{X}[cc]{$x_2/\delta$} 
\psfrag{Y}[bc]{$\langle u_i'^{2+}\rangle^{1/2}$}
\subfloat[]{\includegraphics[width=0.48\textwidth]{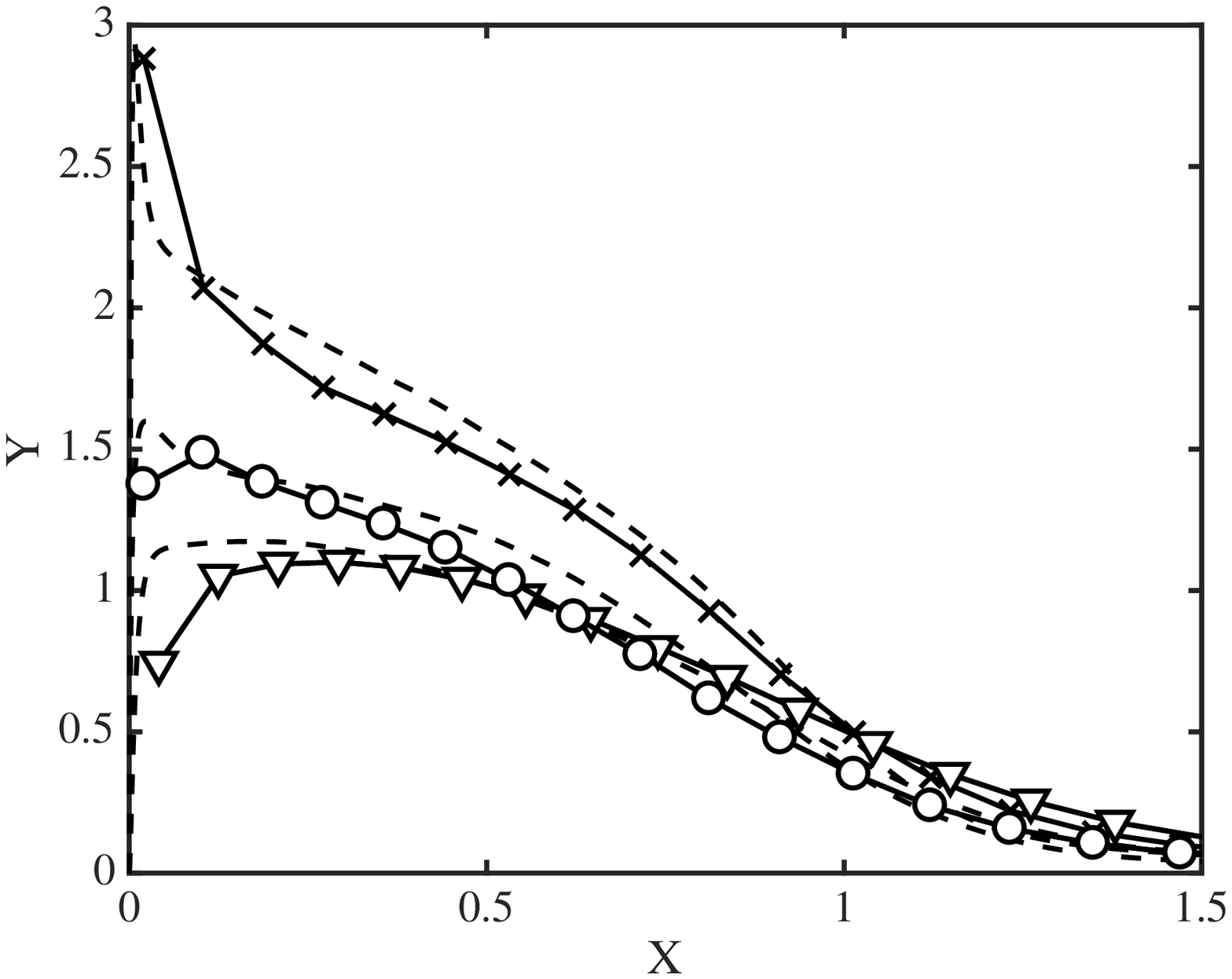}}
} 
\caption{ (a) Mean streamwise velocity profile and (b) rms streamwise
($\times$), spanwise ($\circ$), and wall-normal ($\triangledown$)
fluctuation profiles at $\Rey_\theta\approx6500$.
Symbols are for LES. DNS from \citet{Sillero2013} (dashed line).
\label{fig:bl_U_rms}}
\end{figure}

% Umean and rms profiles
The mean streamwise velocity and the three rms velocity fluctuations
at $\Rey_\theta\approx6500$ ($\Rey_\tau \approx 1989.5$)
are shown in figure \ref{fig:bl_U_rms} and compared with
\citet{Sillero2013}. As expected, the mean DNS and LES velocities
match in the wake region, as the correct $C_f$ in the LES is imposed.
The shape of the profile is also well predicted. The rms velocities
are reasonably well reproduced at this Reynolds number, with no
over-prediction of the streamwise rms velocity and under-prediction of
the other two components close to the wall, consistent with the
analysis in section \ref{sec:UV_RMS}.
%
%-----------------------------------------------------------%
%-----------------------------------------------------------%
%
Overall, these results along with those from LES of channel flow in
the previous sections show that the slip boundary condition
successfully reproduces the one-point statistics of the flow
as long as the slip lengths accurately reflect the correct mean wall
stress.

% mass flow
Two more cases (not shown) were run to test the effect of the slip
boundary condition on the net mass flow through the wall. In the first
case, a slip boundary condition was imposed such that the net mass
flow through the wall is positive (incoming flow through the wall)
such that $\langle \bar{u}_2|_w\rangle_w \approx 0.01 U_\infty$. In this case,
the boundary layer thickness grew five times faster than the reference
DNS. On the contrary, when the simulation was run with net negative
mass flow through the walls ($\langle \bar{u}_2|_w\rangle_w \approx -0.01
U_\infty$), the flow remained laminar. The results are consistent with
observations in previous studies on blowing and suction of boundary
layers \citep{Simpson1969,Antonia1988,Chung2001,Yoshioka2006} and
highlight the relevance of imposing a correct zero net mass flow
through the walls in order to faithfully predict the boundary layer
growth.

%====================================================================%
%\subsection{Stability of the slip boundary
%condition}\label{sec:stability}
%====================================================================%

%Although a systematic analysis of the stability of the slip boundary
%condition was not performed, all channel flow calculations included
%in section \ref{sec:numerics} were stable starting from a random
%initial condition for different grids and Reynolds numbers. However,
%this was not the case for LES of a flat-plate turbulent boundary
%layer in section \ref{sec:transpiration}. In the latter, a flow field
%generated from a previous coarse no-slip calculation was necessary as
%an initial condition for the simulation to run stably. The literature
%involving the slip boundary condition with transpiration is limited,
%but \citet{CartondeWiart2017} reported that compressible channel flow
%calculations with the slip boundary condition became unstable at high
%Reynolds numbers. 

%=====================================================================
\section{Dynamic wall models for the slip boundary condition}
\label{sec:dynamicWM}
%=====================================================================
% goal of wall models 
It is pertinent to discuss first the expected role of
wall models in LES. From section \ref{sec:pred_log} and previous
analysis in the literature \citep{Lee2013}, the most important
requirement for a wall model is to supply accurate mean tangential
stress at the wall.  This requirement must be accompanied by an
effective SGS model responsible for generating correct turbulence
statistics in the outer region, where the wall model plays a
secondary role.  The first requirement is necessary for obtaining the
correct bulk velocity, whereas the last point is crucial to predict
the shape of the mean velocity profile and rms velocity fluctuations
far from the wall (see sections \ref{sec:control} and
\ref{sec:UV_RMS}).

% traditional wall models: how they work
The wall models reviewed in the introduction are capable of meeting
the first requirement by assuming a specific state of the boundary
layer and relying on empirical parameters consistent with such state.
In this regard, most traditional wall models assume quasi-equilibrium
turbulence in the vicinity of the wall and encode explicitly or
implicitly information about the law of the wall, which cannot be
derived from first principles but can only be extracted from DNS or wind
tunnel experiments, such as the values of $\kappa$ and $B$. Despite the
equilibrium-turbulence assumption, current wall-modelling approaches
have been successful in predicting numerous flow configurations up to
date, although their performance in some regimes such as transitional
or separated flows as well as non-equilibrium turbulence is still
open to debate.

% dynamic wall models: the challenge
The main purpose of a dynamic wall model is similar to that of
traditional wall models, i.e., the estimation of accurate wall stress
$\tau_w$. However, the objective is to achieve this goal without prior
assumptions regarding the state of the boundary layer or embedded
empirical parameters. Instead, dynamic wall models aim to use only the current (local)
state of the LES velocity field and universal modelling assumptions
valid across different flow scenarios. Note that the task outlined
above is an outstanding challenge, since without any empirical
coefficients there is no explicit reference to how the near wall flow
should behave in different situations. Moreover, the instantaneous
velocity field is intertwined with the effects of the LES grid
resolution, Reynolds number, and SGS model choice as documented in
previous sections. Additionally, numerical errors are amplified at the
wall, and discretisation schemes are expected to play an important
role as well. Dynamic models must encompass these factors in
order to be of practical use, and whether this can be accomplished for
arbitrary flow configurations remains to be demonstrated. 

Despite the aforementioned difficulties, we provide below a
dynamic slip wall model that shows the ability to adapt to different
grid resolutions and Reynolds numbers as well as flow configurations,
provided an SGS model.  For a slip boundary condition of the form
(\ref{eq:slip_bc}), the problem of estimating $\tau_w$ can be
reformulated as finding the value of slip parameters that provides the
correct mean wall stress. The relationship between $l$ and $\tau_w$ was
shown in sections \ref{sec:control} and \ref{sec:transpiration} for
channels and boundary layers. Moreover, for the slip boundary
condition to be used as a predictive tool in wall-modelled LES, the
computed $l_i$ and $v_i$ should comply with the observations discussed
in the previous sections.

%====================================================================%
\subsection{Previous dynamic models}\label{sec:previous}
%====================================================================%

\citet{Bose2014} introduced a dynamic wall model based on the slip
boundary condition free of any \emph{a priori} parameters. The slip
length, assumed to be equal for the three spatial directions, is
computed via a modified form of the Germano's identity
\citep{Germano1991},
\begin{equation}
l^2\left(\Delta_R^2 \frac{\partial\hat{\bar{u}}_i}{\partial
n}\frac{\partial\hat{\bar{u}}_j}{\partial n} -
\frac{\partial\bar{u}_i}{\partial n}\frac{\partial\bar{u}_j}{\partial
n}\right) + T_{ij}^\text{SGS} - \widehat{\tau_{ij}}^\text{SGS} =
\widehat{\bar{u}_i\bar{u}}_j - \bar{u}_i\bar{u}_j, 
\label{eq:bose} 
\end{equation}
where $l$ is the slip length, $(\hat{\cdot})$ is the test filter,
$\Delta_R$ is the filter size ratio between the test and grid filters,
$\tau_{ij}^\text{SGS}$ and $T_{ij}^\text{SGS}$ represent the grid and
test filter SGS tensors, respectively. Equation  (\ref{eq:bose}) is
then solved for $l$ by using least-squares.

In \citet{Bose2014}, the model was tested for a series of LES of
turbulent channel flow and NACA 4412 airfoil. However, our
attempts to reproduce the channel flow results did not perform as
expected with our current implementation, which uses a different SGS
model and numerical discretisation.  The discrepancies motivated a
deeper study of the slip boundary condition and investigation of
alternative dynamic wall models as the one presented in the next
section.

%====================================================================%
\subsection{Wall-stress invariant dynamic wall
model}\label{sec:new_model}
%====================================================================%

% Outline:
% - Goal: dynamics wall model based on slip bs and invariance of T_ij combination
% - Define T_ij and provide model
% - Motivation of the model reasoning with 12 component
% - Apply slip boundary condition to the model (final model) 
% - Plots explainig the mechanisim

% intro model: slip and grid and test level
We present a dynamic wall model formulated for the slip boundary
condition based on the invariance of wall stress under test filtering.
We will refer to this new model as wall-stress invariant model or
WSIM. We will assume a single slip length $l_1=l_2=l_3=l$ and neglect any
potential contribution from the terms $v_i$. The goal is to define a
dynamic procedure to obtain $l$ at each time step such that the mean
wall stress obtained from (\ref{eq:wall_stress}) is an accurate
representation of the stress resulting from solving the Navier-Stokes
equations with DNS resolution. 

% Reynolds stress at the wall
Let us consider the wall stress operator $\mathcal{T}_{ij}$ given by
\begin{equation}
\mathcal{T}_{ij}(\boldsymbol{u}) = -R_{ij}(\boldsymbol{u})|_w - \tau^\text{SGS}_{ij}(\boldsymbol{u})|_w 
+ 2 \nu S_{ij}(\boldsymbol{u})|_w - p(\boldsymbol{u})|_w\delta_{ij},
\label{eq:wall_stresses}
\end{equation}
where terms $R_{ij}$, $\tau_{ij}^\text{SGS}$, $S_{ij}$ and
$p\delta_{ij}$ are the Reynolds stress, subgrid stress, strain-rate,
and pressure tensors, respectively, computed from the specified
velocity field $\boldsymbol{u}$. In our formulation of
$\mathcal{T}_{ij}$, the wall is assumed to be smooth, but the wall
stress can be extended to encode information about the type of wall
(smooth, rough, hydrophobic, etc.) by adding the appropriate drag
terms into right-hand side of (\ref{eq:wall_stresses}).
The modelling choice for the dynamic wall model is
\begin{align}
\mathcal{T}_{ij}(\boldsymbol{\bar{u}}) - \mathcal{T}_{ij}(\boldsymbol{\hat{\bar{u}}}) &= 0, \label{eq:E1} \\
\mathcal{T}_{ij}(\boldsymbol{\hat{\hat{\bar{u}}}})  - \widehat{\mathcal{T}_{ij}}(\boldsymbol{\hat{\bar{u}}}) &= 0,  \label{eq:E2}
\end{align}
where the different wall stresses, $\mathcal{T}_{ij}$, are obtained by
either computing the stress of the filtered velocity field or by
filtering the total stress.
Condition (\ref{eq:E1}) enforces the
invariance of wall stress under test filtering and allows the wall
model to predict the same wall stress regardless of the grid
resolution (or filter). A similar approach was also adopted by
\citet{Anderson2011} for rough walls.  Condition (\ref{eq:E2}) is
analogous to the Germano's identity for the total stress of the
filtered velocity and can be interpreted as a consistency condition
between the wall stress and filter operator.  The proposed
model is given by combining (\ref{eq:E1}) and (\ref{eq:E2}) such that
\begin{equation} 
\mathcal{F}_{ij} = \mathcal{T}_{ij}(\boldsymbol{\bar{u}}) -
\mathcal{T}_{ij}(\boldsymbol{\hat{\bar{u}}})
+\mathcal{T}_{ij}(\boldsymbol{\hat{\hat{\bar{u}}}}) -
\widehat{\mathcal{T}}_{ij}(\boldsymbol{\hat{\bar{u}}}) = 0.
\label{eq:invar_ws}
\end{equation}
%Our modelling choice is to assume that the correct wall stress is
%predicted when 
%%
%\begin{equation} 
%\mathcal{T}^1_{ij} + \mathcal{T}^4_{ij} = \mathcal{T}^2_{ij} +
%\mathcal{T}^3_{ij}. 
%\label{eq:invar_ws}
%\end{equation}
%%
The rationale for this particular combination of \eqref{eq:E1} and 
\eqref{eq:E2} is given by considering
$\mathcal{F}_{12}$, the dominant shear stress component in a
boundary layer.  This term can be simplified
by test filtering only in the wall-normal direction with a box filter
with filter size $\hat{\Delta}+x_2$ at $x_2 < \hat{\Delta}$. Assuming
that the subgrid stress tensor is given by an eddy viscosity, i.e.,
$(\tau_{ij}^\text{SGS}-1/3\delta_{ij}\tau_{kk}^\text{SGS})(\boldsymbol{\bar{u}}) = -2\nu_t
S_{ij}(\boldsymbol{\bar{u}})$ and that $\nu_t$ is constant in the
near-wall region, the first order approximation of $\mathcal{F}_{12}$
can be shown to be
\begin{equation}
\mathcal{F}_{12} =
\frac{\hat\Delta}{2}\left(\frac{\partial\bar{u}_1\bar{u}_2}{\partial
x_2}-\frac{\partial}{\partial x_2}\left[(\nu+\nu_t)|_w \frac{\partial
\bar{u}_1}{\partial x_2}\right]\right) + \mathcal{O}(\hat\Delta^2).
\label{shear_F}
\end{equation}
Hence, $\mathcal{F}_{12}=0$ implicitly enforces the well-established 
constant stress
layer across the wall-normal direction.  
Note that (\ref{eq:E1}) and (\ref{eq:E2}) may be combined
differently to produce alternative versions of (\ref{eq:invar_ws}),
but not all these groupings lead to a first order approximation
consistent with the form reported in \eqref{shear_F}. For example,
enforcing only (\ref{eq:E1}) would not be consistant with \eqref{shear_F}. Nevertheless,
wall models using different variants of (\ref{eq:invar_ws}) 
may be constructed based on similar principles; a
broader family of dynamic models postulated on stress-invariant
principles was investigated in \citet{Lozano2017}.

Next we introduce an explicit dependence on the slip length condition
in (\ref{eq:invar_ws}). 
We will assume that the slip boundary condition also applies for the
test-filtered velocity field,
\begin{equation} 
\hat{\bar{u}}_i = \hat{l} \frac{\partial \hat{\bar{u}}_i}{\partial n},
\label{eq:bc_wm_test}
\end{equation}
where $\hat{l}$ is the slip length at the test filter level. We will
further suppose that a linear functional dependence of the slip length
with the filter size of the form $\hat{l} = \Delta_R l$ holds. This
assumption will be shown to be reasonably well satisfied in figure
\ref{fig:errors}(c), which contains \emph{a posteriori} values for
the optimal slip length (\ref{eq:l_2}) as a function of grid
resolution. Introducing the slip boundary condition at grid- and
test-filter levels for the Reynolds stress terms in
$\mathcal{T}_{ij}(\boldsymbol{\bar{u}})$ and $\mathcal{T}_{ij}(\boldsymbol{\hat{\bar{u}}})$, such that
\begin{equation} 
\bar{u}_i\bar{u}_j = l^2 \frac{\partial \bar{u}_i}{\partial
n}\frac{\partial \bar{u}_j}{\partial n},\quad \hat{\bar{u}}_i
\hat{\bar{u}}_j = l^2 \Delta_R^2 \frac{\partial
\hat{\bar{u}}_i}{\partial n}\frac{\partial \hat{\bar{u}}_j}{\partial
n},  
\label{eq:RS_wm}
\end{equation}
the wall-stress invariant model becomes 
\begin{equation} 
l^2 \left(\frac{\partial \bar{u}_i}{\partial
n}\frac{\partial \bar{u}_j}{\partial n} - \Delta_R^2 \frac{\partial
\hat{\bar{u}}_i}{\partial n}\frac{\partial \hat{\bar{u}}_j}{\partial
n}\right)  = \bar{u}_i \bar{u}_j - \hat{\bar{u}}_i \hat{\bar{u}}_j
+\mathcal{T}_{ij}(\boldsymbol{\bar{u}}) - \mathcal{T}_{ij}(\boldsymbol{\hat{\bar{u}}}) +\mathcal{T}_{ij}(\boldsymbol{\hat{\hat{\bar{u}}}}) -
\widehat{\mathcal{T}}_{ij}(\boldsymbol{\hat{\bar{u}}}),
%+\mathcal{T}^1_{ij} - \mathcal{T}^2_{ij} - \mathcal{T}^3_{ij} +
%\mathcal{T}^4_{ij},
\label{eq:gwm} 
\end{equation} 
which can be rewritten as
\begin{equation} 
l^2 \mathcal{M}_{ij} = \mathcal{L}_{ij} + \mathcal{F}_{ij},
\label{eq:gwm_c} 
\end{equation} 
with
\begin{equation} 
\mathcal{M}_{ij} = \frac{\partial \bar{u}_i}{\partial
n}\frac{\partial \bar{u}_j}{\partial n} - \Delta_R^2 \frac{\partial
\hat{\bar{u}}_i}{\partial n}\frac{\partial \hat{\bar{u}}_j}{\partial
n},
\quad 
\mathcal{L}_{ij} = \bar{u}_i \bar{u}_j - \hat{\bar{u}}_i
\hat{\bar{u}}_j,
\label{eq:gwm_c_term} 
\end{equation} 
The system (\ref{eq:gwm_c}) is over-determined and $l$ is computed via
least-squares as
\begin{equation} 
l^2 =
\frac{(\mathcal{L}_{ij}+\mathcal{F}_{ij})\mathcal{M}_{ij}}{\mathcal{M}_{ij}\mathcal{M}_{ij}}
= \frac{\mathcal{L} + \mathcal{F}}{\mathcal{M}},
\label{eq:gwm_c_least} 
\end{equation} 
where repeated indices imply summation and the compact notation
$\mathcal{L} = \mathcal{L}_{ij}\mathcal{M}_{ij}$,
$\mathcal{F}=\mathcal{F}_{ij}\mathcal{M}_{ij}$, and $\mathcal{M} =
\mathcal{M}_{ij}\mathcal{M}_{ij}$ is used. For incompressible flows,
the isotropic part of $\tau_{ij}^\text{SGS}$ is usually not defined by
the SGS models and, since the system is already over-determined, we
exclude the $i=j$ components of (\ref{eq:gwm_c_least}). 
%A similar
%approach founded on the invariance of the wall stress was adopted by
%\citet{Anderson2011} for modelling rough walls, and a broader family
%of dynamic models postulated on stress-invariant principles was
%investigated in \citet{Lozano2017}.

% Regulating mechanism: a priori testing 
%-----------------------------------------------------------%
\begin{figure}
\centerline
\subfloat{{\includegraphics[width=0.47\textwidth]{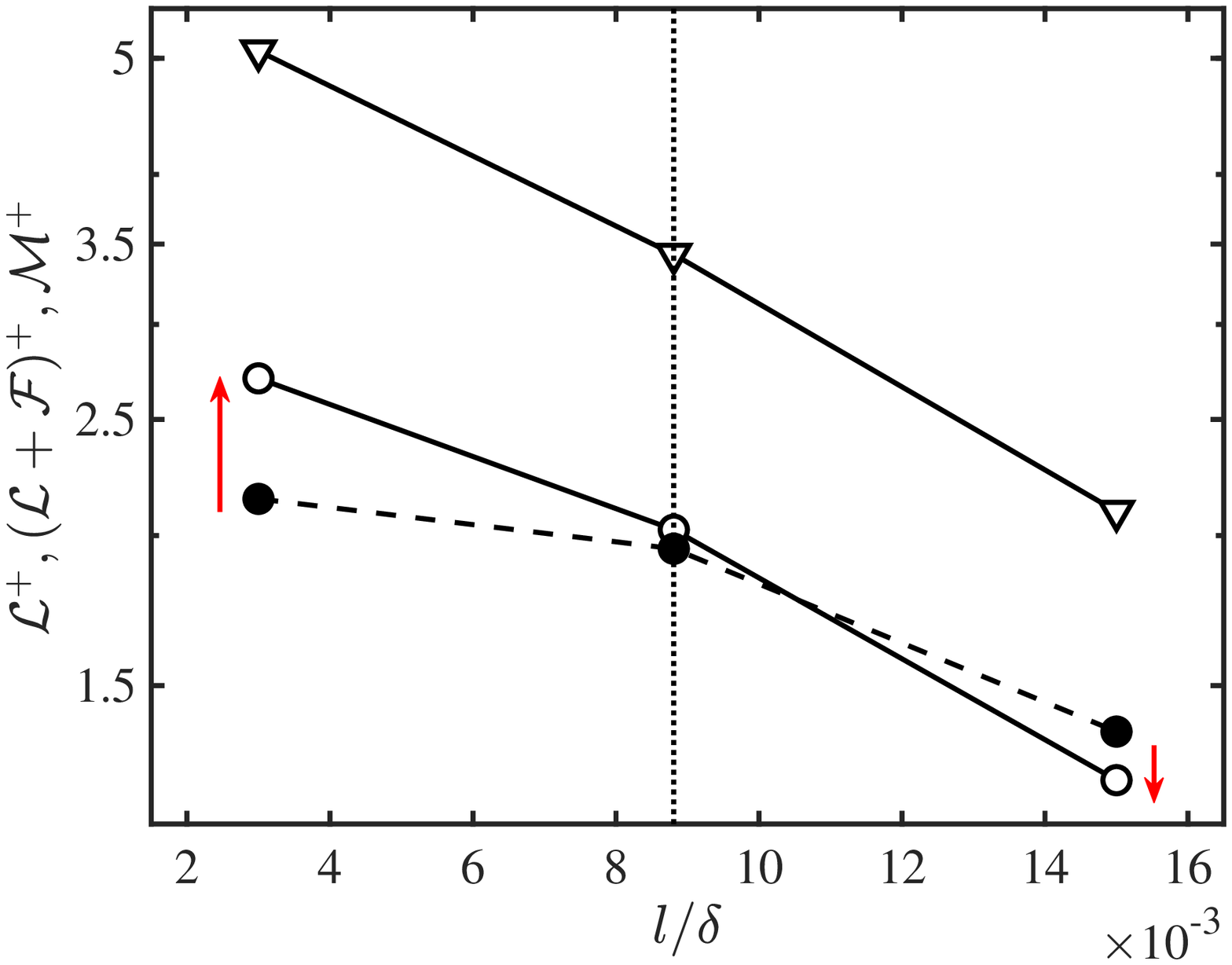}}
\hspace{0.1cm}
\subfloat{{\includegraphics[width=0.47\textwidth]{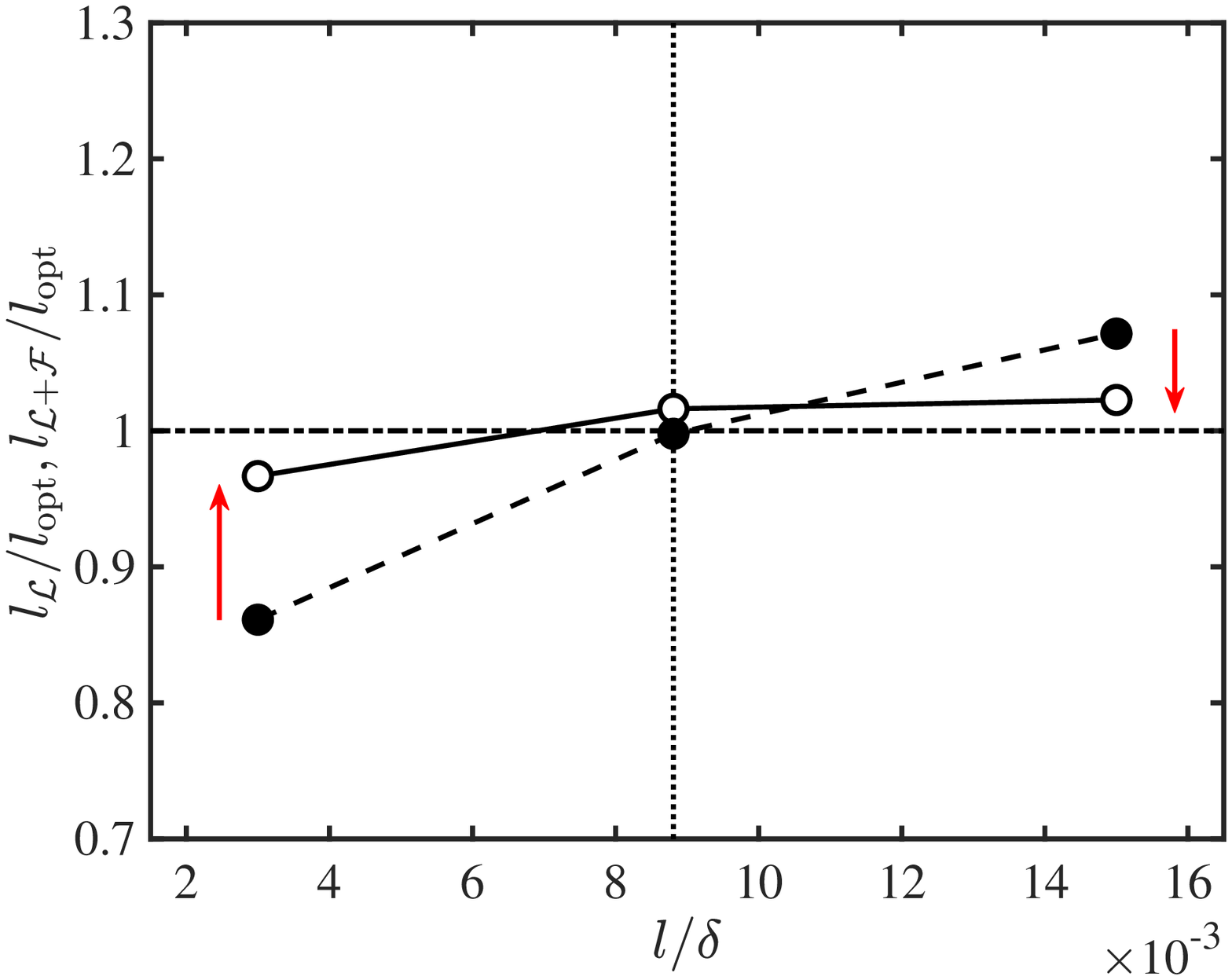}}}}
\caption{ (a) $\mathcal{L}$ ($\bullet$), $\mathcal{L}+\mathcal{F}$
($\circ$), and $\mathcal{M}$ ($\triangledown$) computed from 
LES of channel flow using the slip boundary condition with fixed $l$ equal to $l=0.35
l_\text{opt}=0.003\delta$, $l=l_\text{opt}=0.009\delta$, and $l=1.70
l_\text{opt}=0.015\delta$.  (b) The slip lengths $l_\mathcal{L}$
($\bullet$), $l_{\mathcal{L}+\mathcal{F}}$ ($\circ$) normalised by the optimal slip
length. $\Delta_R$ was assigned to be $1.6$ (see section
\ref{sec:test_cases_wm}). The vertical dotted lines are
$l=l_\text{opt}$. Red arrows highlight the improvement achieved by
including the control term $\mathcal{F}_{ij}$. See text for more
details.
\label{fig:wall_model_testing}}
\end{figure}
%-----------------------------------------------------------%
%
Note that the first part of the right-hand-side of (\ref{eq:gwm}),
$\bar{u}_i\bar{u}_j-\hat{\bar{u}}_i\hat{\bar{u}}_j$
($\mathcal{L}_{ij}$), is the result of applying the boundary condition
at the grid and test filter levels. The remaining terms,
$\mathcal{T}_{ij}(\boldsymbol{\bar{u}}) - \mathcal{T}_{ij}(\boldsymbol{\hat{\bar{u}}}) +\mathcal{T}_{ij}(\boldsymbol{\hat{\hat{\bar{u}}}}) -
\widehat{\mathcal{T}}_{ij}(\boldsymbol{\hat{\bar{u}}})$
%$\mathcal{T}_{ij}^1-\mathcal{T}_{ij}^2-\mathcal{T}_{ij}^3+\mathcal{T}_{ij}^4$
($\mathcal{F}_{ij}$), then act as an effective control such that the
slip length increases if the current wall stress is under-predicted,
and decreases if the wall stress is over-predicted.  This
self-regulating mechanism can be examined by analysing the terms
$\mathcal{M}$, $\mathcal{L}$, and $\mathcal{F}$ for three test cases
of a channel flow at $Re_\tau \approx 4200$ with DSM and grid
$\Delta_1=\Delta_2=\Delta_3=0.05\delta$. The first case is computed
by imposing the optimal slip length, $l = l_\text{opt}=0.009\delta$.
The second and third cases are analogous to the first case but with
$l = 1.70 l_\text{opt}$ and $l = 0.35 l_\text{opt}$, respectively. The
terms $\mathcal{M}$, $\mathcal{L}$, and $\mathcal{F}$ were evaluated
after the cases were run with their corresponding slip lengths fixed
in time until the statistically steady state was reached. Note that
$\mathcal{L}$ can be interpreted as the model prior to
applying the control mechanism, and this allows us to define two slip
lengths, namely, $l_\mathcal{L} = \mathcal{L}/\mathcal{M}$ and
$l_{\mathcal{L}+\mathcal{F}}= (\mathcal{L}+\mathcal{F})/\mathcal{M}$. 

The terms $\mathcal{M}$, $\mathcal{L}$, and $\mathcal{L}+\mathcal{F}$
evaluated from the three test cases are plotted in figure
\ref{fig:wall_model_testing}(a), and the corresponding slip lengths
$l_\mathcal{L}$ and $l_{\mathcal{L}+\mathcal{F}}$ are presented in figure
\ref{fig:wall_model_testing}(b). The results show that application of
WSIM recovers the optimal slip length, and thus the correct wall
stress, through the control mechanism $\mathcal{F}_{ij}$. The analysis
provided here is performed \emph{a priori}, that is, the wall model
was used to predict $l$ at time $t+\Delta t$ for a given flow field at
time $t$, but without an actual dynamic coupling between WSIM and LES.
The remainder of the paper is devoted to test WSIM in LES under
different test scenarios.  

%=====================================================================
\section{Performance of the wall-stress invariant
model}\label{sec:results_wm}
%=====================================================================

%=====================================================================
\subsection{Numerical experiments} \label{sec:test_cases_wm}
%=====================================================================

% test filter
To test the performance of WSIM, three flow configurations are
considered: a statistically steady plane turbulent channel (2--D
channel), a non-equilibrium three-dimensional transient channel (3--D
channel), and a zero-pressure-gradient flat-plate turbulent boundary
layer.  The numerical methods of the simulations are the same as the
ones given in sections \ref{sec:numerics} and \ref{sec:transpiration}.
The 2--D channel flow and the turbulent boundary layer were discussed
in section \ref{sec:first_order_stat}.  The three-dimensional
transient channel flow is a temporally developing turbulent boundary
layer in a planar channel subjected to a sudden spanwise forcing as in
\citet{Moin1990}.

% boundary conditions
The size of the 2--D and 3--D channel domain is $8\pi \delta\times
2\delta\times 3\pi \delta$ in the streamwise, wall-normal and spanwise
directions, respectively.  For both 2--D and 3--D channel flows,
periodic boundary conditions are applied in the streamwise and
spanwise directions.  For the top and bottom walls, we impose either
the no-slip (NS), slip boundary condition for WSIM, or Neumann
boundary condition for cases with the equilibrium wall model (EQWM).
The formulation for the EQWM follows \citet{Kawai2013} with a matching
location at the third grid cell for the streamwise velocity, although
recent studies have shown that the first grid cell may be used
for the EQWM when the velocities are filtered using a spatial or
temporal filter \citep{Yang2017} following the methodology first
introduced for algebraic wall models \citep{Bou-Zeid2004}. 

% size and run time
For the 2--D channel, the flow is driven by imposing a constant mean
pressure gradient and the simulations are started from a random
initial condition run for at least $100\delta/u_\tau$ after
transients. In the case of the 3--D channel, the calculations were
started from a 2--D fully developed plane channel flow driven by a
streamwise mean pressure gradient. The subsequent calculations were
performed with a transverse (spanwise) mean pressure gradient of
$\partial P/\partial x_3 = 10\tau_{w0}/\delta$, where $\tau_{w0}$
is the mean wall shear stress of the unperturbed channel. The 3--D
channel simulations were run for 10$u_{\tau0}/\delta$ and averaged
over seven realisations, where $u_{\tau0}$ is the friction velocity of
the 2--D initial condition.

% boundary layer
For the boundary layer, the setup is identical to the one 
in section \ref{sec:transpiration}. The range of $\Rey_\theta$ is
from 1000 to 10 000. The length, height and width of the simulated box
are $L_x=1060\theta_\text{avg}$, $L_y=18\theta_\text{avg}$, and
$L_z=35\theta_\text{avg}$ with the streamwise and spanwise resolutions
of $\Delta_1/\delta = 0.05$ ($\Delta_1^+=118$) and $\Delta_3/\delta =
0.04$ ($\Delta_3^+=84.3$) at $\Rey_\theta\approx 6500$.  The grid is
slightly stretched in the wall-normal direction with minimum
$\Delta_2/\delta = 0.01$ ($\Delta_2^+ = 20.8$). The inlet, outlet, and
top boundary conditions are as in section \ref{sec:transpiration} with
$x_\text{ref}/\theta_0=890$.  For the wall, we impose the slip
boundary condition for WSIM.

% test filter
It is important to note the details of the filter operation, as
dynamic wall models are particularly sensitive to this choice (see
appendix \ref{sec:appendix}). Test filtering a variable $f$ in
a given spatial direction at point $i$ is computed as $1/6f(i-1) + 2/3
f(i) + 1/6 f(i+1)$ (Simpson's rule). The operation is repeated for all
three directions away from the wall. This corresponds to a
discrete fourth-order quadrature over a cell of size $2\Delta_1\times
2\Delta_2\times 2\Delta_3$ for a uniform grid. At the wall, the same
filtering operation is used in the horizontal directions while the
wall-normal filter is one-sided and given by $2/3f(1) + 1/3f(2)$, with
$f(1)$ and $f(2)$ denoting values at the first and second wall-normal
grid points. This is an integration over a cell of size
$2\Delta_1\times \Delta_2\times 2\Delta_3$.  Also, the definition of
the filter operation fixes the value of $\Delta_R$, which is the ratio
between the grid and test filter sizes at the wall. In this case, the
$\Delta_R$ based on the cell volume is given by
$\sqrt[3]{2\times1\times2}\approx 1.6$.

% list of cases
The cases for the 2--D and 3--D channel are labelled following the convention
([Channel type]-[Wall model]-[Reynolds number]-[Grid]), where the grid
labels G0, G1, and G2 given in table \ref{table:cases}
correspond to $320\times25\times120$ ($\Delta_1= \Delta_2 = \Delta_3=
0.080\delta $), $512\times40\times192$ ($\Delta_1= \Delta_2 = \Delta_3
=0.050\delta $), and $1024\times80\times384$ ($\Delta_1= \Delta_2 =
\Delta_3 = 0.025\delta$), respectively. The wall model applied are labelled NS,
EQWM, or WSIM.  Additional cases with
anisotropic grids, different values of $\Delta_R$,
test-filtering operations, or SGS models were run to study the
sensitivity of the model to these choices. They are discussed in
Appendix A but not included in the table.
\begin{table}
\begin{center}
\setlength{\tabcolsep}{10pt}
\begin{tabular}{l c c c} 
Grid label & $\Delta_1/\delta$ & $\Delta_2/\delta$ & $\Delta_3/\delta$\\
\hline
\hline
G0         & $0.080$           & $0.080$           & $0.080$          \\
G1         & $0.050$           & $0.050$           & $0.050$          \\
G2         & $0.025$           & $0.025$           & $0.025$          \\  
\hline
\end{tabular}
\end{center}
\caption{\label{table:cases} 
Grid resolutions in outer units. The first column
contains the label used to name LES cases for the 2--D and 3--D
channel flow simulations computed with different grids. The second,
third, and fourth columns are the streamwise, wall-normal, and
spanwise grid resolutions, respectively.}
%Tabulated list of cases for 2--D and 3--D channel flow simulations.} 
\end{table}

% comparison
The 2--D channel results are compared with DNS data from
\citet{Hoyas2006} and \citet{Lozano-Duran2014} for
$\Rey_\tau\approx2000$ and $4200$, \citet{Yamamoto2018} for
$\Rey_\tau\approx8000$, and with the law-of-the wall for
$\Rey_\tau>8000$. For the boundary layer, the resulting friction
coefficient is compared to the empirical $C_f$ from \cite{White2006},
and the mean velocity profiles are compared with the DNS data from
\citet{Sillero2013} at $\Rey_\theta \approx 6500$ and the experimental
data from \citet{Osterlund1999} at $\Rey_\theta \approx 8000$.

The performance of the WSIM in laminar flows has not
been studied.  However, in the limit of fine grids, the dynamic
procedure of WSIM should produce zero slip lengths and revert to the
no-slip boundary condition. Thus, for laminar cases with enough grid
resolution to resolve the near-wall structures, we expect the dynamic
wall model to naturally switch off.

%=====================================================================
\subsection{Statistically steady two-dimensional channel flow}
\label{sec:2d_channel_results}
%=====================================================================

% Definition of error
We assess the performance of WSIM compared to EQWM and NS. The results
are discussed in terms of the error in the streamwise mean velocity
profile across the logarithmic region. This
choice was necessary in order to include higher Reynolds number cases
where the corresponding DNS was not available and the law of the wall
is used instead. Restricting the error to be evaluated only in the
logarithmic layer is justified as wall models mainly impact the solution by
vertically shifting the mean velocity profile and do not alter its
shape for the range of grid resolutions tested as shown in section
\ref{sec:pred_log}. In particular, the error is measured as the
normalised $L_2$ error of the streamwise mean velocity between the
second grid point and $0.2\delta$,
\begin{equation} 
\mathcal{E} = \left[\frac{\int_{\Delta_2}^{0.2\delta}
\left(\langle \bar{u}_1\rangle - \langle
u_1^\text{DNS}\rangle\right)^2
\mathrm{d}x_2}{\int_{\Delta_2}^{0.2\delta} \left(\langle
u_1^\text{DNS}\rangle\right)^2 \mathrm{d}x_2}\right]^{1/2}.
\end{equation}
In the case where the corresponding DNS does not exist, $\langle
u_1^\text{DNS}\rangle$ is replaced by the law of the wall,
\begin{equation}
\langle u_1^{\text{DNS}+}\rangle = \frac{1}{\kappa} \log x_2^+ + B,
\end{equation}
with $\kappa = 0.392$ and $B = 4.48$ \citep{Luchini2017}.

\begin{figure} 
\begin{center}
\subfloat[]{\includegraphics[width=0.48\textwidth]{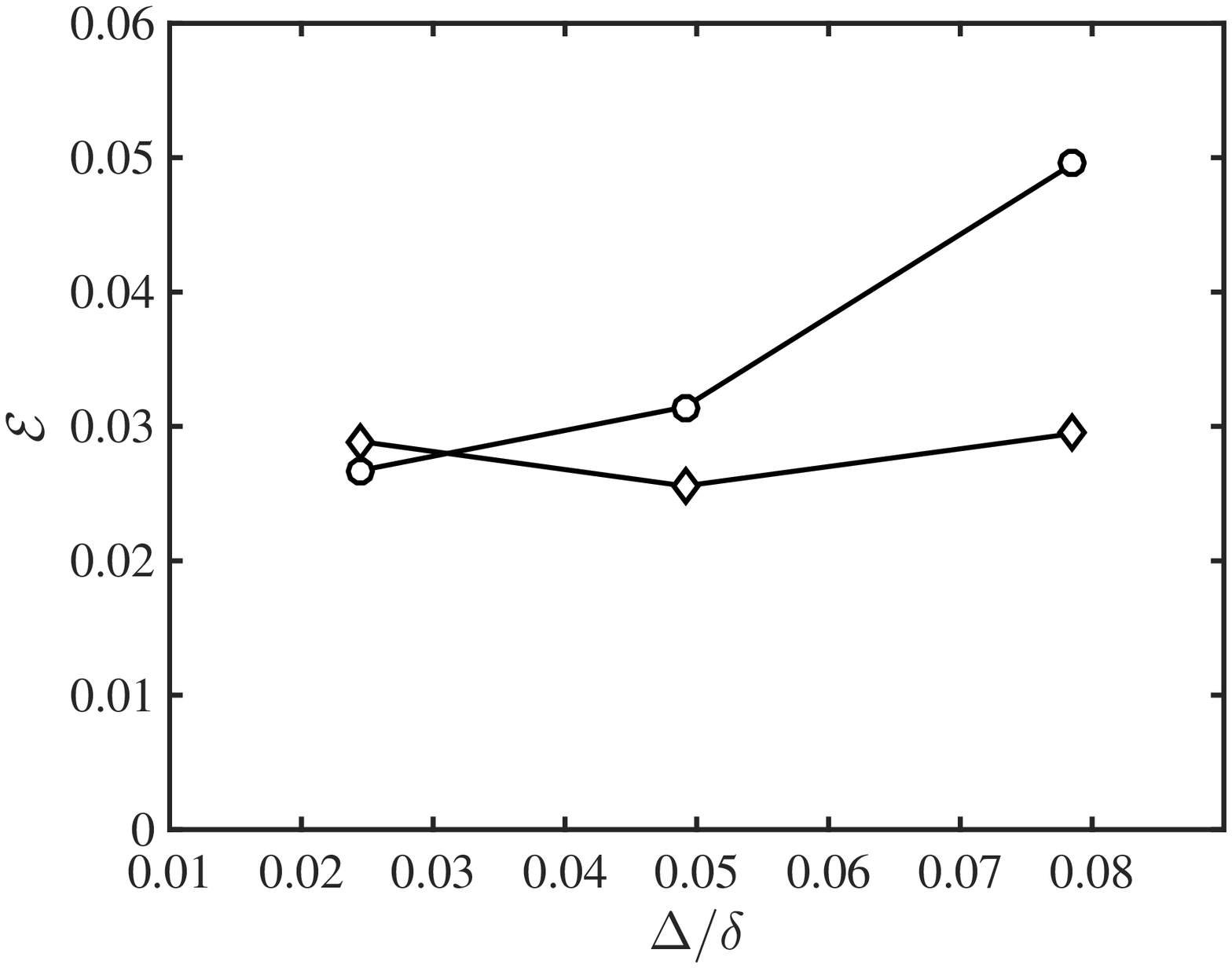}}
\hspace{0.2cm}
\subfloat[]{\includegraphics[width=0.48\textwidth]{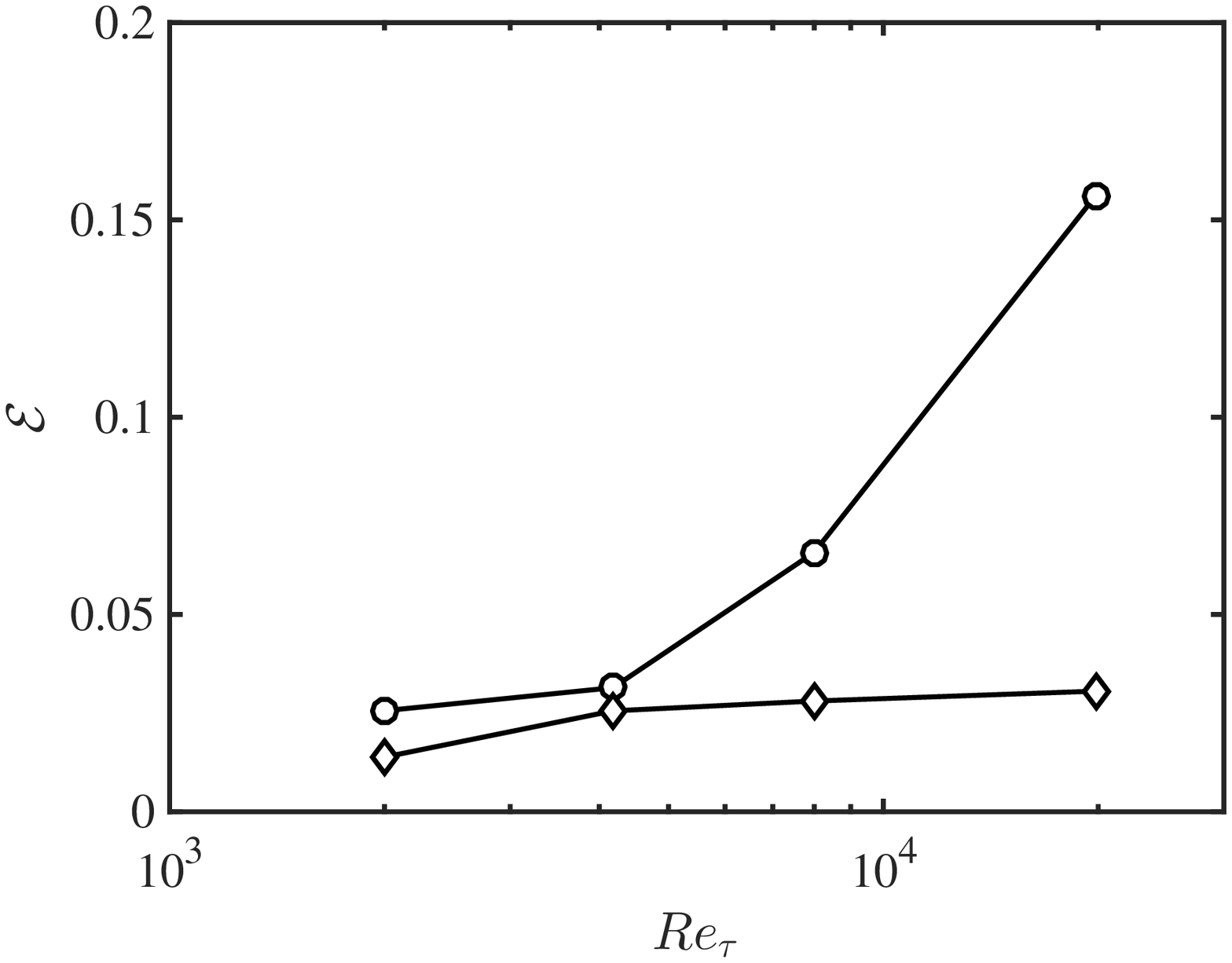}}\\
\subfloat[]{\includegraphics[width=0.48\textwidth]{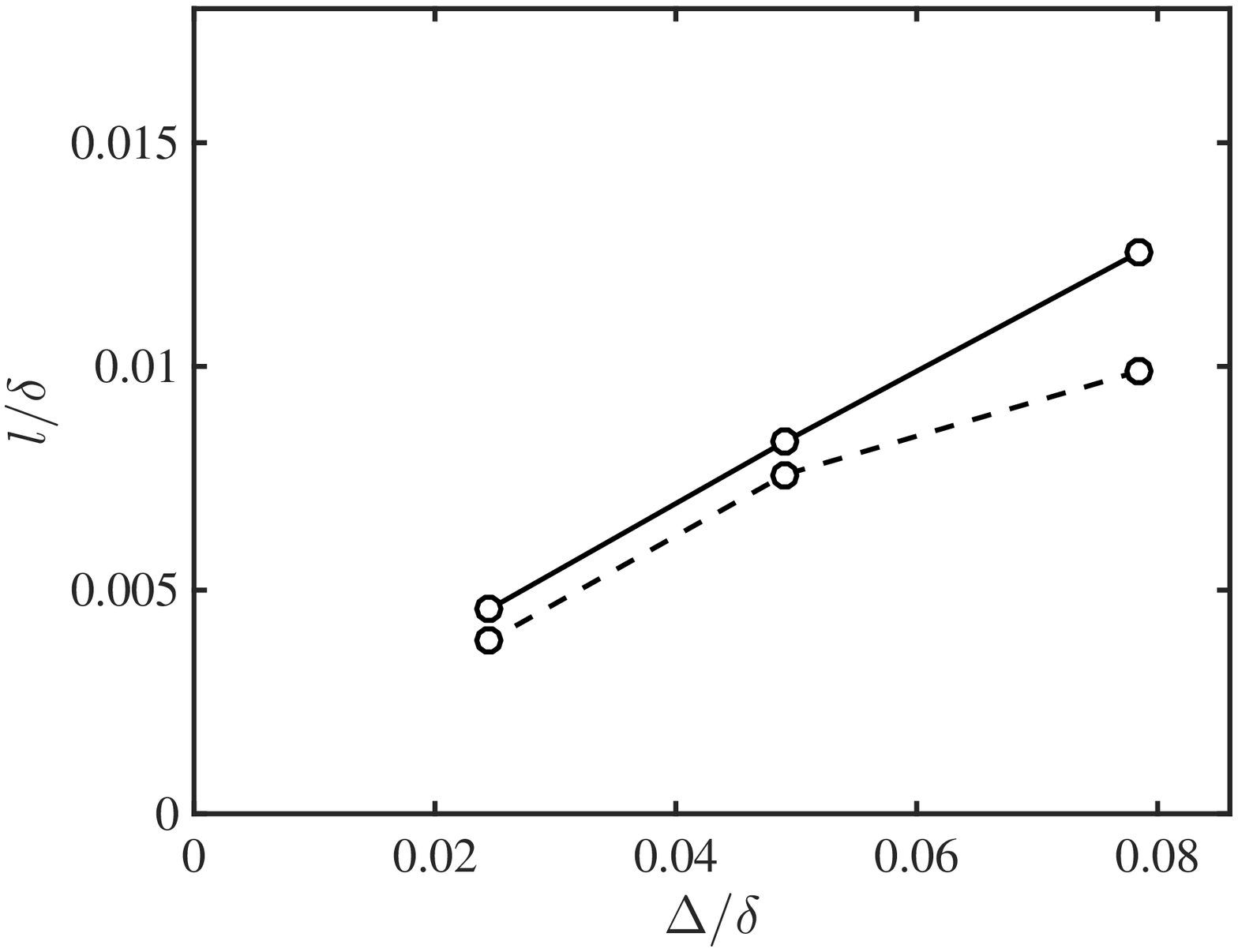}}
\hspace{0.2cm}
\subfloat[]{\includegraphics[width=0.48\textwidth]{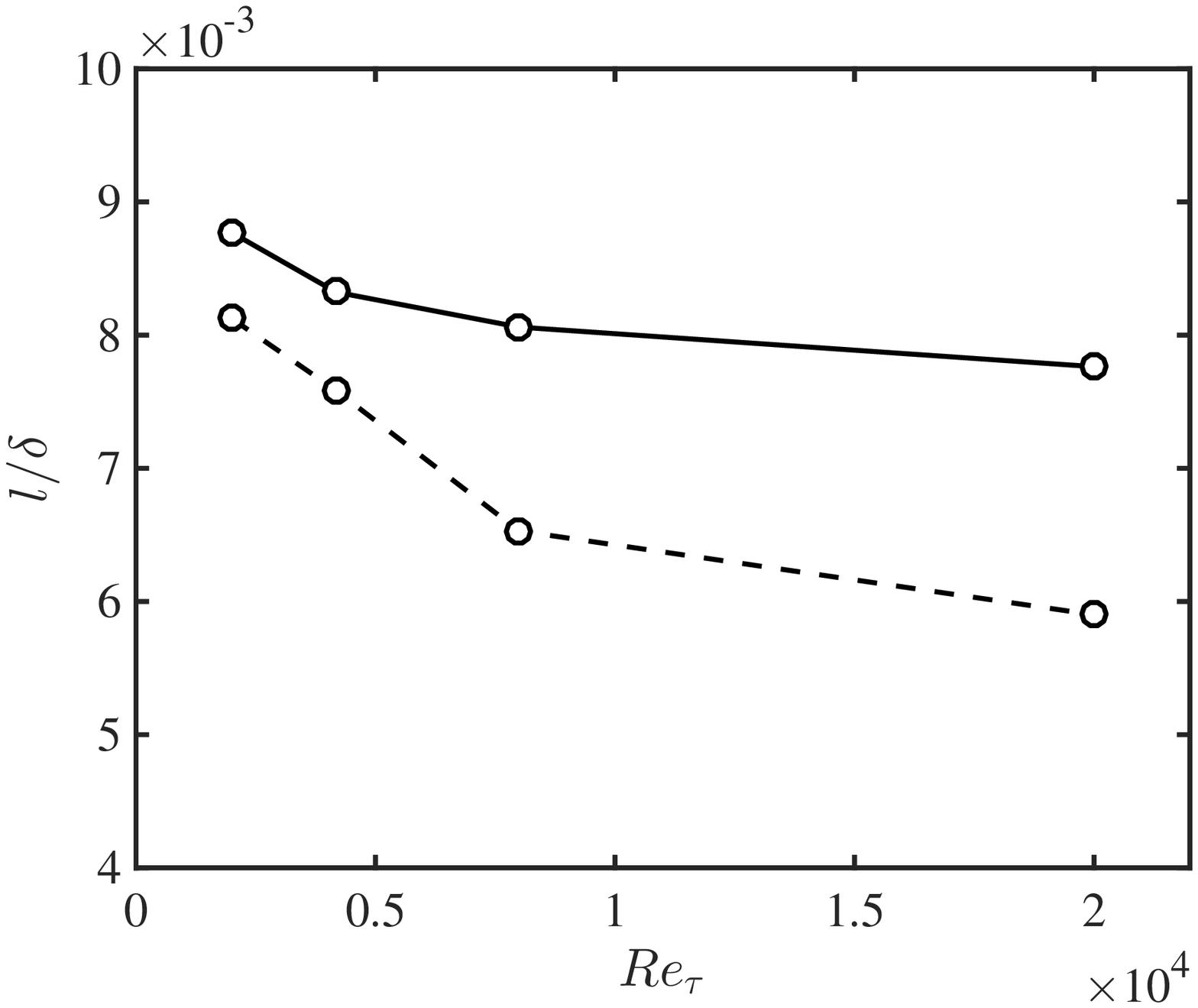}}
\caption{Error in the streamwise mean velocity profile, $\mathcal{E}$,
as a function of (a) grid size (for $\Rey_\tau\approx4200$) and (b) Reynolds
number (for grid G1). WSIM ($\circ$) and EQWM ($\lozenge$).  The
predicted slip lengths $l/\delta$ for WSIM (solid lines) and optimal
slip lengths (dashed lines) as a function of (c) grid resolution for
$\Rey_\tau\approx4200$ and (d) Reynolds number for grid G1.
\label{fig:errors}}
\end{center}
\end{figure}

\begin{figure}
\begin{center}
\subfloat[]{\includegraphics[width=0.48\textwidth]{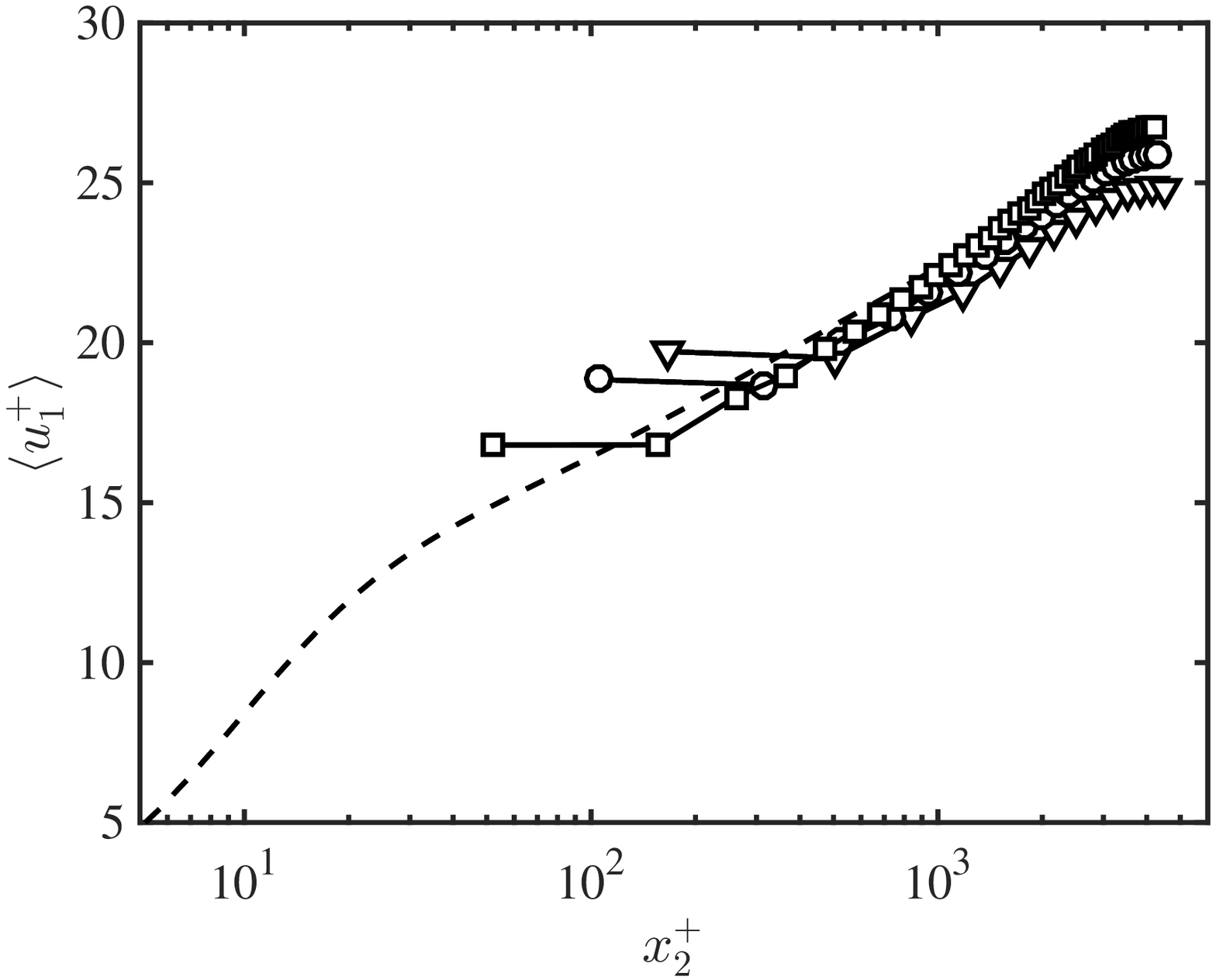}}
\hspace{0.2cm}
\subfloat[]{\includegraphics[width=0.48\textwidth]{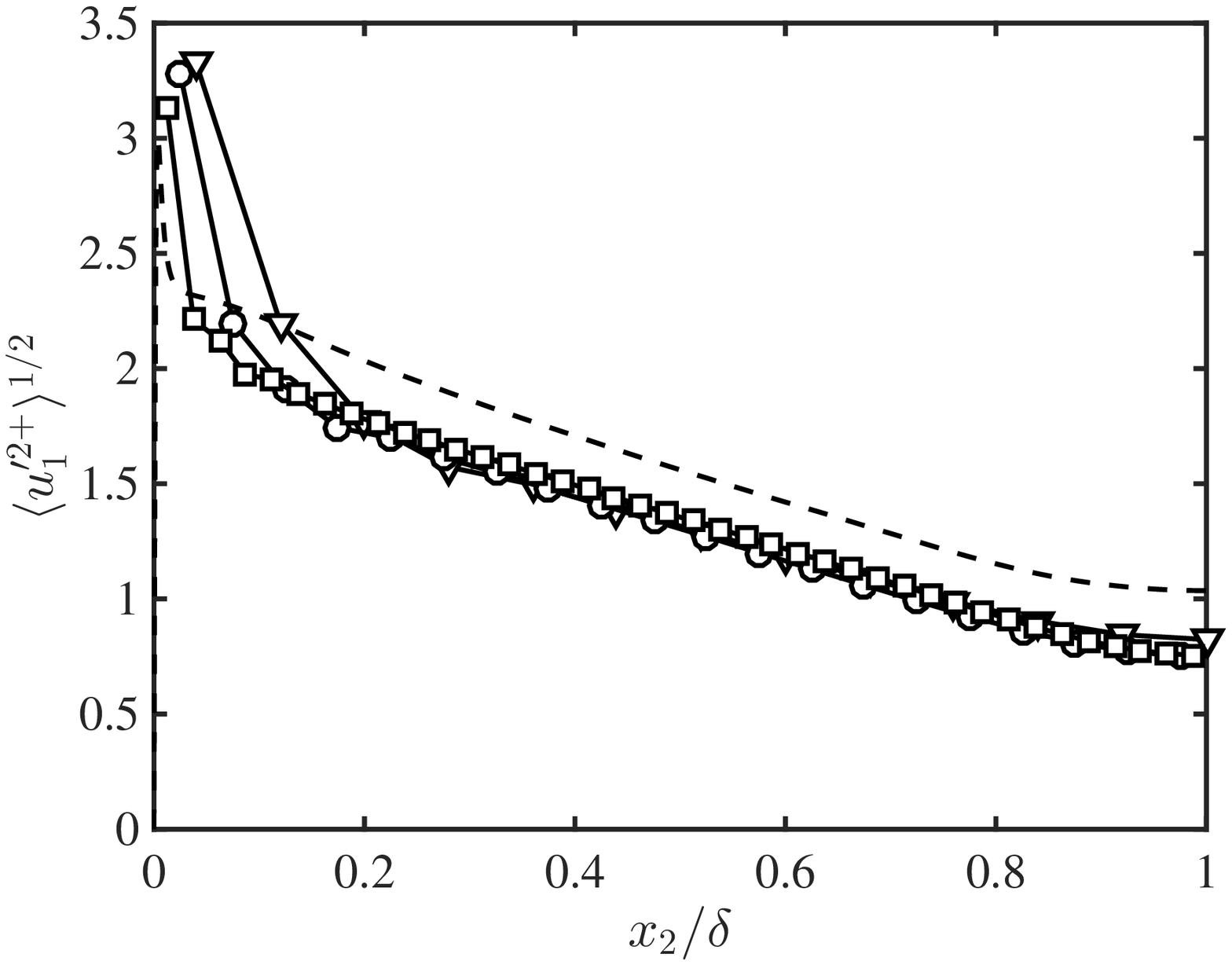}}\\
\subfloat[]{\includegraphics[width=0.48\textwidth]{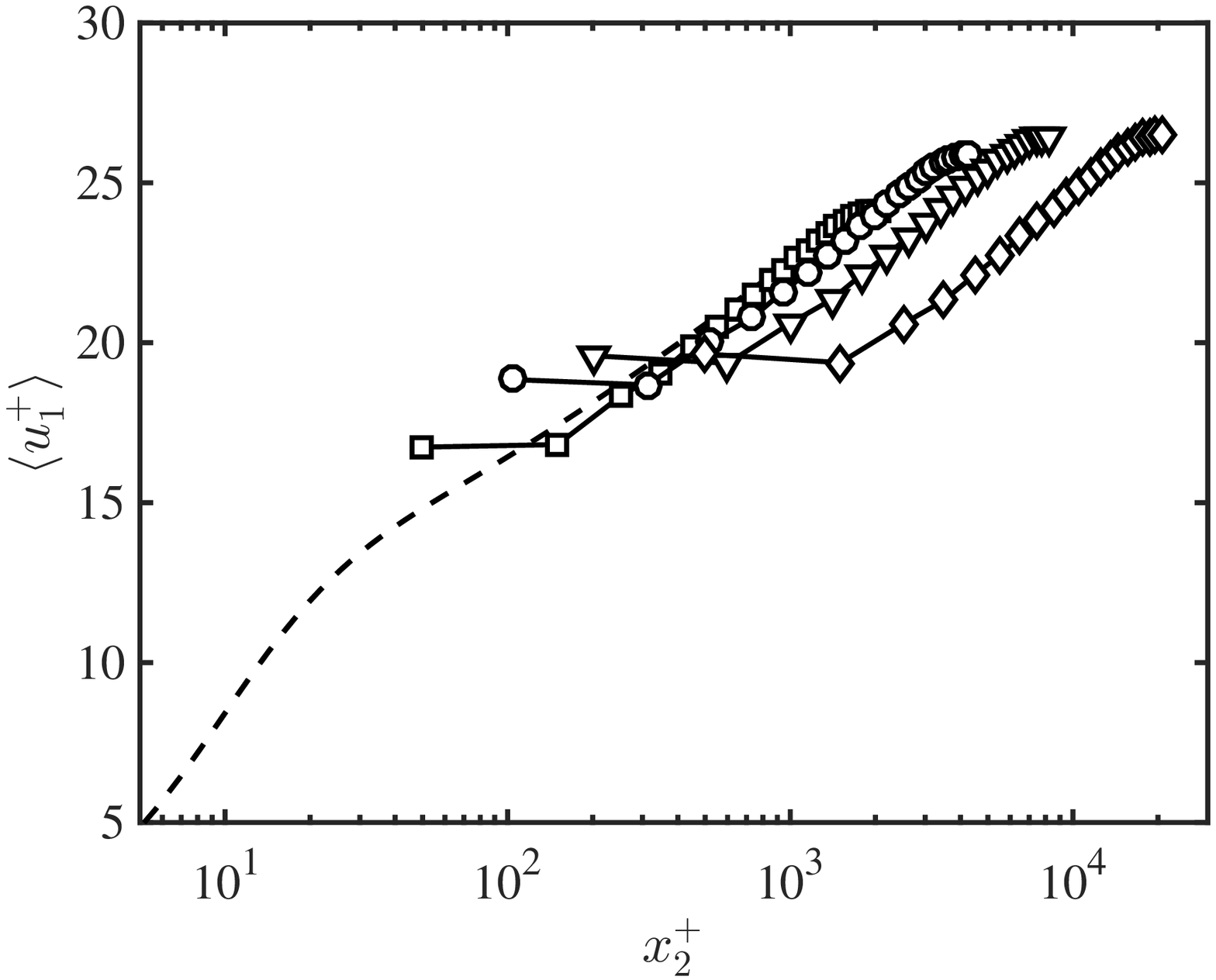}}
\hspace{0.2cm}
\subfloat[]{\includegraphics[width=0.48\textwidth]{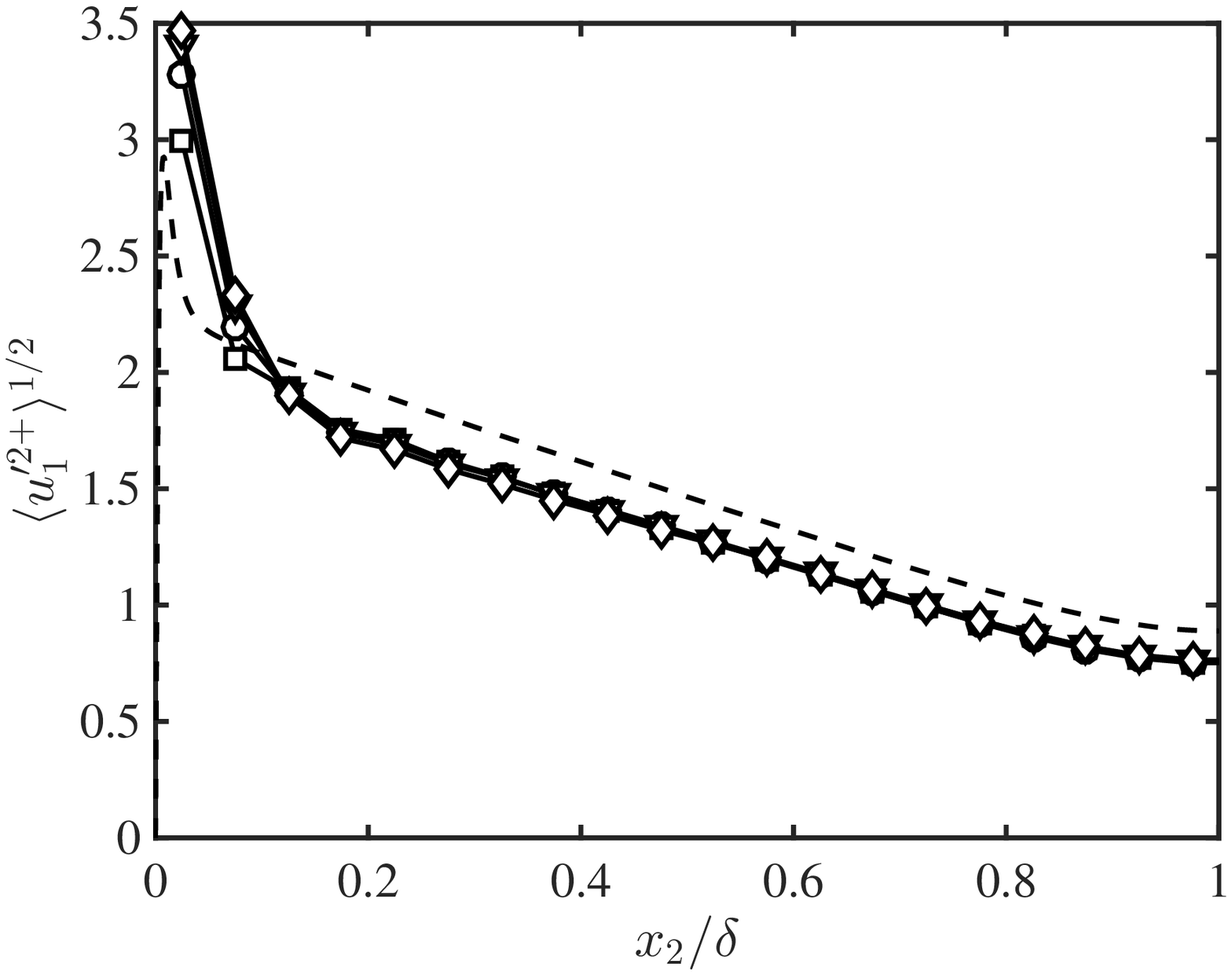}}
\caption{(a) Mean velocity profiles and (b) streamwise rms velocity
fluctuations for WSIM at $\Rey_\tau\approx4200$ for grid G0
($\triangledown$), G1 ($\circ$), and G2 ($\square$). DNS for
$\Rey_\tau\approx4200$ (dashed line). (c) Mean velocity profiles and
(d) streamwise rms velocity fluctuations for WSIM at
$\Rey_\tau\approx2000$ ($\square$), $4200$ ($\circ$), $8000$
($\triangledown$), and 20 000 ($\lozenge$) for grid G1. DNS for
$\Rey_\tau\approx4200$ (dashed line).
\label{fig:Umean}}
\end{center}
\end{figure}

% Errors
Figures \ref{fig:errors}(a) and (b) show $\mathcal{E}$ as a function
of grid resolution and Reynolds number. At
moderate Reynolds numbers ($\Rey_\tau < 8000$) and all grid
resolutions, the error for WSIM
($\mathcal{E}\approx$2--6\%) is similar to that of the
EQWM ($\mathcal{E}\approx$2--3\%). With increasing Reynolds number,
the performance degrades (up to $\mathcal{E}\approx$15\% at $\Rey_\tau
\approx$20 000), while the EQWM does not. The accurate
results for EQWM are not surprising as its modelling assumptions are
well satisfied for channel flow settings. The reason for the declining
performance of WSIM at very high Reynolds number can be found in
sections \ref{sec:theory} and \ref{sec:apriori}, where it was argued
that the underlying assumptions for the slip condition are invalidated
for large filter sizes. However, it is worth mentioning that the
errors for an LES with no wall model ($\mathcal{E}\approx$100\% for
2D-NS-4200-G1) are an order of magnitude larger than the errors of
WSIM for all cases. The mean velocity profiles and
streamwise rms velocity fluctuations for WSIM for various cases are
shown in figure \ref{fig:Umean}. Additional sensitivities to
anisotropic grids, different values of $\Delta_R$, test-filtering
operations, or SGS model are discussed in appendix
\ref{sec:appendix}.

% slip length prediction
The slip lengths predicted by WSIM are shown in figure
\ref{fig:errors}(c) and (d) as a function of grid resolution and
Reynolds number and compared to the optimal slip lengths
(\ref{eq:l_2}). It is remarkable that WSIM captures the overall
behaviour of the optimal slip lengths, that is, a strong dependence on
grid resolution and a weak variation with Reynolds number.
%
%-----------------------------------------------------------%
%-----------------------------------------------------------%
%
%-----------------------------------------------------------%
%-----------------------------------------------------------%
%

% Discussion for Boses' model
%Finally, we discuss the results for the model presented in
%\citet{Bose2014}. We tested three Reynolds numbers ($Re_\tau = 950,
%2000, 4200$), three grid resolutions ($\Delta_{x,y,z}/\delta =
%\Delta_i/\delta = 0.080, 0.050, 0.025$), two SGS models (AMD model and
%DSM), three different values for $\Delta_R$ (1.4, 1.6, 1.8), and test
%filters using Simpson's and trapazoidal rules. $T^{SGS}_{ij}$ was
%computed as in Section \ref{sec:new_model}. The ambiguity regarding
%the isotropic part of $\tau_{ij}^{SGS}$ mentioned in Section
%\ref{sec:new_model} was not discussed in \citet{Bose2014}, and this
%component is set to zero in our current implementation.  All of the
%cases simulated yielded an imaginary slip length (clipped to zero)
%when solving Eq. (\ref{eq:bose}), and hence, results from
%\citet{Bose2014} could not be reproduced. One certain deviance from
%\citet{Bose2014} is the numerical discretization (staggered
%second-order finite differences versus collocated second-order finite
%volumes). Other differences are the SGS model, test filter, and
%methodology to compute $T^{SGS}_{ij}$.

%=====================================================================
\subsection{Three-dimensional transient channel flow}
\label{sec:3d_channel_results}
%=====================================================================

% 3D channel flow
The performance of WSIM in non-equilibrium scenarios is assessed in a
three-dimensional transient channel flow \citep{Moin1990}. Note that
in general wall models cannot be assumed to be effective at
transferring information from the inner to the outer layer in
non-equilibrium flows.  Hence, the current flow set up, characterised
by a spanwise boundary layer growing from the wall due to viscous
effects, is expected to be problematic for wall-modelled LES. A plane
channel flow was modified to incorporate a lateral (transverse)
pressure gradient 10 times that of the streamwise pressure gradient.
The details of the simulations were given in section
\ref{sec:test_cases_wm}.

\begin{figure}
\begin{center}
\subfloat[]{\includegraphics[width=0.48\textwidth]{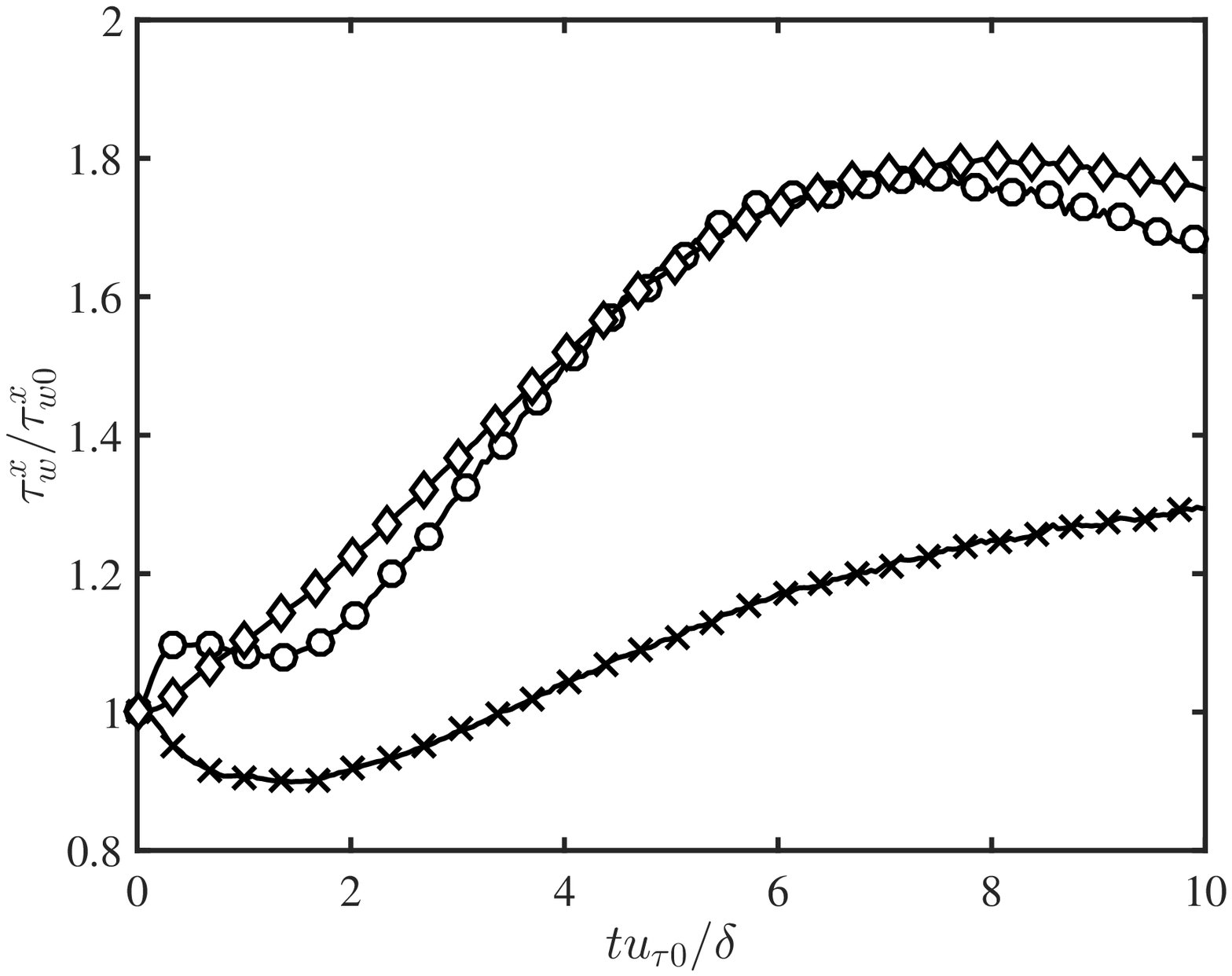}}
\hspace{0.4cm}
\subfloat[]{\includegraphics[width=0.48\textwidth]{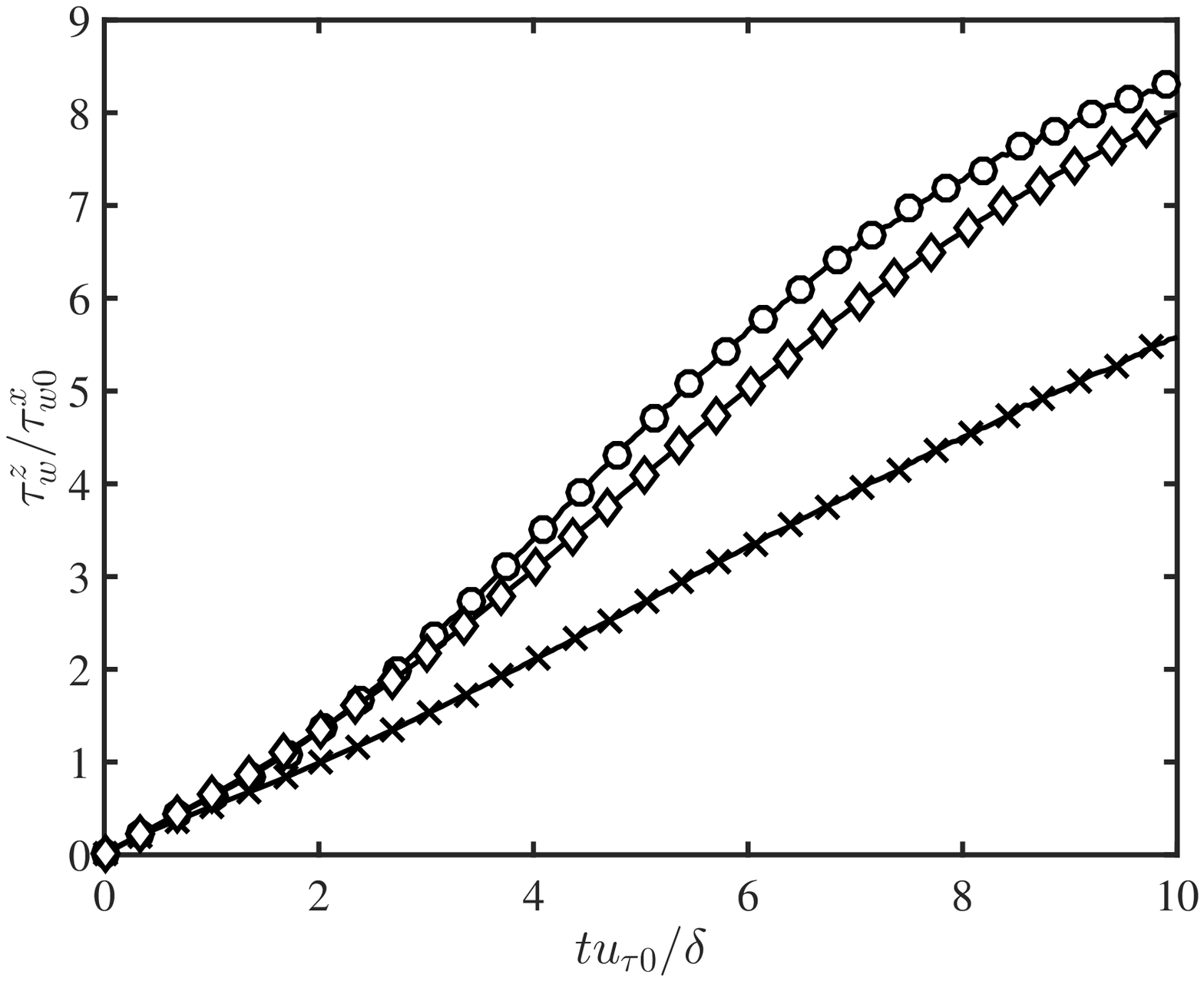}}
\caption{Wall stress in (a) streamwise and (b) spanwise directions as a
function of time for WSIM ($\circ$), EQWM ($\lozenge$), and NS
($\times$).
\label{fig:3Dchannel}}
\end{center}
\end{figure}

\begin{figure}
\begin{center}
\subfloat[]{\includegraphics[width=0.48\textwidth]{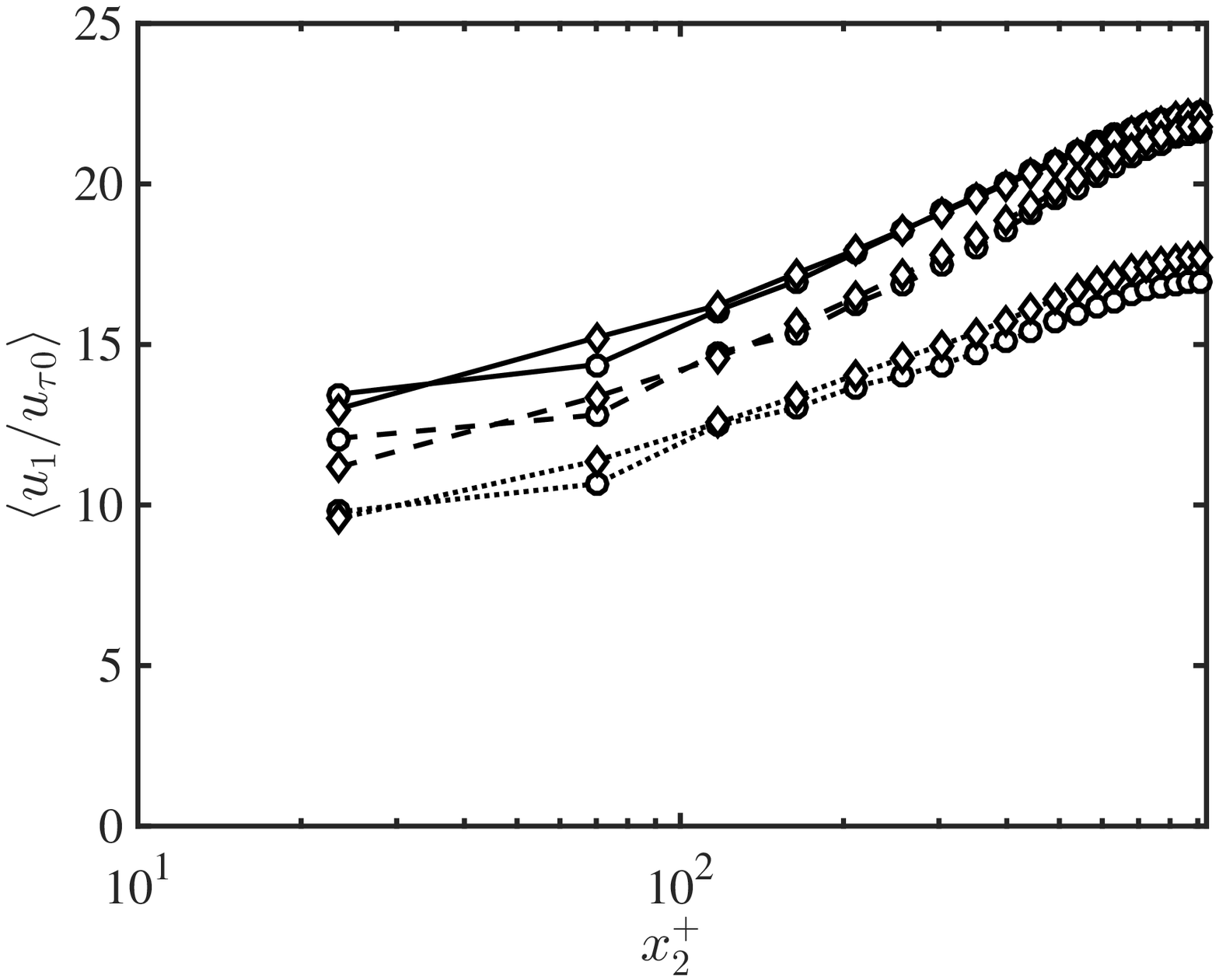}}
\hspace{0.4cm}
\subfloat[]{\includegraphics[width=0.48\textwidth]{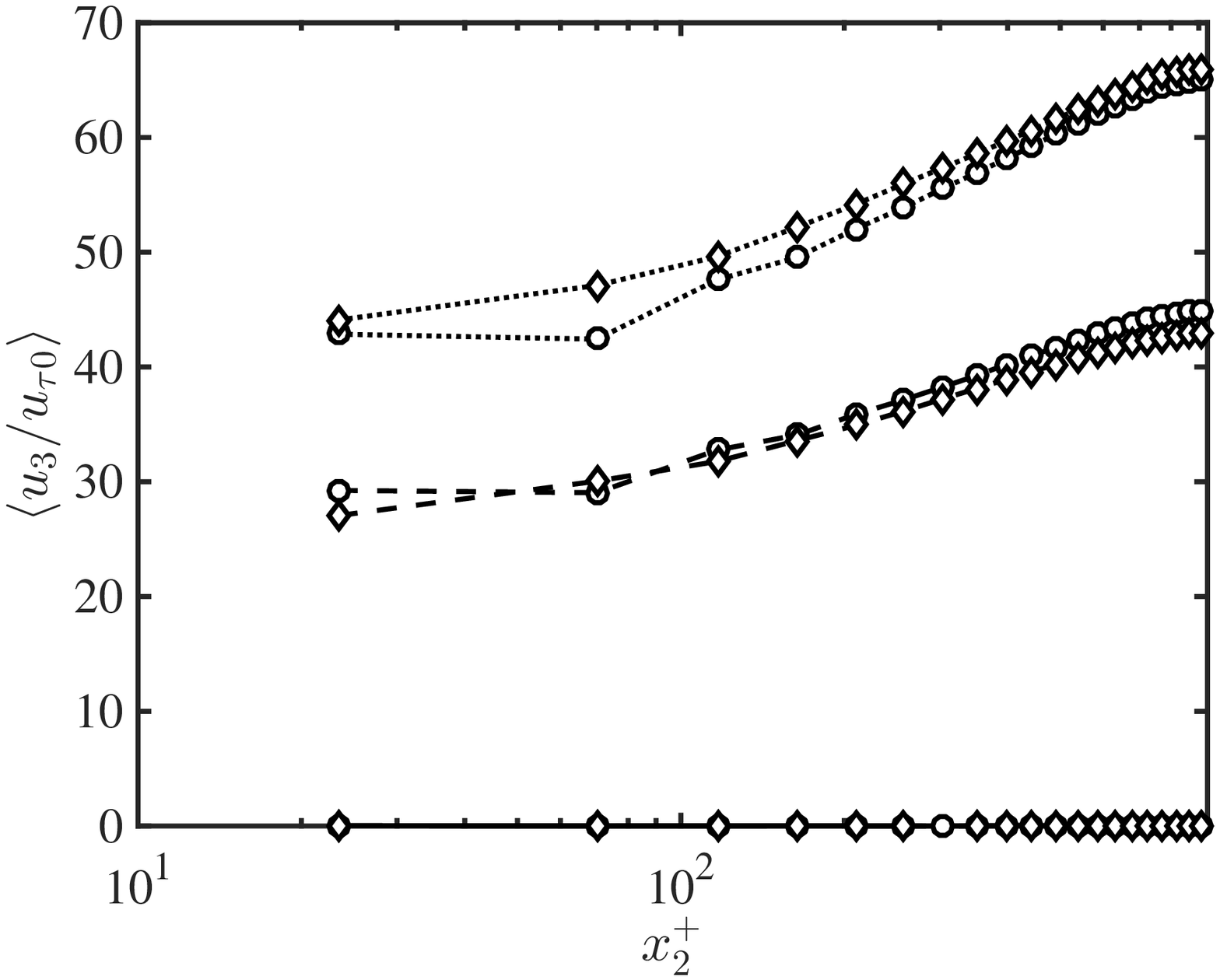}}
\caption{Mean (a) streamwise and (b) spanwise velocity profiles as a
function of $x_2/\delta$ at $tu_{\tau0}/\delta = 0$ (solid lines), 4.5
(dashed lines), and 9 (dotted lines) for WSIM ($\circ$) and EQWM
($\lozenge$).
\label{fig:3Dchannel_Umean}}
\end{center}
\end{figure}

% results
The wall models explored are WSIM and EQWM. A case with the no-slip
boundary condition is used for control, and the figure of merit is the
evolution of the streamwise and spanwise wall stress as a function of
time (figure \ref{fig:3Dchannel}).  Note that the
temporal increment of the wall stress magnitude involves an increase
of the Reynolds number from $Re_\tau \approx 932$ at $t = 0$ to
$Re_\tau \approx 2600$ at $t=10\delta/u_{\tau0}$. The results show that
the performance of WSIM is similar to the EQWM despite its
parameter-free nature.  The streamwise and spanwise mean velocity
profiles at various time instances are given in figure
\ref{fig:3Dchannel_Umean}, which also shows that both WSIM and EQWM
predict similar time evolutions. Although there is no reference DNS
available for the full time span of our simulations, in the limited
time range from $t = 0$ to $1\delta/u_{\tau0}$, \citet{Giometto2017}
showed that the EQWM predicts the evolution of the wall stresses with
less than 10\% deviation in the spanwise wall stress prediction from
DNS, and thus the results from WSIM are also expected to exhibit a
similar error.  Both the EQWM and WSIM entail a quantitative
correction to the prediction provide by the no-slip boundary
condition. Consequently, the computational simplicity and absence of a
secondary mesh makes WSIM an appealing approach at the cost of a
moderate attenuation of the predictive capabilities compared to more
sophisticated wall models. 
%
%-----------------------------------------------------------%
%-----------------------------------------------------------%
%
%-----------------------------------------------------------%
%-----------------------------------------------------------%

%=====================================================================
\subsection{Zero-pressure-gradient flat-plate turbulent boundary layer}
\label{sec:bl_results}
%=====================================================================

\begin{figure}
\begin{center}
\psfrag{X}[cc]{$\Rey_\theta$}
\psfrag{Y}[bc][bc][1][180]{$10^3Cf$}
%\psfrag{Y}[bc]{$10^3Cf$}
\includegraphics[width=0.75\textwidth]{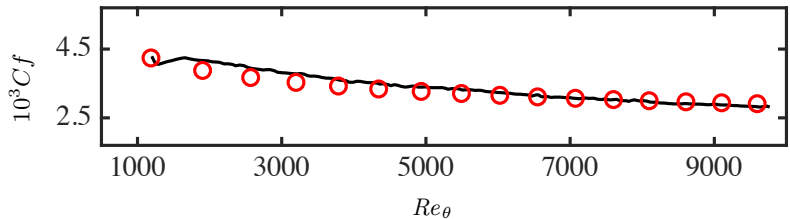}
\caption{Empirical friction coefficient from \citet{White2006} (red
$\circ$) and the friction coefficient from WSIM (black ---)
\label{fig:Cf_bl_wm}}
\end{center}
\end{figure}

\begin{figure}
\begin{center}
\centerline{
\psfrag{X}[cc]{$x_2/\delta$}
\psfrag{Y}[bc]{$\langle u_1^+\rangle$}
\subfloat[]{\includegraphics[width=0.48\textwidth]{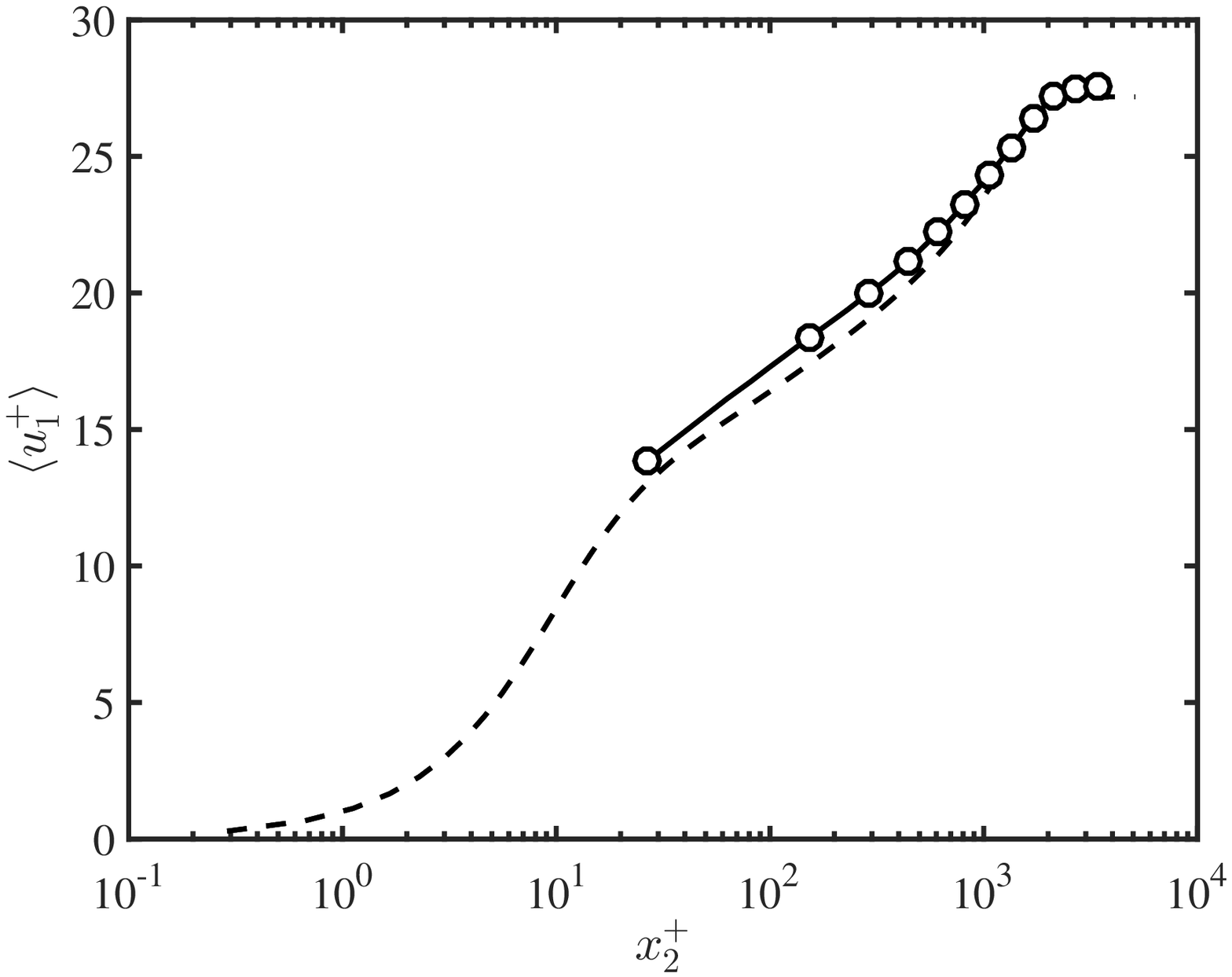}}
\hspace{0.3cm}
\psfrag{X}[cc]{$x_2/\delta$}
\psfrag{Y}[bc]{$\langle u_1^+\rangle$}
\subfloat[]{\includegraphics[width=0.48\textwidth]{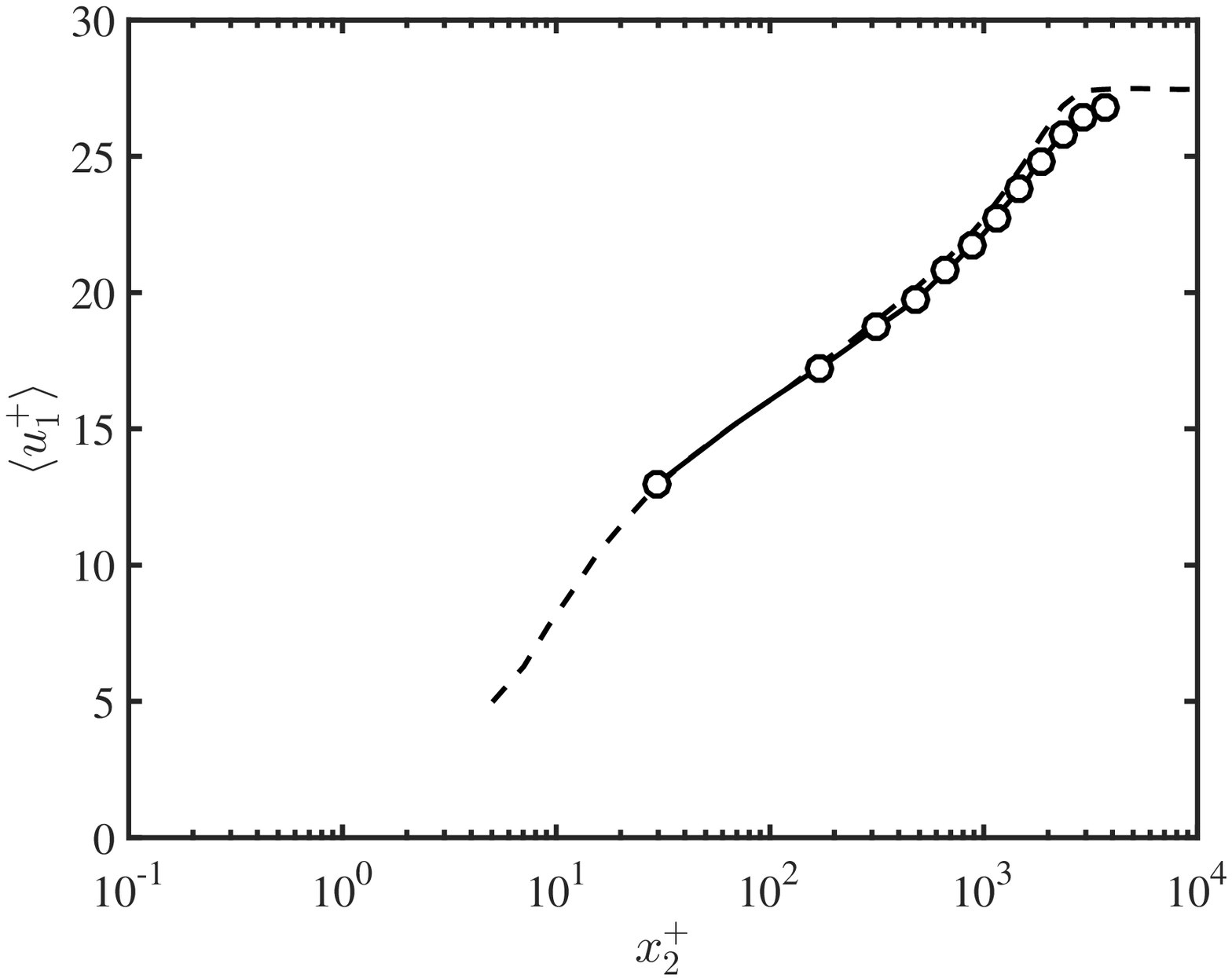}}
}
\centerline{
\psfrag{X}[cc]{$x_2/\delta$}
\psfrag{Y}[bc]{$\langle u_1'^{2+}\rangle^{1/2}$}
\subfloat[]{\includegraphics[width=0.48\textwidth]{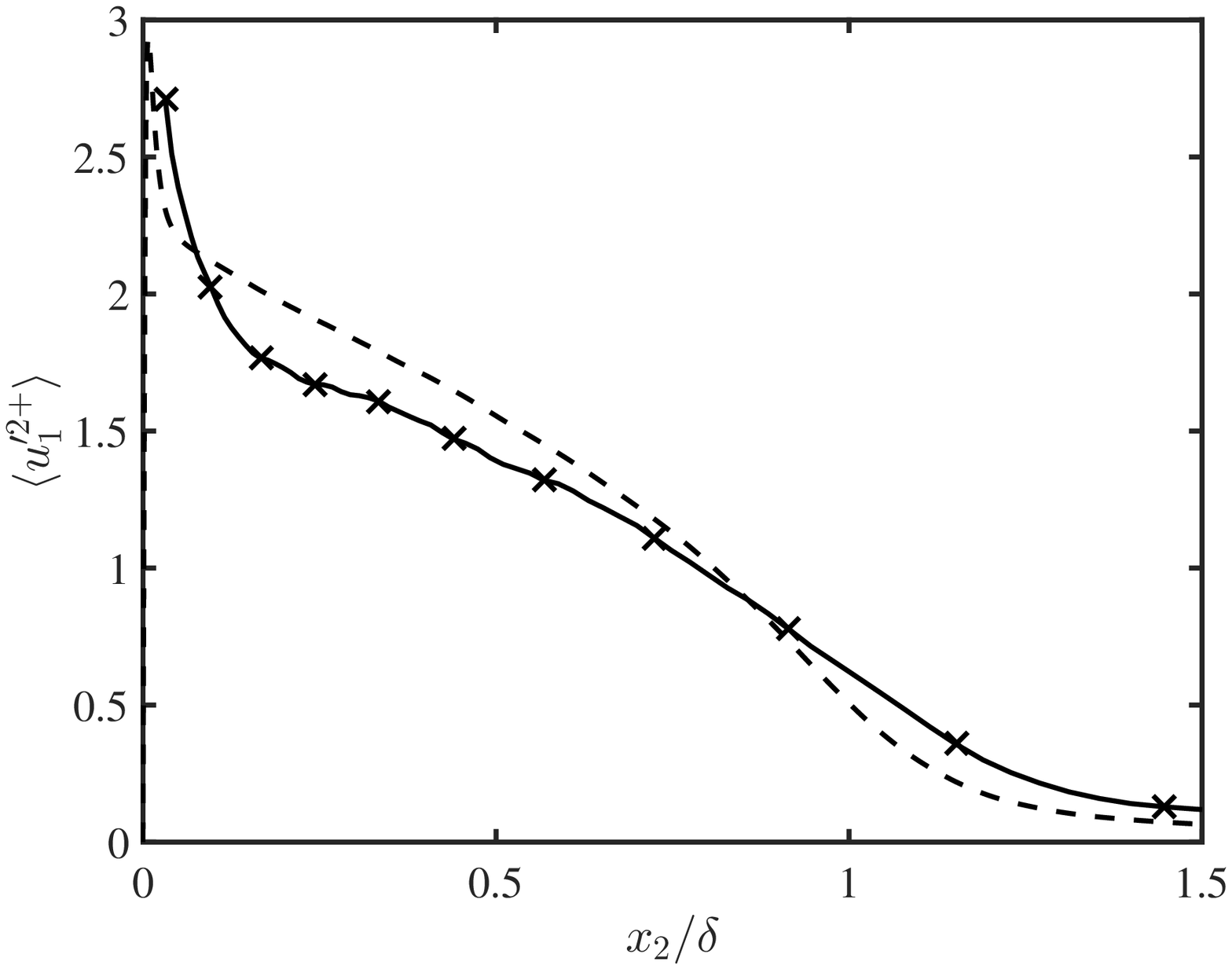}}
\hspace{0.3cm}
\psfrag{X}[cc]{$x_2/\delta$}
\psfrag{Y}[bc]{$\langle u_{2,3}'^{2+}\rangle^{1/2}$}
\subfloat[]{\includegraphics[width=0.48\textwidth]{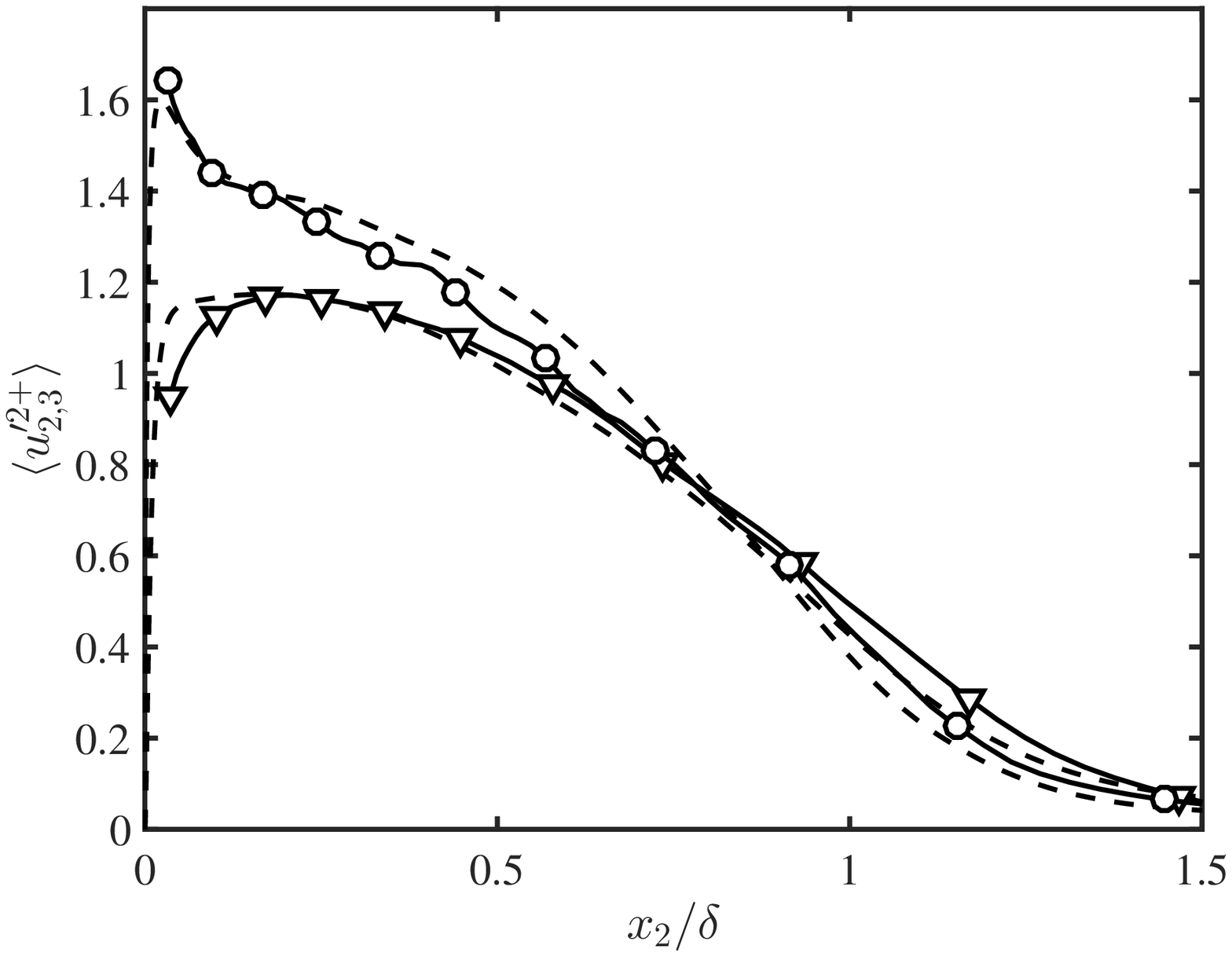}}
}
\caption{Mean streamwise velocity profile for (a)
$\Rey_\theta\approx 6500$ and (b) $\Rey_\theta\approx 8000$, and the
rms (c) streamwise ($\times$), (d) spanwise ($\circ$), and wall-normal
($\triangledown$) fluctuation profiles at $\Rey_\theta \approx 6500$.
WSIM (symbols) and DNS \citep{Sillero2013} or experiment
\citep{Osterlund1999} (dashed line). \label{fig:stats_bl_wm}}
\end{center}
\end{figure}

% Boundary layer
Finally, the performance of WSIM is assessed in a flat-plate turbulent
boundary layer. The details of the numerical set-up were discussed in
section \ref{sec:test_cases_wm}.  The friction
coefficient is shown in figure \ref{fig:Cf_bl_wm} from
$\Rey_\theta\approx1000$ to 10 000.  Note that the recycling scheme of
\citet{Lund1998} imposes an artificial boundary condition at the inlet,
requiring an initial development region for the flow to fully adapt to
the slip boundary condition, which is the reason for the discrepancy
in $C_f$ near the inlet. Consistent with previous test cases, the
results show that WSIM predicts the friction coefficient well within
$4\%$ error for $\Rey_\theta > 6000$. The mean streamwise velocity
profile at $\Rey_\theta \approx 6500$ and $8000$ and the rms velocity
fluctuations at $\Rey_\theta \approx 6500$ are also well predicted as
reported in figure \ref{fig:stats_bl_wm}.
%
%-----------------------------------------------------------%
%-----------------------------------------------------------%
%
%-----------------------------------------------------------%
%-----------------------------------------------------------%

%=====================================================================
\section{Conclusions}\label{sec:conclusions}
%=====================================================================

% intro
Due to the scaling of grid resolution requirements in DNS and
wall-resolved LES, wall-modelled LES stands as the most viable
approach for most engineering applications. In most existing
wall models, the Dirichlet no-slip boundary condition at the wall is
replaced by a Neumann and no-transpiration conditions in the
wall-parallel and wall-normal directions, respectively. In this study,
we have investigated the efficacy of the Robin (slip) boundary
condition, where the velocities at the wall are characterised by the
slip lengths and slip velocities. One novel aspect of this boundary
condition is the non-zero instantaneous wall-normal velocity at the
wall, i.e., transpiration, that opens a new avenue to model near-wall
turbulence in LES.  We have also presented a new dynamic slip wall
model, WSIM, that is free from \emph{a priori} tunable RANS
parameters, which most traditional wall models for LES rely on. The
model is based on the invariance of the wall stress under test
filtering and is effectively applied through the slip boundary
condition.

% Motivation 
We have provided theoretical support for the use of the slip condition
in wall-modelled LES instead of the widely applied Neumann boundary
condition with no-transpiration. \emph{A priori} testing was performed
to assess the validity of the slip condition in the context of
filtered DNS data.
 
% Advantages 
The slip boundary condition was implemented in LES of channel flow in
order to gain a better insight into its capabilities and shortcomings.
One of the key properties, made possible by transpiration, is that the
correct wall stress can always be achieved by an appropriate
combination of slip lengths. This property is crucial when the grid
resolution in the near-wall region does not capture the buffer and
logarithmic layer dynamics, which may result in an under- or
over-prediction of the wall stress.  We have derived the consistency
conditions for coupling the wall stress with the slip boundary
condition in channel flows and flat-plate boundary layers, and showed
that such constraints are sufficient to guarantee the correct wall
stress. Another advantage emanates from the non-zero Reynolds stress
at the wall. This is not only consistent with the filtered velocity
fields, but also alleviates the well-known problem of wall-stress
under-estimation by commonly used SGS models. We have also assessed
the sensitivities of one-point statistics to grid refinements, changes
in $\Rey_\tau$, and different SGS models. The role of imposing zero
mean mass flow through the wall by proper calculation of the slip
parameters has also been shown to be a key component of the model.

% testing
Finally, we have tested the performance of a dynamic slip wall model,
WSIM, in an LES of a plane turbulent channel flow at various Reynolds
numbers and grid resolutions. The model was able to correctly capture
the overall behaviour of the optimal slip length for a wide range of
grid resolutions and Reynolds numbers. The results have been compared
with those from the EQWM and the no-slip boundary condition. In all
cases, WSIM performed substantially better than the no-slip, and the
predictive error in the mean velocity profile was found to be below
10\% for $\Rey_\tau <$ 10 000 and all grid resolutions investigated.
The model was also tested for an LES of three-dimensional transient
channel flow, where the performance was similar to that of the EQWM,
and zero-pressure-gradient flat-plate turbulent boundary layer at
$\Rey_\theta$ up to 10 000, where the error in the friction
coefficient was less than 4\%.
  
%However, the model is
%  sensitive to the choice of SGS model and numerical filter operators,
%  and alternate formulations of the wall-stress-invariant condition
%  may be required to improve robustness. Moreover, the application of
%  spatial test filtering may deter the extension of the model to
%  complex geometries, and additional advancements are necessary to
%  formulate a robust dynamic wall model with applications to complex
%  geometry.

The present work has established the foundations and underlying principles 
for using the slip boundary condition for dynamic wall-modelled LES, free 
of \emph{a priori} tunable parameters. 
Based on the principles laid out in this paper, new formulations of the 
wall-stress-invariant condition can be explored to account for the 
dependency on SGS models and numerical filtering operations documented in 
appendix \ref{sec:appendix}.

\begin{appendices}
\section{Wall-stress invariant model: additional sensitivity analysis}
\label{sec:appendix}
% sensitivity
Four additional cases were computed to analyse the sensitivity of the
WSIM to $\Delta_R$, grid anisotropy, shape of the test filter, and
choice of SGS model. The effect of $\Delta_R$ turned out to be
negligible for the plausible range of values $\Delta_R = [1.4,1.8]$,
and the measured difference in $\mathcal{E}$ was less than 1\%.
Regarding grid anisotropy, coarsening case 2D-WSIM-4200-G1 by a factor
of two in the streamwise and spanwise directions had a negligible
effect on $\mathcal{E}$. While coarsening in both the streamwise and
spanwise directions simultaneously by a factor of two had a larger
effect with $\mathcal{E}$ increasing to $\approx$8\%. The error trend
for anisotropic grids also follows the results shown in figure
\ref{fig:errors}(a) when scaled with grid size based on cell volume,
$\Delta=(\Delta_1\Delta_2\Delta_3)^{1/3}$. This shows that the wall
model is robust to mild grid anisotropies.  

On the contrary, the test filter shape and SGS model highly impacted
the prediction of the mean flow. Case 2D-WSIM-4200-G1 was repeated
using test filter based on the trapezoidal rule, and the error
increased from 2.5\% to 32\%.  When 2D-WSIM-4200-G1 was run using the
anisotropic minimum-dissipation (AMD) model \citep{Rozema2015}, the
stress provided by the SGS model $\tau_{12}^\text{SGS}$ was larger
than $u_\tau^2$, and the slip length prediction by WSIM was clipped to
zero due to the excess of wall stress, reverting the boundary
condition to no-slip. Although this is consistent with the fact that
$\tau_{12}^\text{SGS}>u_\tau^2$ (see figure
\ref{fig:nu_t}a), it also implies that the correct stress at the
wall can never be obtained through the slip boundary condition with a
single slip length in this case. It was shown in section
\ref{sec:first_order_stat} that the slip lengths in the wall-normal
direction must be larger than the wall-parallel ones in order to drain
the excess of stress supplied by the SGS model. This suggests that
WSIM should be generalised to a formulation with a different slip
length in each spatial direction to overcome this limitation. It also
remains to study the near wall behaviour of various SGS models in the
wall-modelled grid limits in more detail. 
%
%-----------------------------------------------------------%
\begin{figure}
\vspace{0.2cm}
\centering
\includegraphics[width=0.48\textwidth]{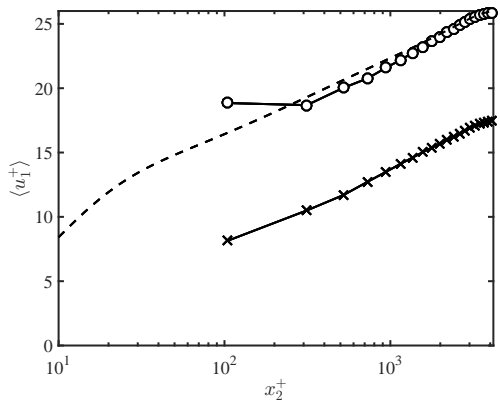}
\caption{Mean streamwise velocity profile for WSIM for filter using
Simpson's rule ($\circ$) and trapezoidal rule ($\times$). DNS
(dashed line). See section \ref{sec:test_cases_wm} for details
regarding the test filter operators.} 
\end{figure}
%-----------------------------------------------------------%

\end{appendices}

%=====================================================================
\section*{Acknowledgement}
%=====================================================================
This work was supported by NASA under the Transformative Aeronautics
Concepts Program, grant no. NNX15AU93A. We are thankful to Kevin
P. Griffin for his helpful comments on the manuscript.
%=====================================================================

\bibliography{slip_wm}
\bibliographystyle{jfm}

\end{document}